%% file: workflow_iga_fem_embeddedmesh_arxiv.tex
\documentclass[a4paper]{article}
\usepackage{setspace}

\usepackage[pages=all, color=black, position={current page.south}, placement=bottom, scale=1, opacity=1, vshift=5mm]{background}
\SetBgContents{
}      

\usepackage[margin=1in]{geometry} 

\usepackage{amsmath}
\usepackage{amsthm}
\usepackage{amssymb}

\usepackage{import}

\usepackage[utf8]{inputenc}
\usepackage{hyperref}
\hypersetup{
	unicode,
	pdfauthor={Author One, Author Two, Author Three},
	pdftitle={An Embedded Mesh Approach for Isogeometric Boundary Layers in Contact Mechanics},
	pdfsubject={An Embedded Mesh Approach for Isogeometric Boundary Layers in Contact Mechanics},
	pdfkeywords={article, template, simple},
	pdfproducer={LaTeX},
	pdfcreator={pdflatex}
}

\usepackage{graphicx, color}
\usepackage{subcaption}
\usepackage{multirow}
\usepackage{xcolor}

\usepackage{algorithm, algpseudocode} 
\usepackage{mathrsfs} 

\usepackage{enumitem} 

\usepackage{lipsum}

\usepackage{caption,subcaption}
\usepackage{booktabs}

\usepackage{tikz} 
\usepackage{pgfmath}
\usepackage{pgfplots}   
\usetikzlibrary{calc}
\usetikzlibrary{shapes.geometric, positioning, math}  

\newif\ifreviewerversion
\reviewerversionfalse   

\newcommand{\responsetoview}[1]{%
    \ifreviewerversion
        \textcolor{blue}{#1}%
    \else
        #1%
    \fi
}

\theoremstyle{remark}
\newtheorem{remark}{Remark}[section] 

\makeatletter
\def\th@remark{%
  \normalfont 
  \thm@headfont{\bfseries} 
}
\makeatother

\title{An Embedded Mesh Approach for Isogeometric Boundary Layers in Contact Mechanics}
\author{Eugenia Gabriela Loera Villeda $^a$ 
		\and Ivo Steinbrecher $^a$\footnote{Corresponding author: \\Email addresses: \texttt{gabriela.loera@unibw.de} (Eugenia Gabriela Loera Villeda), \texttt{ivo.steinbrecher@unibw.de} (Ivo Steinbrecher), \texttt{alexander.popp@unibw.de} (Alexander Popp)} \and Alexander Popp$^a$}

\date{
	$^a$Institute for Mathematics and Computer-Based Simulation,\\
	University of the Bundeswehr Munich, Werner-Heisenberg-Weg 39, 85577, Neubiberg, Bavaria, Germany%
}

\begin{document}
	\input{macros}
	
	\maketitle
	
	\input{section/abstract_arxiv}

	\input{section/introduction}

	\graphicspath{{fig/discretization_workflow}}
	\input{section/discretization_workflow}

	\input{section/problem_formulation}

	\graphicspath{{fig/spatial_discretization}}
	\input{section/spatial_discretization}

	\graphicspath{{fig/numerical_examples}}
	\input{section/numerical_examples}

	\input{section/concluding_remarks}

	\input{section/acknowledgements}

	\input{section/declaration_generative_ai}

	\bibliography{workflow_iga_fem_embeddedmesh_arxiv}
	\bibliographystyle{ieeetr}

\end{document}

%% file: macros.tex
\newcommand{\cf}{c.f.}
\newcommand{\dof}{DOF}
\newcommand{\dofs}{DOFs}
\newcommand{\eg}{e.g.,}
\newcommand{\etal}{et al.}
\newcommand{\ie}{i.e.,}
\newcommand{\notApplicable}{None}
\newcommand{\vs}{vs.}
\newcommand{\fem}{FEM}
\newcommand{\iga}{IGA}
\newcommand{\cad}{CAD}
\newcommand{\nurbs}{NURBS}
\newcommand{\brep}{B-rep}

\newcommand{\xvector}{\ensuremath{\mathbf{e}_{1}}}
\newcommand{\yvector}{\ensuremath{\mathbf{e}_{2}}}
\newcommand{\zvector}{\ensuremath{\mathbf{e}_{3}}}
\newcommand{\xaxis}{\ensuremath{x}}
\newcommand{\yaxis}{\ensuremath{y}}
\newcommand{\zaxis}{\ensuremath{z}}

\newcommand{\slaveindex}{\mathcal{S}}
\newcommand{\masterindex}{\mathcal{M}}
\newcommand{\layerindex}{\mathcal{L}}
\newcommand{\backgroundindex}{\mathcal{B}}
\newcommand{\interfaceindex}{\mathcal{I}}
\newcommand{\cuteleindex}{\mathcal{C}}
\newcommand{\restofDOFSindex}{\mathcal{N}}
\newcommand{\contactsubscript}{\mathrm{co}}

\newcommand{\kronecker}[1]{\ensuremath{\delta_{#1}}}
\newcommand{\bodyref}[1]{\ensuremath{\Omega^{(#1)}_0}}
\newcommand{\realspace}{\ensuremath{\mathbb{R}}}
\newcommand{\defmapping}[1]{\ensuremath{\mathbf{\varphi}^{(#1)}}}
\newcommand{\bodycurrent}[1]{\ensuremath{\Omega^{(#1)}}}
\newcommand{\bodysurfref}[1]{\ensuremath{\partial \Omega^{(#1)}_0}}
\newcommand{\pospointref}[1]{\ensuremath{\mathbf{X}^{#1}}}
\newcommand{\pospointcurrent}[1]{\ensuremath{\mathbf{x}^{#1}}}
\newcommand{\dispvector}[1]{\ensuremath{\mathbf{u}^{(#1)}}}
\newcommand{\dispscalar}[2]{\ensuremath{{u}_{#1}^{#2}}}
\newcommand{\dddispvector}[1]{\ensuremath{\mathbf{\ddot{u}}^{(#1)}}}
\newcommand{\prescribeddisp}[1]{\ensuremath{\mathbf{\hat{u}}^{(#1)}}}
\newcommand{\firstPK}[1]{\ensuremath{\mathbf{P}^{(#1)}}}
\newcommand{\densityref}[1]{\ensuremath{\rho^{(#1)}_{0}}}
\newcommand{\secondPK}[1]{\ensuremath{\mathbf{S}^{#1}}}
\newcommand{\defgrad}[1]{\ensuremath{\mathbf{F}^{(#1)}}}
\newcommand{\bodyforce}[1]{\ensuremath{\mathbf{b}^{(#1)}_0}}
\newcommand{\anoutwardsnormalvec}[1]{\ensuremath{\mathbf{n}^{#1}}}
\newcommand{\anoutwardsnormalvecrefconf}[1]{\ensuremath{\mathbf{N}^{#1}}}
\newcommand{\prescribedtraction}[1]{\ensuremath{\mathbf{\hat{t}}^{(#1)}_0}}
\newcommand{\strainenergyfunction}{\ensuremath{\Psi}}
\newcommand{\greenlagstrain}{\ensuremath{\mathbf{E}}}
\newcommand{\gapfunction}{\ensuremath{g_{\mathrm{n}}}}
\newcommand{\gapfunctiondiscrete}{\ensuremath{\tilde{g}_{\mathrm{n}}}}
\newcommand{\contacttraction}[1]{\ensuremath{\mathbf{t}^{(#1)}_\mathrm{c}}}
\newcommand{\normalcontactpressure}{\ensuremath{p_{\mathrm{n}}}}
\newcommand{\normalcontactforce}{\ensuremath{\mathbf{t}_{\mathrm{n}}}}
\newcommand{\tangentialcontactforce}{\ensuremath{\mathbf{t}_{\tau}}}
\newcommand{\solutionspace}[1]{\ensuremath{\boldsymbol{\mathcal{U}}^{(#1)}}}
\newcommand{\testingspace}[1]{\ensuremath{\boldsymbol{\mathcal{V}}^{(#1)}}}
\newcommand{\tracetestingspace}[1]{\ensuremath{\boldsymbol{\mathcal{W}}^{#1}}}
\newcommand{\lagmultvector}[1]{\ensuremath{\boldsymbol{\lambda}_{#1}}}
\newcommand{\testlagmultvector}[1]{\ensuremath{\boldsymbol{\mu}_{#1}}}
\newcommand{\testlagmultvectornormal}{\ensuremath{\mu_{\mathrm{n}}}}
\newcommand{\lagmultscalarnormal}{\ensuremath{\lambda_{\mathrm{n}}}}
\newcommand{\lagmultvectortang}{\ensuremath{\boldsymbol{\lambda}_{\tau}}}
\newcommand{\lagmultsolutionspace}{\ensuremath{\boldsymbol{\mathcal{M}}}}
\newcommand{\lagmultembeddedmeshsolutionspace}{\ensuremath{\boldsymbol{\mathcal{L}}}}
\newcommand{\kinenergy}{\ensuremath{\delta \mathcal{W}_{\mathrm{kin}}}}
\newcommand{\intextenergy}{\ensuremath{\delta \mathcal{W}_{\mathrm{int,ext}}}}
\newcommand{\kinintextenergy}{\ensuremath{\delta \mathcal{W}_{\mathrm{kin,int,ext}}}}
\newcommand{\contactenergy}[1]{\ensuremath{\delta \mathcal{W}_{\contactsubscript #1}}}
\newcommand{\constraintenergy}{\ensuremath{\delta \mathcal{W}_{\lambda}}}
\newcommand{\sobolevspace}[1]{\ensuremath{H^{1}(\bodyref{#1})}}
\newcommand{\lameparamlambda}{\ensuremath{\lambda}}
\newcommand{\lameparammu}{\ensuremath{\mu}}
\newcommand{\dualitypairing}[2]{\ensuremath{\langle #1, #2 \rangle }}
\newcommand{\scalarfunction}{\ensuremath{w}}

\newcommand{\afirstknotvector}{\ensuremath{\Xi}}
\newcommand{\asecondknotvector}{\ensuremath{H}}
\newcommand{\athirdknotvector}{\ensuremath{Z}}
\newcommand{\aparameterspace}[1]{\ensuremath{\xi_{#1}}}
\newcommand{\grevilleabscissae}{\ensuremath{\Xi_{G}}}
\newcommand{\agrevillevalue}[1]{\ensuremath{\xi_{G,#1}}}

\newcommand{\interfacereference}[2]{\ensuremath{\Gamma_{#1}^{#2}}}
\newcommand{\interfacespatial}[2]{\ensuremath{\gamma_{#1}^{#2}}}
\newcommand{\boundarylayerdomain}[1]{\ensuremath{\Omega_0^{(#1\layerindex)}}}
\newcommand{\bulkdomain}[1]{\ensuremath{\Omega_0^{(#1\backgroundindex)}}}
\newcommand{\interfacestar}[2]{\ensuremath{\Gamma_{* #2}^{#1}}}
\newcommand{\lambdastar}[2]{\ensuremath{\boldsymbol{\lambda}_{* #2}^{#1}}}
\newcommand{\dispboundarylayer}[1]{\ensuremath{\mathbf{u}^{(#1,\layerindex)}}}
\newcommand{\dispbulk}[1]{\ensuremath{\mathbf{u}^{(#1,\backgroundindex)}}}
\newcommand{\couplinginterenergy}[2]{\ensuremath{\delta \mathcal{W}^{#1}_{* #2}}}
\newcommand{\couplingconstraintenergy}[2]{\ensuremath{\delta \mathcal{W}^{#1}_{\lambda_{*} #2}}}

\newcommand{\boundarylayerdomainnumexample}{\ensuremath{\Omega^{(\layerindex)}}}
\newcommand{\bulkdomainnumexample}{\ensuremath{\Omega^{(\backgroundindex)}}}
\newcommand{\elesizeboundarylayer}{\ensuremath{h^{(\layerindex)}}}
\newcommand{\elesizebulkdomain}{\ensuremath{h^{(\backgroundindex)}}}

\newcommand{\elepos}[2]{\ensuremath{\mathbf{X}^{(#1, #2)}_h}}
\newcommand{\eledis}[2]{\ensuremath{\mathbf{u}^{(#1, #2)}_h}}
\newcommand{\elevirtualdis}[2]{\ensuremath{\delta \mathbf{u}^{(#1, #2)}_h}}

\newcommand{\eleposlayer}[1]{\elepos{#1}{\layerindex}}
\newcommand{\eleposbackground}[1]{\elepos{#1}{\backgroundindex}}
\newcommand{\eledislayer}[1]{\eledis{#1}{\layerindex}}
\newcommand{\eledisbackground}[1]{\eledis{#1}{\backgroundindex}}
\newcommand{\elevirtualdislayer}[1]{\elevirtualdis{#1}{\layerindex}}
\newcommand{\elevirtualdisbackground}[1]{\elevirtualdis{#1}{\backgroundindex}}

\newcommand{\nodalpos}[2]{\ensuremath{\mathbf{x}^{(#1)}_{#2}}}
\newcommand{\nodaldis}[2]{\ensuremath{\mathbf{d}^{#1}_{#2}}}
\newcommand{\nodaldisslave}[2]{\ensuremath{\mathbf{d}^{\slaveindex#1}_{#2}}}
\newcommand{\nodaldismaster}[2]{\ensuremath{\mathbf{d}^{\masterindex#1}_{#2}}}
\newcommand{\nodaldislayer}[2]{\ensuremath{\mathbf{d}^{\layerindex#1}_{#2}}}
\newcommand{\nodaldisbackground}[2]{\ensuremath{\mathbf{d}^{\backgroundindex#1}_{#2}}}
\newcommand{\nodaldisinterface}[2]{\ensuremath{\mathbf{d}^{\interfaceindex#1}_{#2}}}
\newcommand{\nodaldiscut}[2]{\ensuremath{\mathbf{d}^{\cuteleindex#1}_{#2}}}
\newcommand{\nodaldisnocontactnoembedded}{\ensuremath{\mathbf{d}^{\restofDOFSindex}}}
\newcommand{\nodalvirtualdis}[2]{\ensuremath{\delta \mathbf{d}^{(#1)}_{#2}}}
\newcommand{\contactdiscretedisplacementvector}{\ensuremath{\mathbf{d}^{\contactsubscript}}}
\newcommand{\emdiscretedisplacementvector}{\ensuremath{\mathbf{d}^{*}}}
\newcommand{\globaldiscretedisplacementvector}{\ensuremath{\mathbf{d}}}
\newcommand{\ashapefunction}[1]{\ensuremath{\Psi_{#1}}}
\newcommand{\nurbsshapefunction}[1]{\ensuremath{R_{#1}}}
\newcommand{\absplineshapefunction}[3]{\ensuremath{B^{#1}_{#2}(#3)}}
\newcommand{\anurbsshapefunction}[3]{\ensuremath{R^{#1}_{#2}#3}}
\newcommand{\weightnurbs}[1]{\ensuremath{w_{#1}}}
\newcommand{\lagrangeshapefunction}[1]{\ensuremath{N_{#1}}}
\newcommand{\numnodesdiscret}[2]{\ensuremath{n^{(#1, #2)}}}
\newcommand{\numnodesslavebl}[1]{\ensuremath{n_{sl}^{#1}}}
\newcommand{\numnodesmasterbl}[1]{\ensuremath{n_{ma}^{#1}}}
\newcommand{\numnodesinterfacestar}{\ensuremath{n_{*}^{\interfaceindex}}}
\newcommand{\numnodescutelements}{\ensuremath{n_{*}^{\cuteleindex}}}
\newcommand{\alagmultshapefunction}[1]{\ensuremath{\Phi_{#1}}}
\newcommand{\alagmultshapefunctionstar}[1]{\ensuremath{\Phi^*_{#1}}}
\newcommand{\nodallagmult}[2]{\ensuremath{\boldsymbol{\lambda}_{#1}^{#2}}}
\newcommand{\nodallagmultstar}[2]{\ensuremath{\boldsymbol{\lambda}_{*#1}^{#2}}}
\newcommand{\numnodeswithlagrange}{\ensuremath{m}}
\newcommand{\numinterfacenodes}[1]{\ensuremath{m_{*}^{#1}}}
\newcommand{\localmortaroperatorD}[1]{\ensuremath{\boldsymbol{D} #1}}
\newcommand{\localmortaroperatorM}[1]{\ensuremath{\boldsymbol{M} #1}}
\newcommand{\localmortaroperatorDstar}[2]{\ensuremath{\boldsymbol{D}^{#1}_* #2}}
\newcommand{\localmortaroperatorMstar}[2]{\ensuremath{\boldsymbol{M}^{#1}_* #2}}
\newcommand{\identitymatrix}{\ensuremath{\boldsymbol{I}}}
\newcommand{\contactforce}{\ensuremath{\boldsymbol{f}_{\contactsubscript}}}
\newcommand{\couplingforce}[1]{\ensuremath{\boldsymbol{f}_{*}^{#1}}}
\newcommand{\couplingforcelayer}[1]{\ensuremath{\boldsymbol{f}_{*}^{\layerindex#1}}}
\newcommand{\couplingforcebackground}[1]{\ensuremath{\boldsymbol{f}_{*}^{\backgroundindex#1}}}
\newcommand{\internalforce}[1]{\ensuremath{\boldsymbol{f}_{\mathrm{int}}^{#1}}}
\newcommand{\internalforcelayer}[1]{\ensuremath{\boldsymbol{f}_{\mathrm{int}}^{\layerindex#1}}}
\newcommand{\internalforcebackground}[1]{\ensuremath{\boldsymbol{f}_{\mathrm{int}}^{\backgroundindex#1}}}
\newcommand{\externalforcelayer}[1]{\ensuremath{\boldsymbol{f}_{\mathrm{ext}}^{\layerindex#1}}}
\newcommand{\externalforcebackground}[1]{\ensuremath{\boldsymbol{f}_{\mathrm{ext}}^{\backgroundindex#1}}}
\newcommand{\externalforce}[1]{\ensuremath{\boldsymbol{f}_{\mathrm{ext}}^{#1}}}
\newcommand{\couplingconstraint}[1]{\ensuremath{\boldsymbol{g}_{*}^{#1}}}
\newcommand{\zerovector}{\ensuremath{\boldsymbol{0}}}
\newcommand{\transpose}{\mathsf{T}}
\newcommand{\divergence}{\ensuremath{\nabla \cdot}}
\newcommand{\projectionsurfcetovolume}{\ensuremath{\chi^*_h}}
\newcommand{\projectionsurfacetosurface}{\ensuremath{\chi_h}}

\newcommand{\stiffnessmatrixlayer}{\ensuremath{\boldsymbol{K}_{\layerindex\layerindex}}}
\newcommand{\stiffnessmatrixbackground}{\ensuremath{\boldsymbol{K}_{\backgroundindex\backgroundindex}}}
\newcommand{\stiffnessmatrixinterface}{\ensuremath{\boldsymbol{K}_{\interfaceindex\interfaceindex}}}
\newcommand{\stiffnessmatrixcut}{\ensuremath{\boldsymbol{K}_{\cuteleindex\cuteleindex}}}
\newcommand{\stiffnessmatrixcontact}{\ensuremath{\boldsymbol{K}_{\contactsubscript}}}
\newcommand{\stiffnessmatrixembedded}{\ensuremath{\boldsymbol{K}_{*}}}
\newcommand{\stiffnessmatrixtangential}[1]{\ensuremath{\boldsymbol{K}_{\mathrm{stiff}#1}}}
\newcommand{\zeromatrix}{\ensuremath{\boldsymbol{0}}}

\newcommand{\residualnocontactnoem}{\ensuremath{\boldsymbol{r}^{\restofDOFSindex}}}
\newcommand{\residualcontact}{\ensuremath{\boldsymbol{r}^{\contactsubscript}}}
\newcommand{\residuallambda}{\ensuremath{\boldsymbol{r}^{\lambda}}}
\newcommand{\residualembedded}{\ensuremath{\boldsymbol{r}^{*}}}
\newcommand{\residualinterface}{\ensuremath{\boldsymbol{r}^{\interfaceindex}}}
\newcommand{\residualcutele}{\ensuremath{\boldsymbol{r}^{\cuteleindex}}}
\newcommand{\residual}{\ensuremath{\boldsymbol{r}}}

\newcommand{\complementarityfunction}[1]{\ensuremath{C_{#1}}}

\newcommand{\numtetrahedra}{\ensuremath{n_T}}
\newcommand{\numgpstetrahedra}{\ensuremath{n_{gp}}}
\newcommand{\gpstetrahedra}{\ensuremath{\widetilde{\mathbf{\eta}}_{g}}}
\newcommand{\weightstetrahedra}{\ensuremath{w_{g}}}
\newcommand{\paramspacebackgroundele}{\ensuremath{\mathbf{\xi}}}
\newcommand{\stiffcutele}{\ensuremath{\boldsymbol{K}^{e}_{\cuteleindex\cuteleindex}}}
\newcommand{\jacobianbackgroundele}{\ensuremath{J^e_{\mathrm{cut}}}}

\newcommand{\abodyi}{\ensuremath{(i)}}
\newcommand{\abodynum}[1]{\ensuremath{(#1)}}
\newcommand{\anyinterfacestar}[1]{\ensuremath{\Gamma_{* #1}}}
\newcommand{\lininterfacestar}{\ensuremath{\Gamma_{*, h}^{\mathrm{lin}}}}

\newcommand{\penaltyparam}{\ensuremath{\epsilon}}
\newcommand{\scalingmatrix}[1]{\ensuremath{\boldsymbol{\kappa}^{#1}}}

\newcommand{\offsetdistance}{\ensuremath{\ell}}
\newcommand{\evalpoints}[1]{\ensuremath{\mathbf{x}_{o #1}}}
\newcommand{\basepoints}[1]{\ensuremath{\mathbf{x}_{#1}}}

\newcommand{\acurve}[1]{\ensuremath{\mathbf{C}(#1)}}
\newcommand{\acurvederivative}[1]{\ensuremath{\mathbf{C}'(#1)}}
\newcommand{\acurveoffset}[1]{\ensuremath{\mathbf{C}_{o}#1}}
\newcommand{\acurveoffsetapproximation}{\ensuremath{\mathbf{C}_{o, approx}}}
\newcommand{\asurface}[1]{\ensuremath{\mathbf{S}(#1)}}
\newcommand{\avolume}[1]{\ensuremath{\mathbf{V}(#1)}}

\newcommand{\aninwardsnormalvec}[1]{\ensuremath{\mathbf{n}(#1)}}
\newcommand{\aninwardsnormalvecwithindex}[1]{\ensuremath{\mathbf{n}_{#1}}}
\newcommand{\controlpoints}[1]{\ensuremath{\mathbf{P}_{#1}}}
\newcommand{\controlpointgrid}{\ensuremath{\mathbf{P}}}
\newcommand{\controlpointsoffset}[1]{\ensuremath{\mathbf{P}_{o#1}}}
\newcommand{\nurbsmatrix}{\ensuremath{\mathbf{N}}}
\newcommand{\itergd}{\ensuremath{j}}
\newcommand{\nurbssecondordertwodim}{\ensuremath{nurbs9}}
\newcommand{\nurbssecondorderthreedim}{\ensuremath{nurbs27}}
\newcommand{\lagrangefirstordertwodim}{\ensuremath{quad4}}
\newcommand{\lagrangeseconddordertwodim}{\ensuremath{quad8}}
\newcommand{\lagrangefirstdorderthreedim}{\ensuremath{hex8}}
\newcommand{\lagrangeseconddorderthreedim}{\ensuremath{hex20}}

\newcommand{\normalcontacttraction}{\ensuremath{p_{\mathrm{c}}}}
\newcommand{\maxnormalcontacttraction}{\ensuremath{p_{\mathrm{c,max}}}}
\newcommand{\widthcontactzone}{\ensuremath{b}}

\newcommand{\normalatpatchedge}[1]{\ensuremath{\mathbf{n}_{P#1}}}
\newcommand{\Youngsmodulus}{\ensuremath{E}}
\newcommand{\Poissonratio}{\ensuremath{\nu}}
\newcommand{\Density}{\ensuremath{\rho}}

\newcommand{\errorinfinite}{\ensuremath{e_{\infty}}}
\newcommand{\errorrms}{\ensuremath{e_{L^2}}}

%% file: section/abstract_arxiv.tex
\rule{\textwidth}{0.4pt}
\begin{abstract}

This paper proposes a novel discretization workflow for contact problems in which the discretization of the contact interface is decoupled from that of the bulk domain. This separation enables independently tailored meshes for the contact interface and the bulk volume, allowing local requirements—such as element type and mesh resolution—to be addressed efficiently. Exploiting the boundary representation of CAD models, the contact interface of each body is discretized using a NURBS-based boundary layer mesh. This provides a smooth geometric description of the contact surface and enhanced inter-element continuity. The bulk domain is discretized using a structured Cartesian grid. To couple the resulting non-matching discretizations, an embedded mesh approach based on a mortar-type constraint formulation is employed. The paper describes in detail the proposed discretization workflow for generating both the isogeometric boundary layer and the structured Cartesian grid, and presents several strategies for constructing NURBS-based boundary layer meshes. Finally, a set of numerical examples is provided to validate the proposed approach.

\vspace{0.2cm}
\end{abstract}
\rule{\textwidth}{0.4pt}

%% file: section/introduction.tex
\section{Introduction} \label{sec:intro}

Contact problems are of fundamental importance, as they arise in a wide range of physical and 
biological systems. They are characterized by complex interface phenomena, including friction, wear, 
lubrication, and thermomechanical effects. Over the past decades, advances in numerical contact 
formulations, together with increasing computational resources, have enabled significant 
technological progress. Notable examples include improvements in vehicle safety through crash 
simulations and the assessment of wear-induced damage in machine components~\cite{Dong2017}. 
In the medical field, sophisticated contact models have been developed for the planning of 
maxillofacial surgery~\cite{Obaidellah2008} and, more recently, for the development of 
endovascular treatment strategies for cerebral aneurysms~\cite{Frank2024}, with an 
emphasis on incorporating patient-specific geometries.

In the following, we focus on contact formulations based on the finite element method (\fem{}), 
for which substantial research efforts have been devoted to improving accuracy and robustness. 
A central line of research concerns the development of contact discretizations for 
non-matching meshes, including node-to-segment (NTS)~\cite{Wriggers1990, Papadopoulos1992, 
ZavariseDeLorenzis2009} and segment-to-segment (STS)~\cite{Zavarise1998, Fischer2005, Puso2004} techniques. 
A particular class of STS approaches is formed by mortar formulations~\cite{Belgacem2000, Wohlmuth2003, 
Popp2009}, which have gained increasing relevance in recent years, as they are especially well suited 
for finite deformation contact problems and finite sliding. Mortar methods enforce contact 
constraints in a weak sense and enable a variationally consistent treatment of non-penetration and 
frictional sliding conditions, thereby guaranteeing optimal convergence rates~\cite{DeLorenzis2017}. 
In~\cite{Wohlmuth2012}, it was shown that, for unilateral contact, the maximum expected convergence 
order under uniform mesh refinement is $O(h^{3/2})$ for shape functions of polynomial order $p\geqslant 2$. 
This result makes the use of higher-order interpolation questionable for contact problems unless 
combined with adaptive mesh refinement techniques. For a comprehensive overview of computational 
contact mechanics based on \fem{}, the reader is referred to the monograph by~\cite{DeLorenzis2017}.

Classical finite element discretizations based on Lagrange polynomials exhibit significant drawbacks, as their 
$C^0$ interelement continuity leads to a discontinuous normal vector field along the contact interface. 
This is particularly critical because, regardless of the chosen contact discretization technique, 
normal projections between the contacting bodies are required throughout the contact evaluation. 
A commonly adopted strategy to alleviate this issue is the construction of the normal field 
by averaging the normals of the surface elements surrounding a node. Nevertheless, numerical instabilities may arise, 
especially in problems involving large sliding, due to the non-smooth variation of contact variables 
as the slave node slides over the master interface~\cite{DeLorenzis2014, Temizer2012}. 
To mitigate these oscillations, geometric smoothing techniques 
based on Hermite, Bezier, spline, and NURBS basis functions have been proposed. 
The underlying idea is to employ higher-order interpolations on the contact interface to 
improve the accuracy of the contact tractions, while keeping the bulk discretization 
unchanged~\cite{Padmanabhan2001, Wriggers2001, AlDojayli2002, Tur2012, Neto2017}. 
Although such approaches reduce these oscillations, 
they do not necessarily eliminate them entirely~\cite{Temizer2012}.

To address these drawbacks, contact formulations based on isogeometric analysis (\iga{}) have been 
developed, with many established \fem{}-based contact formulations adapted to the \iga{} framework. 
A comprehensive overview of computational contact formulations within 
\iga{} is provided in~\cite{DeLorenzis2014}. The central concept of \iga{} is 
the integration of computer-aided design (\cad{}) and engineering analysis 
by employing the same basis functions used in \cad{} for numerical simulations~\cite{Hughes2005, Cottrell2009}. 
In~\cite{Evans2009, Grossmann2012}, it was shown that \iga{} can deliver increased robustness and 
accuracy per degree of freedom compared to \fem{}, as it relies on B-spline-based basis functions 
such as non-uniform rational B-splines (\nurbs{}) or T-splines. 
These basis functions possess particularly attractive properties for contact problems, as 
already emphasized in the first publication on \iga{}~\cite{Hughes2005}: they enable an accurate and 
smooth representation of the contact interface and guarantee an interelement continuity 
of at least $C^1$, thereby yielding a continuous normal vector field. Nevertheless, 
while the higher-order continuity of \nurbs{} offers clear advantages for contact algorithms, 
it may not be required throughout the bulk of the domain, as it does not necessarily improve 
spatial convergence rates and may instead lead to increased computational cost.

In comparison to finite element mesh generation, 
mesh generation in \iga{} is not as well established and can be particularly challenging, 
especially for complex geometries in three dimensions~\cite{Kuraishi2024, Wobbes2024}. 
Although the isogeometric boundary representation is provided by a \cad{} model, 
it lacks a true volumetric description and is therefore essentially ``hollow''. 
To address this limitation, a volume parametrization conformal to the given 
boundary representation must be constructed, which can be a demanding task~\cite{Martin2008, Akhras2016, Zheng2022}.

\begin{figure}
	\centering
	
	\begin{subfigure}{0.35\textwidth}
		\def\svgwidth{\textwidth}
        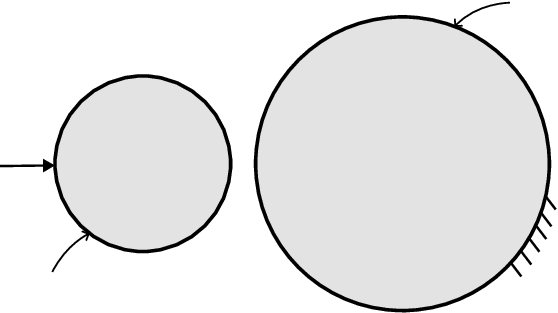
		\caption{}
		\label{subfig:contact_two_bodies}
	\end{subfigure}
	\hspace{1.5cm}
	\begin{subfigure}{0.35\textwidth}
		\includegraphics[width=\textwidth]{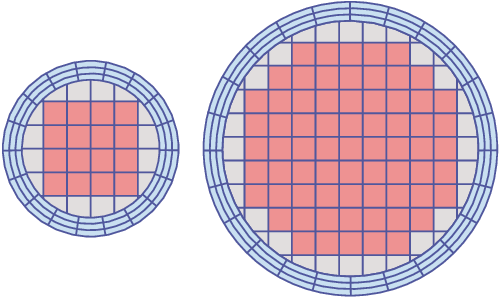}
		\caption{}
		\label{subfig:contact_two_bodies_novel_dis}
	\end{subfigure}
	
	\caption{Two-dimensional contact problem between two bodies: 
	(a) boundary conditions;
	(b) proposed discretization, in which each body is discretized with a NURBS boundary layer mesh (blue) and an underlying Cartesian mesh (red). 
	Some elements of the Cartesian mesh, referred to as cut elements, are intersected by the boundary layer mesh (gray).}
	\label{fig:contact_two_bodies}
\end{figure}

Building on the previous observations, we propose a novel discretization approach 
for contact problems. The approach is illustrated in Figure~\ref{fig:contact_two_bodies} for 
a two-dimensional contact configuration; however, it is applicable to both 
two- and three-dimensional contact problems. The procedure starts from the boundary 
representation (also referred to as \brep{}) of the bodies involved in contact. 
For the two-dimensional setup depicted in Figure~\ref{subfig:contact_two_bodies}, 
the \brep{} is constructed using NURBS curves, whereas for three-dimensional configurations 
it is given by NURBS surfaces. In the proposed approach, the discretization of the 
contact interface is decoupled from that of the bulk domain. Starting from the isogeometric 
\brep{}, a boundary layer mesh is generated by offsetting the NURBS \brep{} inward into 
the body, exploiting the tensor-product structure of NURBS. The remaining bulk domain 
is subsequently discretized using a simple and easy-to-generate Cartesian mesh. 
The resulting discretization is shown in Figure~\ref{subfig:contact_two_bodies_novel_dis}. 
In this configuration, a subset of the Cartesian mesh elements is intersected by 
the boundary layer mesh; these elements are referred to as \textit{cut elements}.

The simplified mesh generation strategy underlying the proposed approach follows the idea of immersed boundary methods, in which a \brep{} is 
embedded into a simple background mesh, thereby eliminating the need for a conforming volume 
parametrization. Immersed boundary methods (also referred to as non-boundary-fitted or fictitious domain methods) comprise a 
broad class of techniques, including the finite cell method~\cite{Rank2012}, CutFEM~\cite{Burman2015}, XFEM~\cite{Ji2004, Möes2006}, 
and embedded mesh methods~\cite{Puso2012}. These approaches have been successfully applied to a wide range of 
complex multiphysics problems, such as fluid–structure interaction (FSI)~\cite{Mayer2009, Schott2016}, 
crack propagation~\cite{Moes1999}, and, more recently, interfacial damage and debonding in composite materials~\cite{Groeneveld2025}. 
A related discretization strategy was introduced in~\cite{Wei2021} under the name of the immersed boundary-conformal method (IBCM) 
for the solution of linear elliptic problems. Furthermore, in~\cite{Guarino2024a}, this approach was employed for the analysis of 
linear elastic Kirchhoff–Love and Reissner–Mindlin shell structures to incorporate geometric features, such as holes, into shell 
models. To the authors’ knowledge, IBCM has not yet been extended to three-dimensional problems nor applied to unilateral contact problems.

\responsetoview{Related approaches combining spline-based contact or boundary discretizations with classical finite element discretizations of the bulk domain have previously been proposed. In~\cite{Corbett2014}, the contact surface was locally enriched with NURBS-based surface elements, while the bulk domain was discretized using standard linear Lagrangian elements. This methodology was subsequently extended to frictional contact and mixed-mode debonding problems in~\cite{Corbett2015}, where a meshing strategy based on an isogeometric surface description was proposed. An enriched hexahedral boundary layer is constructed by inward projection of a T-spline boundary representation and connected to a conforming bulk mesh. This meshing strategy was further discussed and extended in~\cite{Harmel2017}.
Another related method was proposed in~\cite{Otto2019}, where a NURBS surface is introduced between the two contacting bodies. Rather than being in direct contact with each other, both bodies interact with the intermediate NURBS surface. One body is tied to the NURBS surface, while the second body is brought into contact with it.}

In this work, the boundary layer mesh is said to \textit{overlap} the \textit{underlying} Cartesian mesh, and the two 
discretizations are coupled by means of a mortar-based method. It is, however, well known that the application of 
mortar methods to embedded mesh discretizations may lead to an unstable Lagrange multiplier field and can induce 
locking effects~\cite{Sanders2012}. To address this issue in the context of two-dimensional XFEM problems, 
the construction of so-called vital vertices was introduced in~\cite{Bechet2009, Nistor2009} to ensure 
the stability of the Lagrange multiplier field. In the present work, special attention is given to 
the design of the discrete Lagrange multiplier space in order to avoid such instabilities for the 
problems considered here. With these considerations in mind, the applicability of the proposed approach 
to contact problems is demonstrated in Section~\ref{section:numerical_examples}.

The proposed approach leverages the advantages of \iga{} while simplifying mesh generation. 
First, the contact algorithm benefits from the high interelement continuity of isogeometric 
shape functions, which ensures a continuous normal vector field. Consequently, the drawbacks 
associated with $C^0$-continuous discretizations are naturally avoided. Furthermore, the 
boundary layer mesh and the bulk mesh can be tailored independently, allowing for local 
refinement in regions where contact is expected without affecting the Cartesian background 
mesh. Finally, the element technology can be selected independently for the boundary layer 
and the Cartesian meshes. Depending on the problem characteristics and the available software 
resources, one may choose a specific interpolation technology, such as an \iga{}-based 
discretization or classical Lagrangian shape functions, or employ more advanced element 
technologies to mitigate locking effects, including mixed or hybrid formulations~\cite{Malkus1978} 
and enhanced assumed strain (EAS) elements~\cite{Simo1992}.

The remainder of this paper is organized as follows. Section~\ref{section:discretization_workflow} 
presents the mesh generation workflow and investigates three approaches for constructing 
NURBS-based boundary layers. Section~\ref{section:problem_formulation} introduces the governing 
equations of the contact problem and derives the embedded mesh formulation used to couple 
the isogeometric boundary layer with the Cartesian background mesh. Section~\ref{section:spatial_discretization} 
describes the spatial discretization and provides details on the numerical integration of cut elements. 
Section~\ref{section:numerical_examples} presents several numerical examples to demonstrate the applicability of 
the proposed method to both two- and three-dimensional problems. Finally, Section~\ref{section:concluding_remarks} concludes the paper.

%% file: fig/introduction/contact_two_bodies_horizontal.eps_tex
\begingroup%
  \makeatletter%
  \providecommand\color[2][]{%
    \errmessage{(Inkscape) Color is used for the text in Inkscape, but the package 'color.sty' is not loaded}%
    \renewcommand\color[2][]{}%
  }%
  \providecommand\transparent[1]{%
    \errmessage{(Inkscape) Transparency is used (non-zero) for the text in Inkscape, but the package 'transparent.sty' is not loaded}%
    \renewcommand\transparent[1]{}%
  }%
  \providecommand\rotatebox[2]{#2}%
  \newcommand*\fsize{\dimexpr\f@size pt\relax}%
  \newcommand*\lineheight[1]{\fontsize{\fsize}{#1\fsize}\selectfont}%
  \ifx\svgwidth\undefined%
    \setlength{\unitlength}{267.9042714bp}%
    \ifx\svgscale\undefined%
      \relax%
    \else%
      \setlength{\unitlength}{\unitlength * \real{\svgscale}}%
    \fi%
  \else%
    \setlength{\unitlength}{\svgwidth}%
  \fi%
  \global\let\svgwidth\undefined%
  \global\let\svgscale\undefined%
  \makeatother%
  \begin{picture}(1,0.56976309)%
    \lineheight{1}%
    \setlength\tabcolsep{0pt}%
    \put(0,0){\includegraphics[width=\unitlength]{contact_two_bodies_horizontal.eps}}%
    \put(-0.1,0.02){\makebox(0,0)[lt]{\lineheight{1.25}\smash{\begin{tabular}[t]{l}B-rep body 1\end{tabular}}}}%
    \put(0.91956059,0.54283747){\makebox(0,0)[lt]{\lineheight{1.25}\smash{\begin{tabular}[t]{l}B-rep body 2\end{tabular}}}}%
  \end{picture}%
\endgroup%

%% file: section/discretization_workflow.tex
\section{Mesh generation workflow} \label{section:discretization_workflow}

In this section, we describe the discretization approach adopted for the bodies 
involved in the contact problem shown in Figure~\ref{subfig:contact_two_bodies_novel_dis}. 
The mesh generation workflow consists of two main steps. First, an isogeometric boundary 
layer mesh is constructed by performing an inward offset of the body’s \brep{} 
representation. Subsequently, a Cartesian mesh is generated to represent the bulk domain. 
In this work, three different methods for offsetting NURBS curves and surfaces are 
implemented and discussed. Finally, a simple algorithm for the construction of the 
Cartesian background mesh is presented.

\subsection{NURBS geometries}

This subsection briefly reviews the basic concepts of NURBS geometries relevant to this work. 
For a comprehensive discussion of this topic, the reader is referred to the monographs~\cite{Piegl1995, Cottrell2009}. 
Given a set of~$n$ B-spline basis functions of degree $p$, the $i$-th NURBS basis function is given by

\begin{equation}
    \anurbsshapefunction{p}{i}{(\aparameterspace{})}
    =
    \frac{
    \absplineshapefunction{p}{i}{\aparameterspace{}}
    \weightnurbs{i}
    }{
    \sum_{a = 1}^{n}
    \absplineshapefunction{p}{a}{\aparameterspace{}}
    \weightnurbs{a}
    },
    \label{eq:nurbs_basis_function}
\end{equation}
where $\absplineshapefunction{p}{i}{\aparameterspace{}}$ denotes the $i$-th B-spline basis function and $\weightnurbs{i}$ its associated weight. 
In most engineering applications, these weights are positive.

Both NURBS and B-spline basis functions are defined over a non-decreasing 
sequence of parameter values, referred to as the knot vector. 
In the one-dimensional case, the knot vector is denoted by~$\afirstknotvector{}~=~\{\aparameterspace{1}, \aparameterspace{2}, \dots, \aparameterspace{n+p+1}\}$ 
and is defined over the parametric domain $\aparameterspace{} \in [\aparameterspace{1},\aparameterspace{n+p+1}]$. The individual entries $\aparameterspace{i}$ 
are called knots, and their multiplicity governs the continuity of the basis functions. 
In particular, a knot with multiplicity $m_i$ yields $C^{p-m_i}$ continuity at the corresponding parameter value \aparameterspace{i}. 
An open knot vector is characterized by a multiplicity of $p+1$ at the first and last knots, which results in 
interpolatory basis functions at the ends of the parameter interval~$[\aparameterspace{1}, \aparameterspace{n+p+1}]$.

B-spline basis functions are constructed recursively using the Cox--de Boor formula~\cite{Cox1972, DeBoor1972}. 
They possess several important properties: they form a partition of unity (\ie{} $\sum_{i=1}^{n} \absplineshapefunction{p}{i}{\aparameterspace{}} = 1$), 
are non-negative and locally supported, and exhibit $C^{p-1}$ continuity in the absence of repeated knots.

NURBS geometries are obtained as linear combinations of NURBS basis functions and control points. 
To this end, each control point $\controlpoints{i}$ is associated with a weight $\weightnurbs{i}$. 
The set of $n$ control points defining the geometry is referred to as the control polygon, 
denoted by $\controlpointgrid{} = \{\controlpoints{1}, \controlpoints{2}, \dots, \controlpoints{n}\}$. 
For control points defined in a spatial dimension $d$, with $\controlpoints{i} \in \mathbb{R}^{d}$, 
a NURBS curve is defined as

\begin{equation}
    \acurve{\aparameterspace{}}
    =
    \sum_{i = 1}^{n}
    \anurbsshapefunction{p}{i}{(\aparameterspace{})}
    \controlpoints{i}.
\end{equation}

Higher-dimensional geometries are constructed using tensor-product formulations. 
Let $\controlpoints{i,j} \in \mathbb{R}^{d}$ with indices $i=1,\dots,n$ and $j=1,\dots,m$, 
polynomial orders $p$ and $q$, and knot vectors 
$\afirstknotvector{} = \{\aparameterspace{1}, \dots, \aparameterspace{n+p+1}\}$ and 
$\asecondknotvector{} = \{\eta_{1}, \dots, \eta_{m+q+1}\}$. 
The resulting NURBS surface is expressed as
\begin{equation}
    \asurface{\aparameterspace{}, \eta}
    =
    \sum_{i = 1}^{n}
    \sum_{j = 1}^{m}
    \anurbsshapefunction{p,q}{i,j}{(\aparameterspace{}, \eta)}
    \controlpoints{i,j},
\end{equation}
where the corresponding bivariate NURBS basis functions are given by
\begin{equation}
    \anurbsshapefunction{p,q}{i,j}{(\aparameterspace{}, \eta)}
    =
    \frac{
    \absplineshapefunction{p}{i}{\aparameterspace{}}
    \absplineshapefunction{q}{j}{\eta}
    \weightnurbs{i,j}
    }{
    \sum_{a = 1}^{n}
    \sum_{b = 1}^{m}
    \absplineshapefunction{p}{a}{\aparameterspace{}}
    \absplineshapefunction{q}{b}{\eta}
    \weightnurbs{a,b}
    }.
\end{equation}

Analogously, a NURBS volume is constructed from control points 
$\controlpoints{i,j,k} \in \mathbb{R}^{d}$ with indices $i=1,\dots,n$, $j=1,\dots,m$, and $k=1,\dots,l$, 
polynomial orders $p$, $q$, and $r$, and three knot vectors 
$\afirstknotvector{} = \{\aparameterspace{1}, \dots, \aparameterspace{n+p+1}\}$, 
$\asecondknotvector{} = \{\eta_{1}, \dots, \eta_{m+q+1}\}$, and 
$\athirdknotvector{} = \{\zeta_{1}, \dots, \zeta_{l+r+1}\}$. 
The resulting volume representation takes the form
\begin{equation}
    \avolume{\aparameterspace{}, \eta, \zeta}
    =
    \sum_{i = 1}^{n}
    \sum_{j = 1}^{m}
    \sum_{k = 1}^{l}
    \anurbsshapefunction{p,q,r}{i,j,k}{(\aparameterspace{}, \eta, \zeta)}
    \controlpoints{i,j,k},
\end{equation}
with the associated basis functions
\begin{equation}
    \anurbsshapefunction{p,q,r}{i,j,k}{(\aparameterspace{}, \eta, \zeta)}
    =
    \frac{
    \absplineshapefunction{p}{i}{\aparameterspace{}}
    \absplineshapefunction{q}{j}{\eta}
    \absplineshapefunction{r}{k}{\zeta}
    \weightnurbs{i,j,k}
    }{
    \sum_{a = 1}^{n}
    \sum_{b = 1}^{m}
    \sum_{c = 1}^{l}
    \absplineshapefunction{p}{a}{\aparameterspace{}}
    \absplineshapefunction{q}{b}{\eta}
    \absplineshapefunction{r}{c}{\zeta}
    \weightnurbs{a,b,c}
    }.
\end{equation}

NURBS geometries inherit the properties of their underlying basis functions. 
Owing to their local support, modifying a single control point influences at most $p+1$ elements in its neighborhood. 
Furthermore, NURBS geometries satisfy a strong convex hull property, meaning that the geometry is entirely contained within 
the convex hull spanned by the control polygon.

\subsection{Generation of an isogeometric boundary layer mesh }

Within the IGA framework, mesh generation begins with the construction of the 
isogeometric \brep{} representation of the body. Depending on the spatial dimension, 
this \brep{} is described by a curve in two-dimensional settings or by a surface 
in three-dimensional problems. In the following, this initial representation is 
referred to as the \textit{base \brep{}}.

The generation of an isogeometric boundary layer mesh exploits the tensor-product 
structure of NURBS surfaces and volumes. The basic idea is to extend the parametric 
representation of the base \brep{} by a distance $\offsetdistance{}$ along its inward 
normal direction. This is achieved by applying an offset operation, resulting in an 
\textit{offset \brep{}}. However, NURBS geometries cannot, in general, be offset analytically. 
This limitation can be illustrated for a NURBS curve $\acurve{\aparameterspace{}}$, 
whose offset curve $\acurveoffset{}$ is defined as
\begin{equation}
    \acurveoffset{(\aparameterspace{})} = \acurve{\aparameterspace{}} + \offsetdistance{} \, \aninwardsnormalvec{\aparameterspace{}},
\end{equation}
where $\aninwardsnormalvec{\aparameterspace{}}$ denotes the inward-facing unit normal vector of 
$\acurve{\aparameterspace{}}$. For a planar curve, this normal vector is given by
\begin{equation}
    \aninwardsnormalvec{\aparameterspace{}} = \frac{(y'(\aparameterspace{}), -x'(\aparameterspace{}))}{\sqrt{x'(\aparameterspace{})^2 + y'(\aparameterspace{})^2}},
    \label{eq:normal_vector_C}
\end{equation}
with $\acurvederivative{\aparameterspace{}} = \partial \acurve{\aparameterspace{}} \ / \ \partial \aparameterspace{} = (x'(\aparameterspace{}), y'(\aparameterspace{}))$. 

Although the regularity of $\acurve{\aparameterspace{}}$ guarantees that 
$\aninwardsnormalvec{\aparameterspace{}}$ is well defined along the curve, 
the normalization in equation~\eqref{eq:normal_vector_C} introduces a square root 
in the denominator. As a consequence, the resulting offset curve is generally 
non-rational. Therefore, approximation techniques must be employed to compute 
offsets of NURBS geometries.

Extensive research has addressed the offsetting of rational curves, as this 
operation plays a central role in CAD/CAM, robotics, and numerous industrial 
applications, including NC machining~\cite{Chuang2006, Elber1997, Shih2008}. 
However, many of the existing algorithms are not directly applicable in the present 
context, since the offset \brep{} is required to preserve the parametric structure 
of the base \brep{}, \ie{} the polynomial degree and knot vectors must remain unchanged.

To this end, three approaches for offsetting the base \brep{} while preserving its 
parametric representation are compared. The resulting offset \brep{} defines the 
coupling interface $\interfacestar{\abodyi}{}$ between the boundary layer mesh and the 
background mesh of body~$i$, where the coupling conditions are imposed. It should be 
emphasized that the considered approximation techniques do not yield an exact offset 
of the base geometry, and a certain approximation error is therefore unavoidable. 
However, the interface $\interfacestar{\abodyi}{}$ constitutes an artificial internal 
boundary separating the boundary layer and bulk discretization regions and does not 
carry any physical meaning; it merely serves as the domain on which the embedded mesh 
coupling conditions are enforced. Consequently, an exact geometric offset is not 
required for the construction of~$\interfacestar{\abodyi}{}$.

\begin{figure}
    \centering 

    \begin{subfigure}{0.45\textwidth}
        \centering
        
        \begin{minipage}{0.5\linewidth}
            \def\svgwidth{\linewidth}
            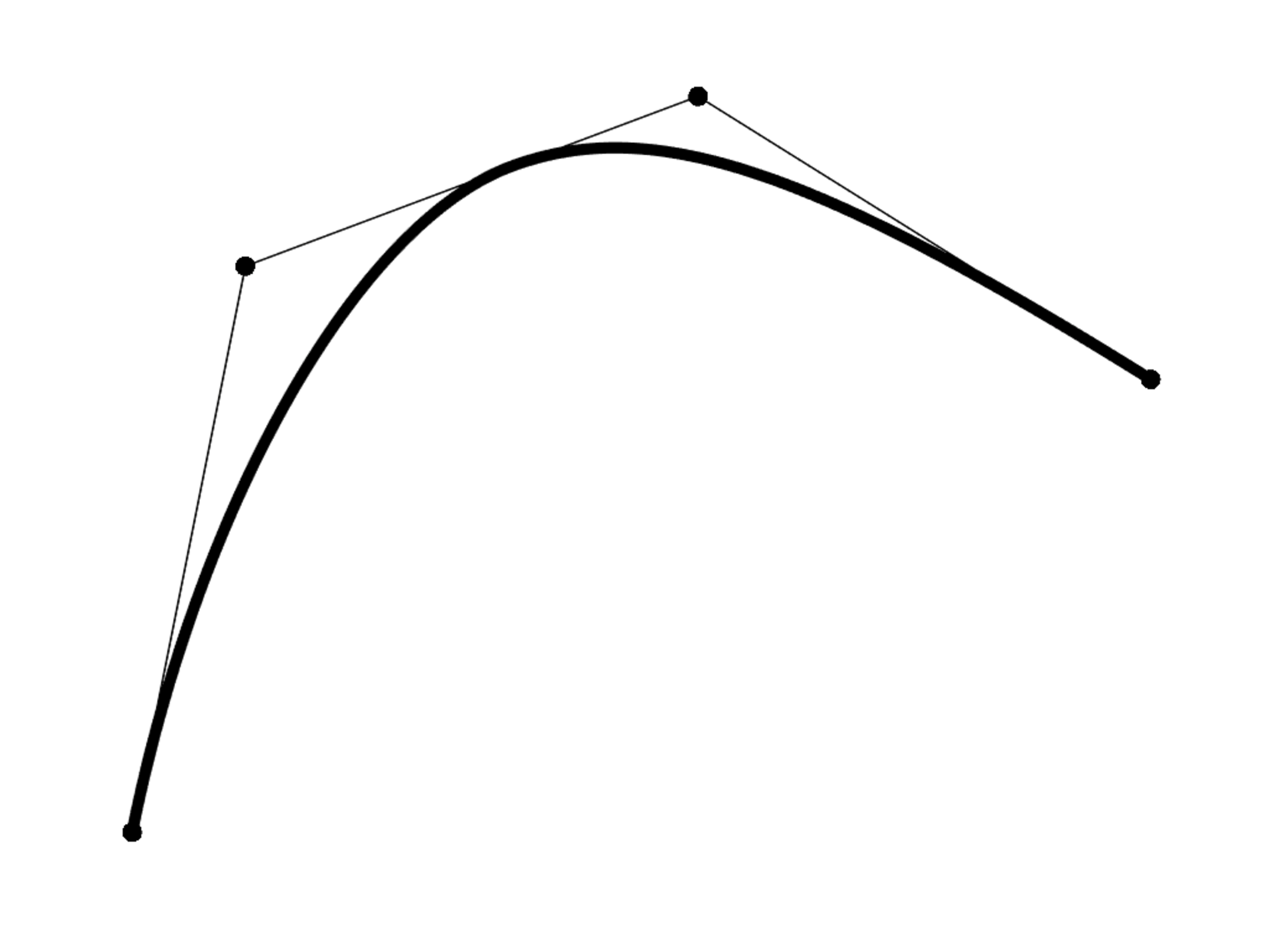
        \end{minipage}

        \caption{}
        \label{fig:base_curve}
    \end{subfigure}
    \hfill
    \begin{subfigure}{0.45\textwidth}
        \centering

        \begin{minipage}{0.5\linewidth}
            \def\svgwidth{\linewidth}
            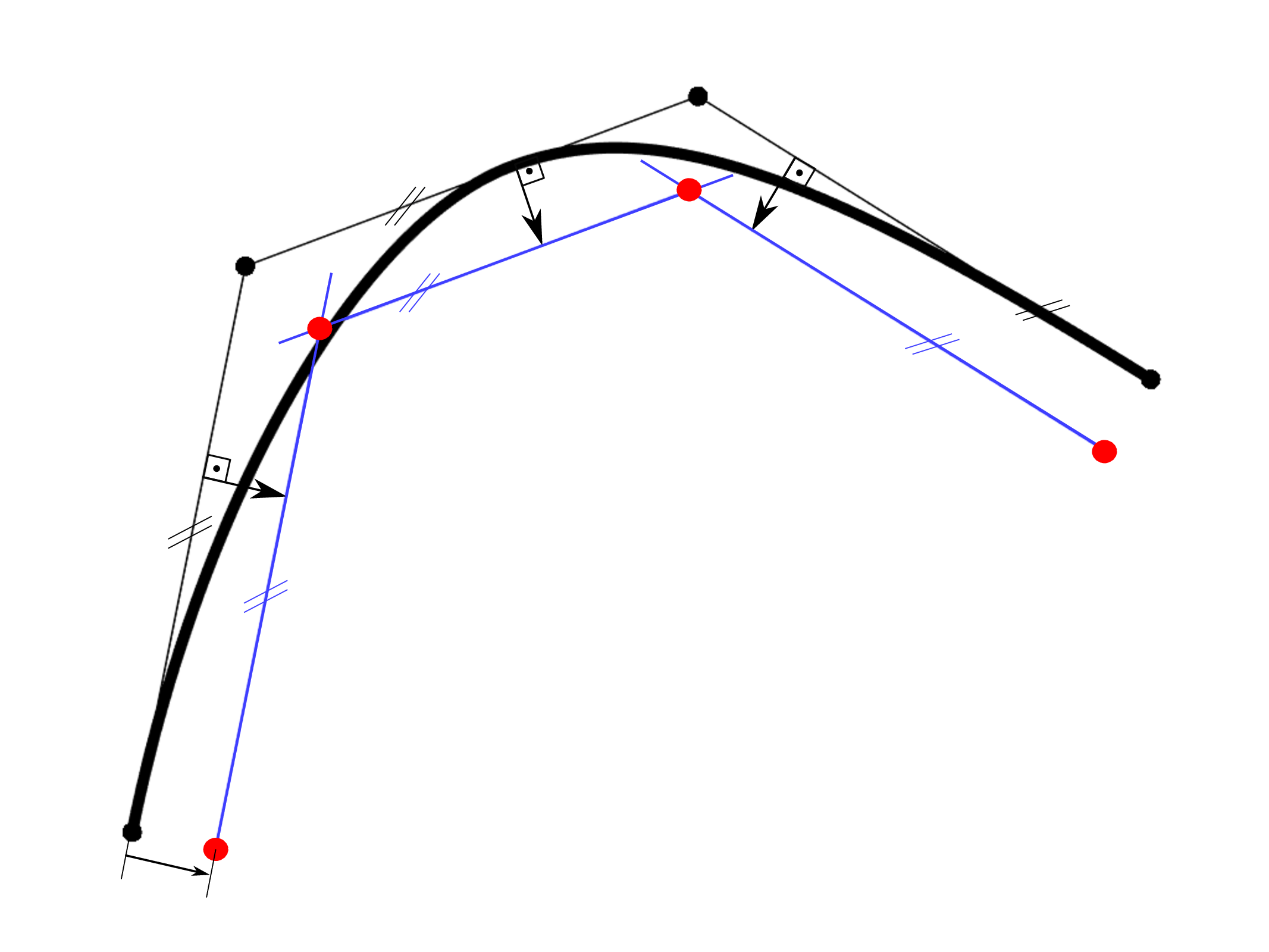
        \end{minipage}%
        \hfill
        \begin{minipage}{0.5\linewidth}
            \def\svgwidth{\linewidth}
            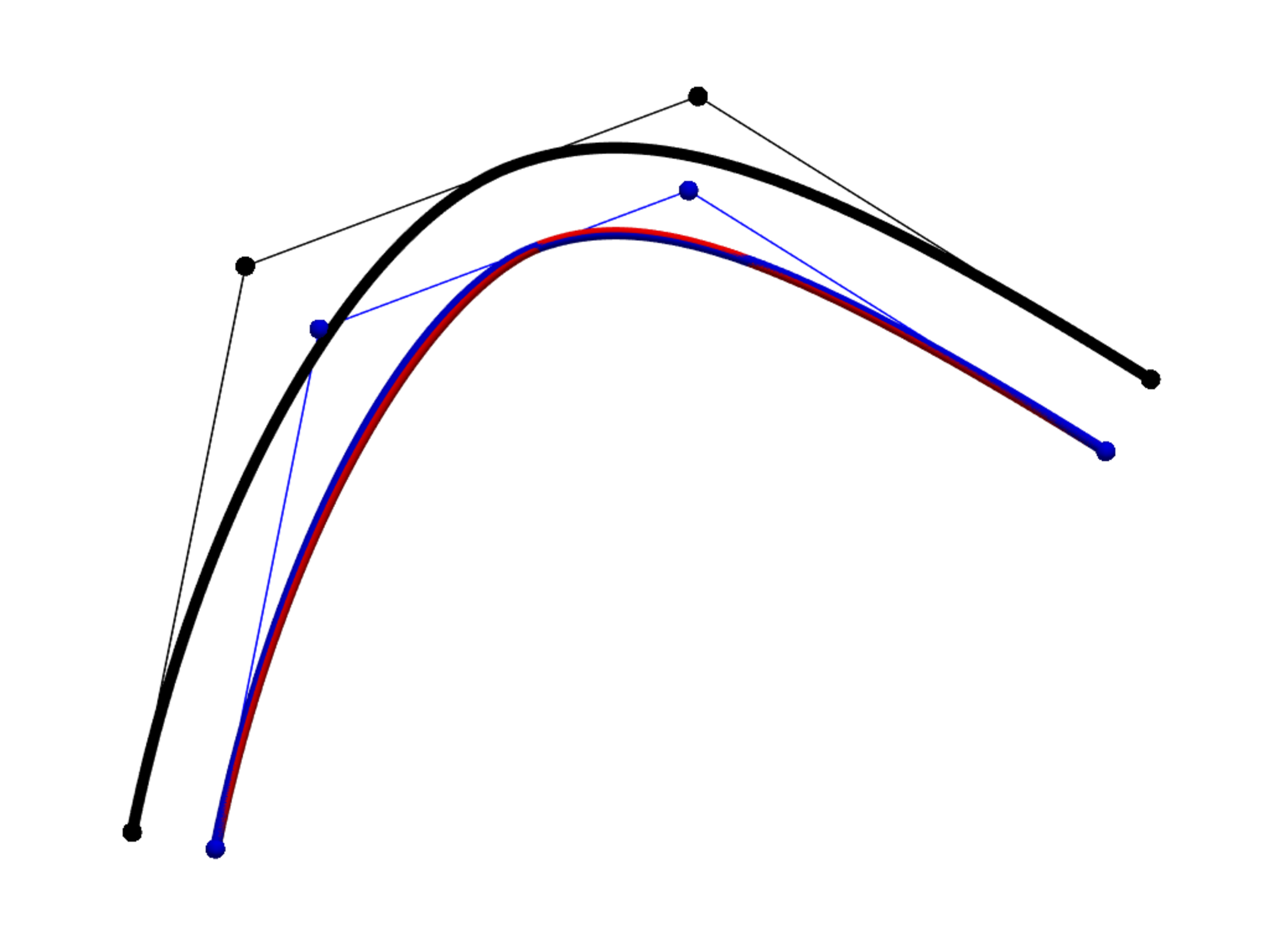
        \end{minipage}

        \caption{}
        \label{fig:translate_cps}
    \end{subfigure}

    \begin{subfigure}{0.45\textwidth}
        \centering

        \begin{minipage}{0.5\linewidth}
            \def\svgwidth{\linewidth}
            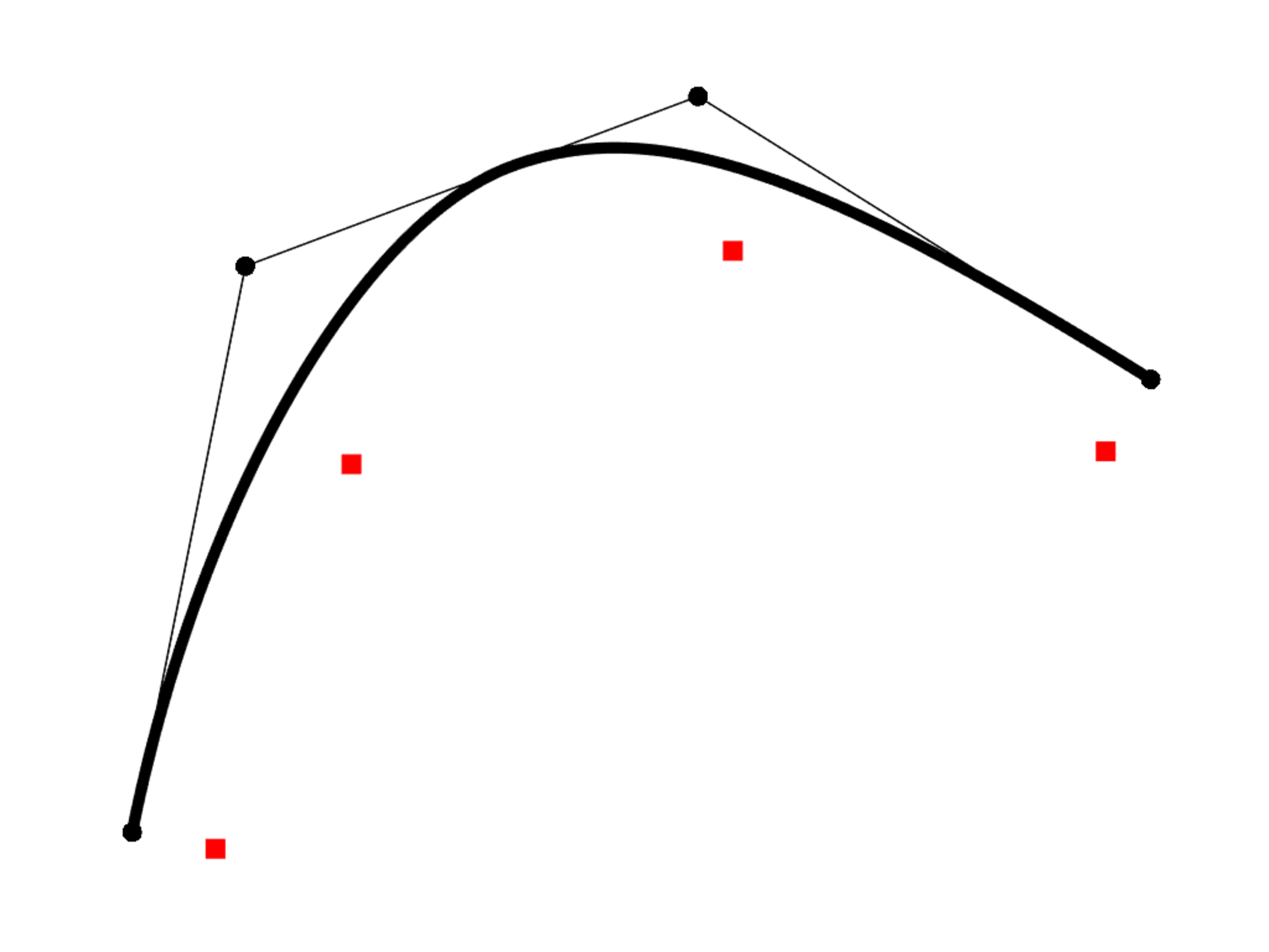
        \end{minipage}%
        \hfill
        \begin{minipage}{0.5\linewidth}
            \def\svgwidth{\linewidth}
            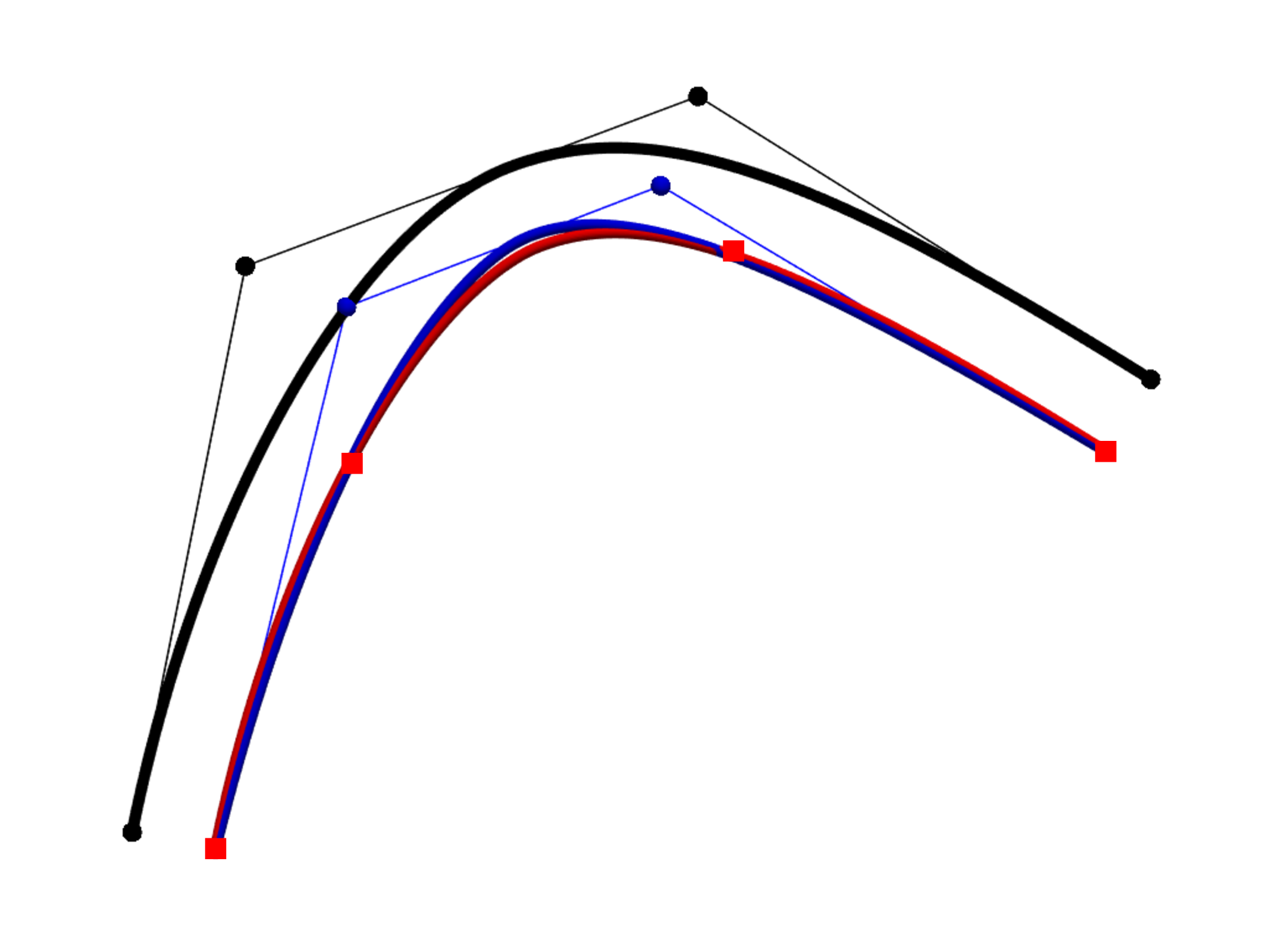
        \end{minipage}

        \caption{}
        \label{fig:solve_sys}
    \end{subfigure}
    \hfill
    \begin{subfigure}{0.45\textwidth}
        \centering

        \begin{minipage}{0.5\linewidth}
            \def\svgwidth{\linewidth}
            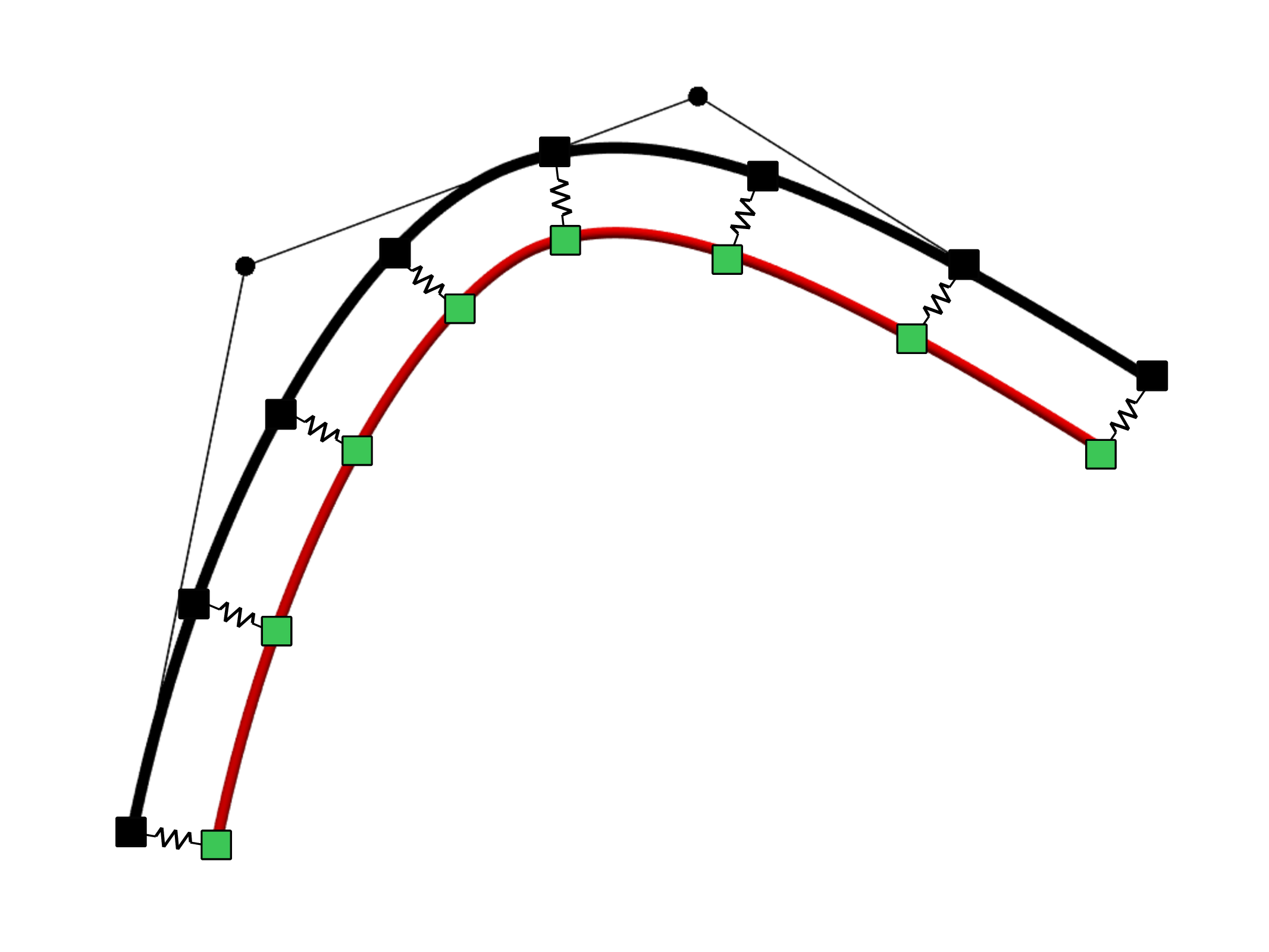
        \end{minipage}%
        \hfill
        \begin{minipage}{0.5\linewidth}
            \def\svgwidth{\linewidth}
            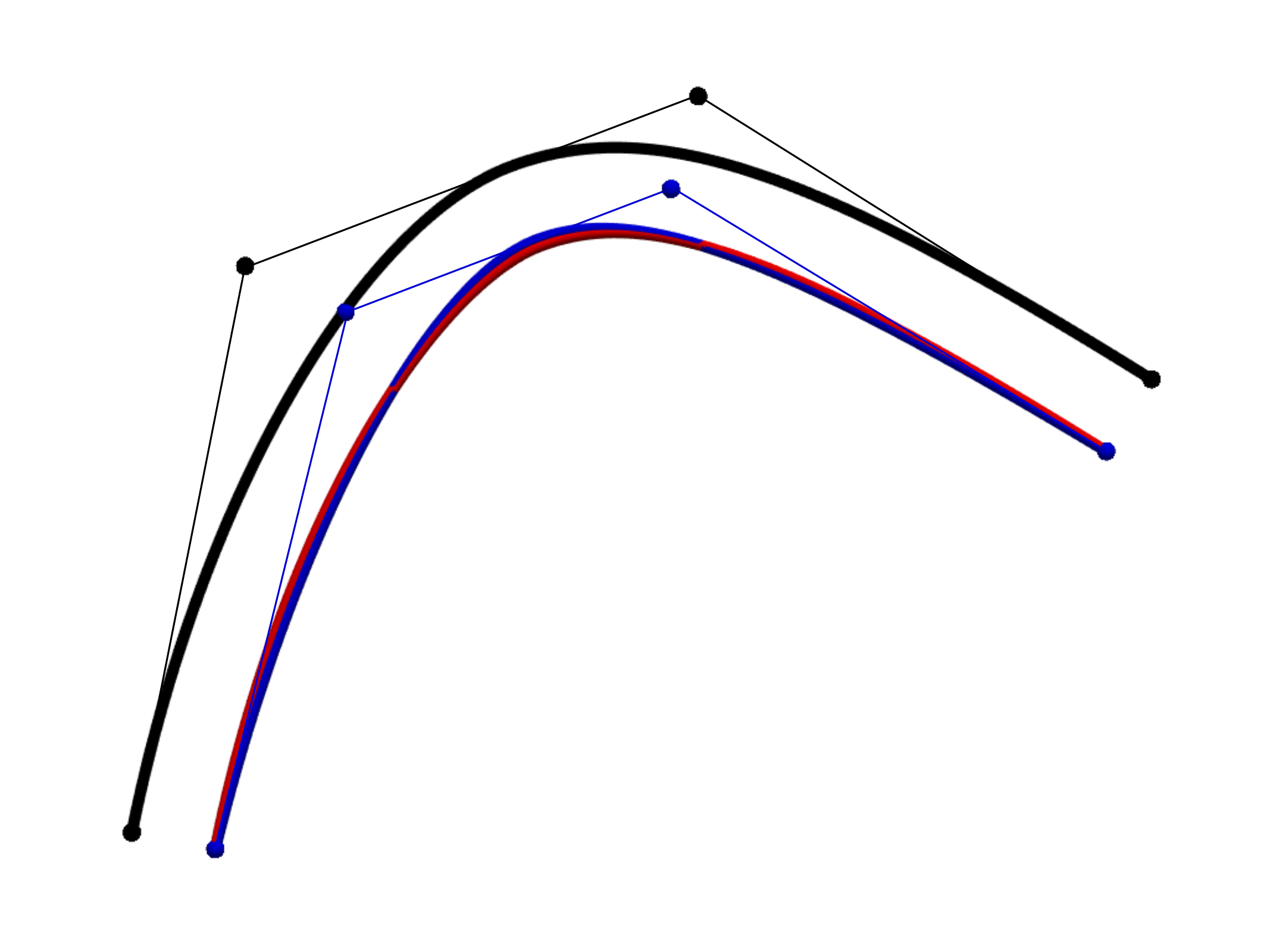
        \end{minipage}

        \caption{}
        \label{fig:grad_des}
    \end{subfigure}

    \caption{Implemented approaches for the offset of a \brep{}. For all methods, the resulting offset (blue) is compared qualitatively with the analytical solution (red). 
    (a) Quadratic NURBS base curve with knot vector $\afirstknotvector{}= \{0,0,0,0.5,1,1,1\}$.  
    (b) Control-polygon translation, where intersections of the line segments are marked by \protect\tikz \protect\draw[red, fill=red] (0,0) circle (0.5ex);.  
    (c) Interpolation method with $\grevilleabscissae{} = \{ 0, 0.33, 0.66, 1\}$. The sampling points~\evalpoints{,i} are denoted by \protect\tikz \protect\draw[fill=red, draw=red] (0,0) rectangle (1ex,1ex);.  
    (d) Optimization of a spring system: sampling points of the base curve \basepoints{i} and offset~\evalpoints{,i} are denoted by \protect\tikz \protect\draw[fill=black, draw=black] (0,0) rectangle (1ex,1ex); and \protect\tikz \protect\draw[fill=green, draw=black] (0,0) rectangle (1ex,1ex);, respectively.  
    Control points of the base \controlpoints{i} and offset \controlpointsoffset{,i} are indicated by \protect\tikz \protect\draw[black, fill=black] (0,0) circle (0.5ex); and \protect\tikz \protect\draw[blue, fill=blue] (0,0) circle (0.5ex);, respectively.}
    \label{fig:offset_approaches}
\end{figure}

In the following, only NURBS geometries represented by open knot vectors are considered. 
Although the extension of the presented approaches to NURBS surfaces is straightforward, 
the methodology is illustrated using a two-dimensional NURBS curve shown in 
Figure~\ref{fig:base_curve}. The curve has polynomial degree~$p = 2$ and is defined by 
the knot vector $\afirstknotvector{} = \{0, 0, 0, 0.5, 1, 1, 1\}$ and a set of control 
points $\controlpoints{i} \in \mathbb{R}^{2}$ and weights $\weightnurbs{i}$, 
with $i=1,\dots,4$. These values are listed in Table~\ref{table:position_cps_curve}.
To avoid self-intersecting offsets, the offset distance $\offsetdistance{}$ is assumed 
to be smaller than the global minimum radius of curvature of the geometry. In all 
approaches, the offset control points $\controlpointsoffset{}$ inherit the same weights 
as their corresponding control points of the base \brep{}.

\begin{table}
    \centering
    \caption{Control points $\controlpoints{i}$ and weights $\weightnurbs{i}$ of a quadratic NURBS curve.}
    \begin{tabular}{ccc}
        \toprule
        i & coordinates &  \weightnurbs{i} \\
        \midrule
        1 & (0.0, 0.0) & 1.0  \\
        2 & (0.2, 1.0) & $1/\sqrt{2}$ \\
        3 & (1.0, 1.3) & 1.0 \\
        4 & (1.8, 0.8) & $1/\sqrt{2}$ \\
        \bottomrule
    \end{tabular}
    \label{table:position_cps_curve}
\end{table}

\begin{enumerate}
    \item Following the approach of \cite{Tiller1984}, the first 
    method operates directly on the control polygon of the base \brep{}. 
    For a NURBS curve, each line segment of the control polygon is translated 
    by the offset distance $\offsetdistance{}$ in the direction normal to the segment, 
    as illustrated in Figure~\ref{fig:translate_cps}. 
    The new control points are then obtained from the intersections of the translated segments.

    For NURBS surfaces, the method is extended by translating the control polygon, 
    which is composed of planar facets. Each control point is determined as the 
    intersection of its adjacent offset planes. Since a control point is connected to 
    four planes, the intersection of any three adjacent planes is sufficient to uniquely 
    determine its position; the specific choice of planes is arbitrary.

    \item The second method is based on the approach presented in \cite{Piegl1999} and 
    illustrated in Figure~\ref{fig:solve_sys}. 
    It determines the offset control points by interpolating a set of points 
    $\evalpoints{} = [ \evalpoints{,1}, \evalpoints{,2}, \dots, \evalpoints{,n} ]^{\transpose}$. 
    These points are obtained by evaluating the offset curve at the Greville abscissae 
    $\grevilleabscissae{} = \{\agrevillevalue{1}, \agrevillevalue{2}, \dots, \agrevillevalue{n}\}$ \cite{Greville1969}, 
    defined as
    \begin{equation}
        \agrevillevalue{i} = \frac{1}{p} \sum_{k = 1}^{p} \aparameterspace{i+k}, \quad i = 1,\dots,n.
    \end{equation}
    The corresponding offset points are computed as
    \begin{equation}
        \evalpoints{,i} = \acurve{\agrevillevalue{i}} + \offsetdistance{}\aninwardsnormalvec{\agrevillevalue{i}}
        \label{eq:offset_point_evaluation}
    \end{equation}
    and subsequently interpolated using the same NURBS shape-function matrix $\nurbsmatrix$ 
    employed for the base curve. The offset control points 
    $\controlpointsoffset{} = [\mathbf{P}_{1}, \dots, \mathbf{P}_{n}]^{\transpose}$ 
    are therefore obtained by solving
    \begin{equation}
        \nurbsmatrix \controlpointsoffset{} = \evalpoints{}.
    \end{equation}

    \item The third method follows the approach proposed in~\cite{Nguyen2014}, 
    which involves solving an optimization problem to determine the position of the control 
    points that best approximate the offset curve. The procedure begins by computing 
    the exact offset position \evalpoints{,i} at a sampling point~$\xi_{,i}$ along the 
    base curve, as done in \eqref{eq:offset_point_evaluation}. The number of sampling 
    points $m$ is user defined. Next, a system of elastic springs is defined, where each 
    spring connects an offset point \evalpoints{,i} to its corresponding base-curve point
     $\basepoints{i} = \acurve{\xi_{,i}}$. The resulting spring system is shown in 
     Figure~\ref{fig:grad_des}. For a $m$ number of springs (with stiffness $k=1$), 
     the total energy of the spring system is defined as 
     \begin{equation} 
        E(\controlpointsoffset{}) = \frac{1}{2}\sum_{i = 1}^{m} (u_{i}(\controlpointsoffset{}))^{2}, 
    \end{equation} 
    where the displacement of a spring $u_{i}(\controlpointsoffset{})$ is defined as 
    \begin{equation} 
        u_{i}(\controlpointsoffset{}) = \| \evalpoints{,i} - \basepoints{i}(\controlpointsoffset{}) \| . 
    \end{equation} 
    The energy of the spring system $E(\controlpointsoffset{})$ serves as the objective function 
    to be minimized using the gradient descent method. During the optimization process, 
    the positions of the control points \controlpointsoffset{} are iteratively 
    updated to minimize the objective function, thereby yielding a curve that 
    closely approximates the offset. 

\end{enumerate}

Figure~\ref{fig:offset_approaches} illustrates the offset curves obtained with 
the different approaches. All methods provide a qualitatively accurate 
approximation of the analytical offset. While such accuracy is sufficient 
for the intended application, we provide a quantitative comparison of the proposed 
approaches in the following to assess ther interpolation quality compared to each other.

To this end, the pointwise distance between the approximated offset curve 
$\acurveoffsetapproximation(\aparameterspace{})$ and the analytical offset is evaluated. 
At a parametric location $\aparameterspace{}$, the distance is defined as
\begin{equation}
    d(\aparameterspace{}) = || \acurveoffsetapproximation{}(\aparameterspace{}) - (\acurve{\aparameterspace{}} + \offsetdistance{}\aninwardsnormalvec{\aparameterspace{}}) ||.
\end{equation}

Local deviations are quantified by the maximum error 
$\errorinfinite{}$, computed from $d(\aparameterspace{})$ by densely sampling 
the parametric domain of $\acurveoffsetapproximation(\aparameterspace{})$. 
To assess the global approximation quality, a length-normalized~$L^2$-type 
error measure is introduced,
\begin{equation}
    \errorrms{} = \left(\frac{1}{s_o} \int_{\Gamma^*} d(\aparameterspace{})^2 ds\right)^{1/2},
    \label{eq:errorrms_curve}
\end{equation}
where $s_o$ denotes the arc-length of $\acurveoffsetapproximation(\aparameterspace{})$.

The error measures are evaluated for the results shown in 
Figure~\ref{fig:offset_approaches} and summarized in 
Table~\ref{table:errors_curve}. For the considered example, the 
optimization-based approach exhibits the smallest local deviations, as 
measured by $\errorinfinite{}$, and achieves the highest global accuracy 
in terms of $\errorrms{}$. It is closely followed by the interpolation-based 
method, whereas the control-polygon translation approach yields larger 
deviations according to both metrics.

\begin{table}
    \centering
    \caption{Error measures for the considered offset methods applied to a NURBS curve.}
    \begin{tabular}{lcc}
        \toprule
        Offset method & \errorinfinite{} & \errorrms{} \\
        \midrule
        Control-polygon translation & $3.599\mathrm{e}{-2}$ & $2.320\mathrm{e}{-2}$ \\
        Interpolation                & $2.646\mathrm{e}{-2}$ & $1.095\mathrm{e}{-2}$ \\
        Optimization                 & $1.386\mathrm{e}{-2}$ & $9.724\mathrm{e}{-3}$ \\
        \bottomrule
    \end{tabular}
    \label{table:errors_curve}
\end{table}

As previously mentioned, the offset approaches can be extended to NURBS surfaces. 
The performance of the methods is demonstrated using a NURBS surface \brep{} 
with the following characteristics: both parametric directions have polynomial 
degrees $p=q=3$, and the corresponding knot vectors 
are~$\afirstknotvector{}~=~\asecondknotvector{} = 
\{0.0, 0.0, 0.0, 0.0, \tfrac{1}{3}, \tfrac{2}{3}, 1.0, 1.0, 1.0, 1.0\}$. 
The coordinates of the control points are listed in Table~\ref{table:position_cps_surf}, 
and all control points have a weight of $\weightnurbs{i,j} = 1.0$. 
The cubic NURBS \brep{} is then offset using an offset thickness 
$\offsetdistance{} = 3$.

\begin{table}
    \centering
    \caption{Control points $\controlpoints{i,j}$ of a cubic NURBS surface.}
    \begin{tabular}{ccc ccc}
        \toprule
        $i$ & $j$ & coordinates & $i$ & $j$ & coordinates \\
        \midrule
        1 & 1 & (-25, -25, -10) & 4 & 1 & (-25, 5, 0)   \\
        1 & 2 & (-15, -25, -8)  & 4 & 2 & (-15, 5, -4)  \\
        1 & 3 & (-5, -25, -5)   & 4 & 3 & (-5, 5, -8)   \\
        1 & 4 & (5, -25, -3)    & 4 & 4 & (5, 5, -8)    \\
        1 & 5 & (15, -25, -8)   & 4 & 5 & (15, 5, -4)   \\
        1 & 6 & (25, -25, -10)  & 4 & 6 & (25, 5, 2)    \\
        2 & 1 & (-25, -15, -5)  & 5 & 1 & (-25, 15, -5) \\
        2 & 2 & (-15, -15, -4)  & 5 & 2 & (-15, 15, -4) \\
        2 & 3 & (-5, -15, -3)   & 5 & 3 & (-5, 15, -3)  \\
        2 & 4 & (5, -15, -2)    & 5 & 4 & (5, 15, -2)   \\
        2 & 5 & (15, -15, -4)   & 5 & 5 & (15, 15, -4)  \\
        2 & 6 & (25, -15, -5)   & 5 & 6 & (25, 15, -5)  \\
        3 & 1 & (-25, -5, 0)    & 6 & 1 & (-25, 25, -10)\\
        3 & 2 & (-15, -5, -4)   & 6 & 2 & (-15, 25, -8) \\
        3 & 3 & (-5, -5, -8)    & 6 & 3 & (-5, 25, -5)  \\
        3 & 4 & (5, -5, -8)     & 6 & 4 & (5, 25, -3)   \\
        3 & 5 & (15, -5, -4)    & 6 & 5 & (15, 25, -8)  \\
        3 & 6 & (25, -5, 2)     & 6 & 6 & (25, 25, -10) \\
        \bottomrule
    \end{tabular}
    \label{table:position_cps_surf}
\end{table}

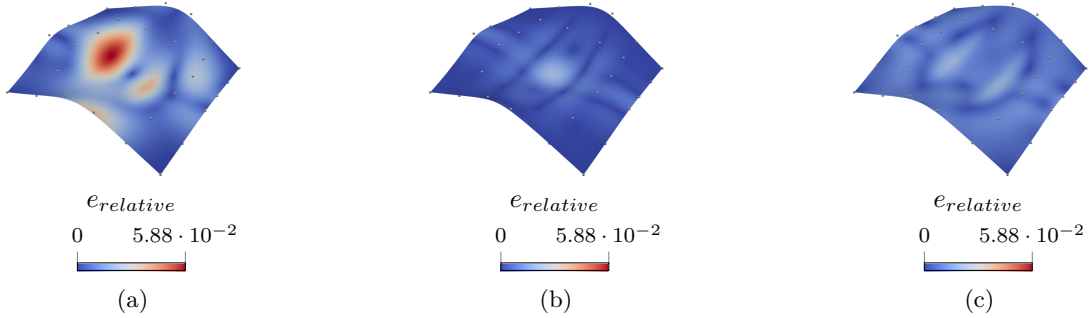
\begin{figure}
	\centering
	\begin{subfigure}{0.3\textwidth}
		\import{fig/discretization_workflow/}{curved_surf_geometric.tex}
		\caption{}
	\end{subfigure}
	\hfill
	\begin{subfigure}{0.3\textwidth}
		\import{fig/discretization_workflow/}{curved_surf_solve_system.tex}
		\caption{}
	\end{subfigure}
    \hfill
	\begin{subfigure}{0.3\textwidth}
		\import{fig/discretization_workflow/}{curved_surf_gradient_descent_method.tex}
		\caption{}
	\end{subfigure}
	
	\caption{Relative error of the offset of a curved surface obtained by different offset methods: (a)~control polygon translation, (b) interpolation method, (c) optimization of a spring system.}
	\label{fig:offset_cases_3d}
\end{figure}

For this geometry, the relative error between the computed offset and the 
analytical solution is shown in Figure~\ref{fig:offset_cases_3d} for the 
different approaches. The largest deviations are observed for the 
control-polygon translation method, although the relative error remains 
below $6\%$. While both the interpolation and optimization approaches 
deliver similar overall accuracy, the optimization method exhibits 
wider regions with locally increased relative error compared to the 
interpolation method.

A more detailed assessment of the local and global accuracy is obtained 
by evaluating the maximum error $\errorinfinite{}$ of the distance 
$d(\aparameterspace{}, \eta)$ and an area-normalized $L^2$-type error, 
analogous to~\eqref{eq:errorrms_curve}. The resulting values are summarized 
in Table~\ref{table:errors_surface}. The optimization method yields the 
smallest local deviation, whereas the interpolation method achieves the 
smallest global error. The control-polygon translation method produces 
the largest deviations in both metrics.

\begin{table}
    \centering
    \caption{Error measures for the considered offset methods applied to a NURBS surface.}
    \begin{tabular}{lcc}
        \toprule
        Offset method & \errorinfinite{} & \errorrms{} \\
        \midrule
        Control-polygon translation & $7.403\mathrm{e}{-1}$ & $2.491\mathrm{e}{-1}$ \\
        Interpolation                & $2.082\mathrm{e}{-1}$ & $4.993\mathrm{e}{-2}$ \\
        Optimization                 & $1.127\mathrm{e}{-1}$ & $6.225\mathrm{e}{-2}$ \\
        \bottomrule
    \end{tabular}
    \label{table:errors_surface}
\end{table}

More accurate offsets can be achieved by refining the base \brep{} 
prior to applying the offset operation. In addition, for the optimization 
approach, the number and distribution of sampling points influence the 
quality of the offset, as discussed in~\cite{Nguyen2014}. However, a 
systematic investigation of these aspects is beyond the scope of the 
present work. For the intended application, the offset method is considered 
sufficient if the approximation error remains below a prescribed 
threshold and does not lead to distorted meshes.

\begin{figure}
    \newcommand{\figextrusionsize}{0.45}
    \centering
  
    \subfloat[]{%
        \def\svgwidth{\figextrusionsize\textwidth}
        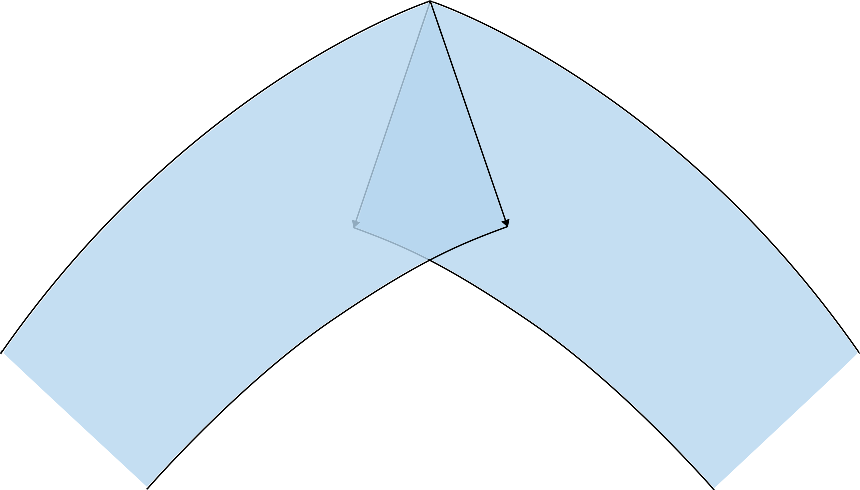
        \label{subfig:average_1}
    }
    \subfloat[]{%
        \def\svgwidth{\figextrusionsize\textwidth}
        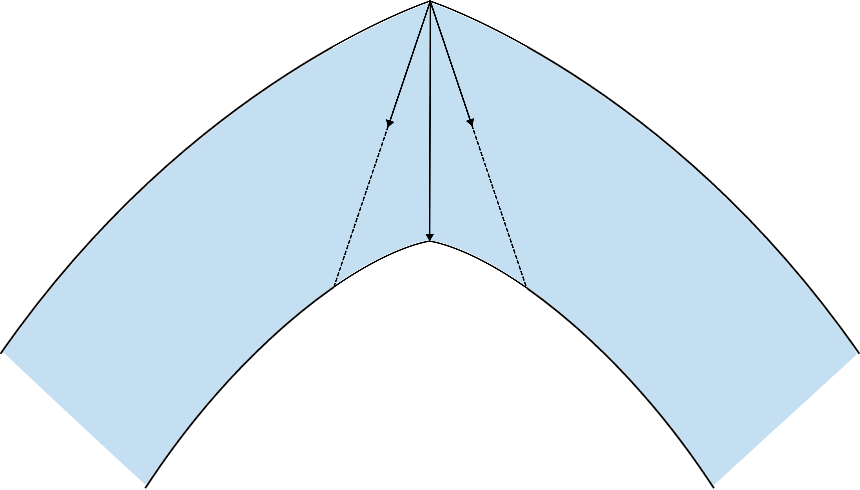
        \label{subfig:average_2}
    }
    \caption{Coupling adjacent patches: (a) initial state after offsetting, (b) averaging of edges. 
    First, the normal \aninwardsnormalvecwithindex{Pi} at each patch is evaluated at the edge and an average 
    normal $\tilde{\aninwardsnormalvecwithindex{}}$ is evaluated. }
    \label{fig:average}
\end{figure}

In general, complex geometries cannot be represented by a single patch 
but instead require multiple patches. In the following, we consider geometries 
composed of multiple patches that are geometrically coincident and 
parametrically conforming at their interfaces, \ie{} they share the same knot 
vectors, polynomial degrees, and number of control points. When adjacent 
patches lack $C^1$ continuity, for instance at sharp corners, their boundary 
normals differ, which leads to mismatched boundaries after offsetting, as 
illustrated in Figure~\ref{subfig:average_1} for two adjacent patches. To 
address this issue, the normals along the common edge 
$\normalatpatchedge{j}$ are averaged across all adjacent patches $j$ 
during the offset operation. This ensures coincident offset boundaries, 
as shown in Figure~\ref{subfig:average_2}.

\begin{figure}
    \begin{subfigure}[]{0.3\textwidth}
        \begin{tikzpicture}
            \node[anchor=south west, inner sep=0] (image) at (0, 0) 
                {\includegraphics[scale=0.55]{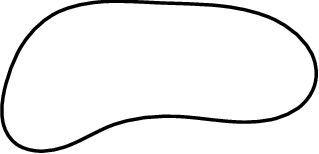}};
        
            \draw[black, dashed] (image.south west) rectangle (image.north east);

            \node[draw, circle, black, inner sep=1pt, label=below:{$(x_{min}, y_{min})$}] (circlesw) at (image.south west) {};
            \node[draw, circle, black, inner sep=1pt, label=below:{$(x_{max}, y_{min})$}] (circlese) at (image.south east) {};
            \node[draw, circle, black, inner sep=1pt, label=above:{$(x_{min}, y_{max})$}] (circlenw) at (image.north west) {};
            \node[draw, circle, black, inner sep=1pt, label=above:{$(x_{max}, y_{max})$}] (circlene) at (image.north east) {};
        \end{tikzpicture}
        \caption{}
        \label{subfig:bounding-box}
    \end{subfigure}
    \hfill
    \begin{subfigure}[]{0.3\textwidth}
        \centering

        \newdimen\imagewidth
        \newdimen\imageheight
        \begin{tikzpicture}
            \definecolor{customred}{RGB}{238,147,146}

            \fill[customred] (image.south west) rectangle (image.north east);
            
            \pgfextractx{\imagewidth}{\pgfpointanchor{image}{east}}
            \pgfextracty{\imageheight}{\pgfpointanchor{image}{north}}
            \draw[xstep=\imagewidth / 6,ystep=\imageheight / 3,black,thin] (image.south west) grid (image.north east);   

            \node[anchor=south west, inner sep=0] (image) at (0, 0) 
            {
                \includegraphics[scale=0.55]{fig/discretization_workflow/gamma_star_.eps}
            };
        \end{tikzpicture}
        \caption{}
        \label{subfig:cartesian-mesh}
    \end{subfigure}
    \hfill
    \begin{subfigure}[]{0.3\textwidth}
        \centering

        \newdimen\imagewidth
        \newdimen\imageheight
        \begin{tikzpicture}
            \definecolor{customred}{RGB}{238,147,146}

            \node[anchor=south west, inner sep=0] (image) at (0, 0) 
            {\includegraphics[scale=0.55]{fig/discretization_workflow/gamma_star_.eps}};

            \fill[customred] (image.south west) rectangle (image.north east);
            
            \pgfextractx{\imagewidth}{\pgfpointanchor{image}{east}}
            \pgfextracty{\imageheight}{\pgfpointanchor{image}{north}}
            \draw[xstep=\imagewidth / 6,ystep=\imageheight / 3,black,thin] (image.south west) grid (image.north east);   
            
            \node[anchor=south west, inner sep=0] (image) at (-0.4, -0.4) 
            {
                \includegraphics[scale=0.55]{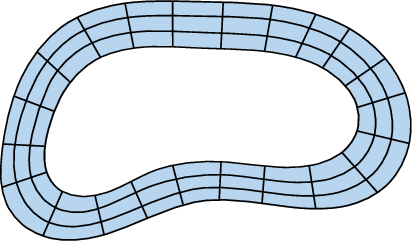}
            };

            \node[anchor=south west, inner sep=0] (image) at (0, 0) 
            {
                \includegraphics[scale=0.55]{fig/discretization_workflow/gamma_star_.eps}
            };
        \end{tikzpicture}
        \caption{}
        \label{subfig:complete-discret}
    \end{subfigure}

    \caption{Generation of a Cartesian mesh: (a) definition of the AABB around $\interfacestar{\abodyi}{}$. (b) Discretization of an AABB using hexahedra-elements. (c) Complete discretization of the domain: boundary layer mesh and Cartesian mesh.}
    \label{fig:discretization-pipeline}
\end{figure}

\begin{remark}
    Non-smooth geometries, such as sharp corners and edges, are generally 
    undesirable in surface-based contact formulations relying on mortar 
    (or other STS) approaches. This is because the weak enforcement of the 
    contact constraints leads to a surface-based weighting of the gap function, 
    which may result in large penetrations at vertices and sharp edges. 
    To address this issue, a robust and accurate mortar-based finite element 
    framework that unifies vertex, edge, and surface contact was proposed 
    in~\cite{Farah2018}. In that framework, three appropriate sets of Lagrange 
    multipliers are defined for the different geometric entities 
    (vertices, edges, and surfaces), together with a variationally consistent 
    discretization. This methodology can, in principle, be extended to the 
    non-smooth geometries arising in NURBS discretizations. However, a detailed 
    discussion of such an extension lies beyond the scope of the present work.
\end{remark}

\subsection{Generation of a Cartesian mesh}

After constructing the boundary layer mesh, the next step is the 
discretization of the bulk domain. For simplicity, a Cartesian mesh 
is employed. It is generated by defining an axis-aligned bounding box 
(AABB) around the interface $\interfacestar{\abodyi}{}$, as illustrated 
for a two-dimensional configuration in Figure~\ref{subfig:bounding-box}. 
The AABB is obtained by determining the minimum and maximum spatial 
coordinates of $\interfacestar{\abodyi}{}$, from which the bounding-box 
limits are defined. The resulting AABB is then discretized in a straightforward manner 
using either classical hexahedral elements or an isogeometric patch, 
as shown in Figure~\ref{subfig:cartesian-mesh}. The complete discretization 
strategy, comprising the boundary layer mesh and the Cartesian background 
mesh, is illustrated in Figure~\ref{subfig:complete-discret}. As part of the coupling 
procedure, the Cartesian mesh is subsequently 
intersected by $\interfacestar{\abodyi}{}$ through geometric cut operations. 
This procedure is discussed in more detail in 
Subsection~\ref{subsection:numerical_integration_cutelements}.

%% file: fig/offset/progenitor_curve.eps_tex
\begingroup%
  \makeatletter%
  \providecommand\color[2][]{%
    \errmessage{(Inkscape) Color is used for the text in Inkscape, but the package 'color.sty' is not loaded}%
    \renewcommand\color[2][]{}%
  }%
  \providecommand\transparent[1]{%
    \errmessage{(Inkscape) Transparency is used (non-zero) for the text in Inkscape, but the package 'transparent.sty' is not loaded}%
    \renewcommand\transparent[1]{}%
  }%
  \providecommand\rotatebox[2]{#2}%
  \newcommand*\fsize{\dimexpr\f@size pt\relax}%
  \newcommand*\lineheight[1]{\fontsize{\fsize}{#1\fsize}\selectfont}%
  \ifx\svgwidth\undefined%
    \setlength{\unitlength}{1024.00001526bp}%
    \ifx\svgscale\undefined%
      \relax%
    \else%
      \setlength{\unitlength}{\unitlength * \real{\svgscale}}%
    \fi%
  \else%
    \setlength{\unitlength}{\svgwidth}%
  \fi%
  \global\let\svgwidth\undefined%
  \global\let\svgscale\undefined%
  \makeatother%
  \begin{picture}(1,0.75)%
    \lineheight{1}%
    \setlength\tabcolsep{0pt}%
    \put(0,0){\includegraphics[width=\unitlength]{progenitor_curve.eps}}%
    \put(0.07376959,0.00634921){\makebox(0,0)[lt]{\lineheight{1.25}\smash{\begin{tabular}[t]{l}$\mathbf{P}_1$\end{tabular}}}}%
    \put(0.08231748,0.55636655){\makebox(0,0)[lt]{\lineheight{1.25}\smash{\begin{tabular}[t]{l}$\mathbf{P}_2$\end{tabular}}}}%
    \put(0.5493477,0.6878549){\makebox(0,0)[lt]{\lineheight{1.25}\smash{\begin{tabular}[t]{l}$\mathbf{P}_3$\end{tabular}}}}%
    \put(0.91029513,0.46832696){\makebox(0,0)[lt]{\lineheight{1.25}\smash{\begin{tabular}[t]{l}$\mathbf{P}_4$\end{tabular}}}}%
  \end{picture}%
\endgroup%

%% file: fig/offset/translate_cps/cp_grid_method_1.eps_tex
\begingroup%
  \makeatletter%
  \providecommand\color[2][]{%
    \errmessage{(Inkscape) Color is used for the text in Inkscape, but the package 'color.sty' is not loaded}%
    \renewcommand\color[2][]{}%
  }%
  \providecommand\transparent[1]{%
    \errmessage{(Inkscape) Transparency is used (non-zero) for the text in Inkscape, but the package 'transparent.sty' is not loaded}%
    \renewcommand\transparent[1]{}%
  }%
  \providecommand\rotatebox[2]{#2}%
  \newcommand*\fsize{\dimexpr\f@size pt\relax}%
  \newcommand*\lineheight[1]{\fontsize{\fsize}{#1\fsize}\selectfont}%
  \ifx\svgwidth\undefined%
    \setlength{\unitlength}{1024.00001526bp}%
    \ifx\svgscale\undefined%
      \relax%
    \else%
      \setlength{\unitlength}{\unitlength * \real{\svgscale}}%
    \fi%
  \else%
    \setlength{\unitlength}{\svgwidth}%
  \fi%
  \global\let\svgwidth\undefined%
  \global\let\svgscale\undefined%
  \makeatother%
  \begin{picture}(1,0.75)%
    \lineheight{1}%
    \setlength\tabcolsep{0pt}%
    \put(0,0){\includegraphics[width=\unitlength]{cp_grid_method_1.eps}}%
    \put(0.23,0.33){\makebox(0,0)[lt]{\lineheight{1.25}\smash{\begin{tabular}[t]{l}$\mathbf{n}_{l1}$\end{tabular}}}}%
    \put(0.37,0.5){\makebox(0,0)[lt]{\lineheight{1.25}\smash{\begin{tabular}[t]{l}$\mathbf{n}_{l2}$\end{tabular}}}}%
    \put(0.55,0.51){\makebox(0,0)[lt]{\lineheight{1.25}\smash{\begin{tabular}[t]{l}$\mathbf{n}_{l3}$\end{tabular}}}}%
    \put(0.1,-0.01433298){\makebox(0,0)[lt]{\lineheight{1.25}\smash{\begin{tabular}[t]{l}$\offsetdistance$\end{tabular}}}}%
  \end{picture}%
\endgroup%

%% file: fig/offset/translate_cps/cp_grid_method_result.eps_tex
\begingroup%
  \makeatletter%
  \providecommand\color[2][]{%
    \errmessage{(Inkscape) Color is used for the text in Inkscape, but the package 'color.sty' is not loaded}%
    \renewcommand\color[2][]{}%
  }%
  \providecommand\transparent[1]{%
    \errmessage{(Inkscape) Transparency is used (non-zero) for the text in Inkscape, but the package 'transparent.sty' is not loaded}%
    \renewcommand\transparent[1]{}%
  }%
  \providecommand\rotatebox[2]{#2}%
  \newcommand*\fsize{\dimexpr\f@size pt\relax}%
  \newcommand*\lineheight[1]{\fontsize{\fsize}{#1\fsize}\selectfont}%
  \ifx\svgwidth\undefined%
    \setlength{\unitlength}{1024.00001526bp}%
    \ifx\svgscale\undefined%
      \relax%
    \else%
      \setlength{\unitlength}{\unitlength * \real{\svgscale}}%
    \fi%
  \else%
    \setlength{\unitlength}{\svgwidth}%
  \fi%
  \global\let\svgwidth\undefined%
  \global\let\svgscale\undefined%
  \makeatother%
  \begin{picture}(1,0.75)%
    \lineheight{1}%
    \setlength\tabcolsep{0pt}%
    \put(0,0){\includegraphics[width=\unitlength]{cp_grid_method_result.eps}}%
  \end{picture}%
\endgroup%

%% file: fig/offset/solve_system/solve_sys_method_1.eps_tex
\begingroup%
  \makeatletter%
  \providecommand\color[2][]{%
    \errmessage{(Inkscape) Color is used for the text in Inkscape, but the package 'color.sty' is not loaded}%
    \renewcommand\color[2][]{}%
  }%
  \providecommand\transparent[1]{%
    \errmessage{(Inkscape) Transparency is used (non-zero) for the text in Inkscape, but the package 'transparent.sty' is not loaded}%
    \renewcommand\transparent[1]{}%
  }%
  \providecommand\rotatebox[2]{#2}%
  \newcommand*\fsize{\dimexpr\f@size pt\relax}%
  \newcommand*\lineheight[1]{\fontsize{\fsize}{#1\fsize}\selectfont}%
  \ifx\svgwidth\undefined%
    \setlength{\unitlength}{1024.00001526bp}%
    \ifx\svgscale\undefined%
      \relax%
    \else%
      \setlength{\unitlength}{\unitlength * \real{\svgscale}}%
    \fi%
  \else%
    \setlength{\unitlength}{\svgwidth}%
  \fi%
  \global\let\svgwidth\undefined%
  \global\let\svgscale\undefined%
  \makeatother%
  \begin{picture}(1,0.75)%
    \lineheight{1}%
    \setlength\tabcolsep{0pt}%
    \put(0,0){\includegraphics[width=\unitlength]{solve_sys_method_1.eps}}%
    \put(0.185,0.09){\makebox(0,0)[lt]{\lineheight{1.25}\smash{\begin{tabular}[t]{l}$\evalpoints{,1}$\end{tabular}}}}%
    \put(0.29774329,0.38170636){\makebox(0,0)[lt]{\lineheight{1.25}\smash{\begin{tabular}[t]{l}$\evalpoints{,2}$\end{tabular}}}}%
    \put(0.57526773,0.49){\makebox(0,0)[lt]{\lineheight{1.25}\smash{\begin{tabular}[t]{l}$\evalpoints{,3}$\end{tabular}}}}%
    \put(0.83425307,0.33){\makebox(0,0)[lt]{\lineheight{1.25}\smash{\begin{tabular}[t]{l}$\evalpoints{,4}$\end{tabular}}}}%
  \end{picture}%
\endgroup%

%% file: fig/offset/solve_system/solve_sys_method_result.eps_tex
\begingroup%
  \makeatletter%
  \providecommand\color[2][]{%
    \errmessage{(Inkscape) Color is used for the text in Inkscape, but the package 'color.sty' is not loaded}%
    \renewcommand\color[2][]{}%
  }%
  \providecommand\transparent[1]{%
    \errmessage{(Inkscape) Transparency is used (non-zero) for the text in Inkscape, but the package 'transparent.sty' is not loaded}%
    \renewcommand\transparent[1]{}%
  }%
  \providecommand\rotatebox[2]{#2}%
  \newcommand*\fsize{\dimexpr\f@size pt\relax}%
  \newcommand*\lineheight[1]{\fontsize{\fsize}{#1\fsize}\selectfont}%
  \ifx\svgwidth\undefined%
    \setlength{\unitlength}{1024.00001526bp}%
    \ifx\svgscale\undefined%
      \relax%
    \else%
      \setlength{\unitlength}{\unitlength * \real{\svgscale}}%
    \fi%
  \else%
    \setlength{\unitlength}{\svgwidth}%
  \fi%
  \global\let\svgwidth\undefined%
  \global\let\svgscale\undefined%
  \makeatother%
  \begin{picture}(1,0.75)%
    \lineheight{1}%
    \setlength\tabcolsep{0pt}%
    \put(0,0){\includegraphics[width=\unitlength]{solve_sys_method_result.eps}}%
  \end{picture}%
\endgroup%

%% file: fig/offset/gradient_descent/gradient_descend_method_1_without_initial_guess.eps_tex
\begingroup%
  \makeatletter%
  \providecommand\color[2][]{%
    \errmessage{(Inkscape) Color is used for the text in Inkscape, but the package 'color.sty' is not loaded}%
    \renewcommand\color[2][]{}%
  }%
  \providecommand\transparent[1]{%
    \errmessage{(Inkscape) Transparency is used (non-zero) for the text in Inkscape, but the package 'transparent.sty' is not loaded}%
    \renewcommand\transparent[1]{}%
  }%
  \providecommand\rotatebox[2]{#2}%
  \newcommand*\fsize{\dimexpr\f@size pt\relax}%
  \newcommand*\lineheight[1]{\fontsize{\fsize}{#1\fsize}\selectfont}%
  \ifx\svgwidth\undefined%
    \setlength{\unitlength}{1024.00001526bp}%
    \ifx\svgscale\undefined%
      \relax%
    \else%
      \setlength{\unitlength}{\unitlength * \real{\svgscale}}%
    \fi%
  \else%
    \setlength{\unitlength}{\svgwidth}%
  \fi%
  \global\let\svgwidth\undefined%
  \global\let\svgscale\undefined%
  \makeatother%
  \begin{picture}(1,0.75)%
    \lineheight{1}%
    \setlength\tabcolsep{0pt}%
    \put(0,0){\includegraphics[width=\unitlength]{gradient_descend_method_1_without_initial_guess.eps}}%
  \end{picture}%
\endgroup%

%% file: fig/offset/gradient_descent/gradient_descend_method_new_version_result.pdf_tex
\begingroup%
  \makeatletter%
  \providecommand\color[2][]{%
    \errmessage{(Inkscape) Color is used for the text in Inkscape, but the package 'color.sty' is not loaded}%
    \renewcommand\color[2][]{}%
  }%
  \providecommand\transparent[1]{%
    \errmessage{(Inkscape) Transparency is used (non-zero) for the text in Inkscape, but the package 'transparent.sty' is not loaded}%
    \renewcommand\transparent[1]{}%
  }%
  \providecommand\rotatebox[2]{#2}%
  \newcommand*\fsize{\dimexpr\f@size pt\relax}%
  \newcommand*\lineheight[1]{\fontsize{\fsize}{#1\fsize}\selectfont}%
  \ifx\svgwidth\undefined%
    \setlength{\unitlength}{1023.9999744bp}%
    \ifx\svgscale\undefined%
      \relax%
    \else%
      \setlength{\unitlength}{\unitlength * \real{\svgscale}}%
    \fi%
  \else%
    \setlength{\unitlength}{\svgwidth}%
  \fi%
  \global\let\svgwidth\undefined%
  \global\let\svgscale\undefined%
  \makeatother%
  \begin{picture}(1,0.75)%
    \lineheight{1}%
    \setlength\tabcolsep{0pt}%
    \put(0,0){\includegraphics[width=\unitlength,page=1]{gradient_descend_method_new_version_result.pdf}}%
  \end{picture}%
\endgroup%

%% file: fig/discretization_workflow/curved_surf_geometric.tex
\begin{tikzpicture}
    \node[anchor=south west,inner sep=-0.2pt] (image) at (0,0) {\includegraphics[width=\textwidth]{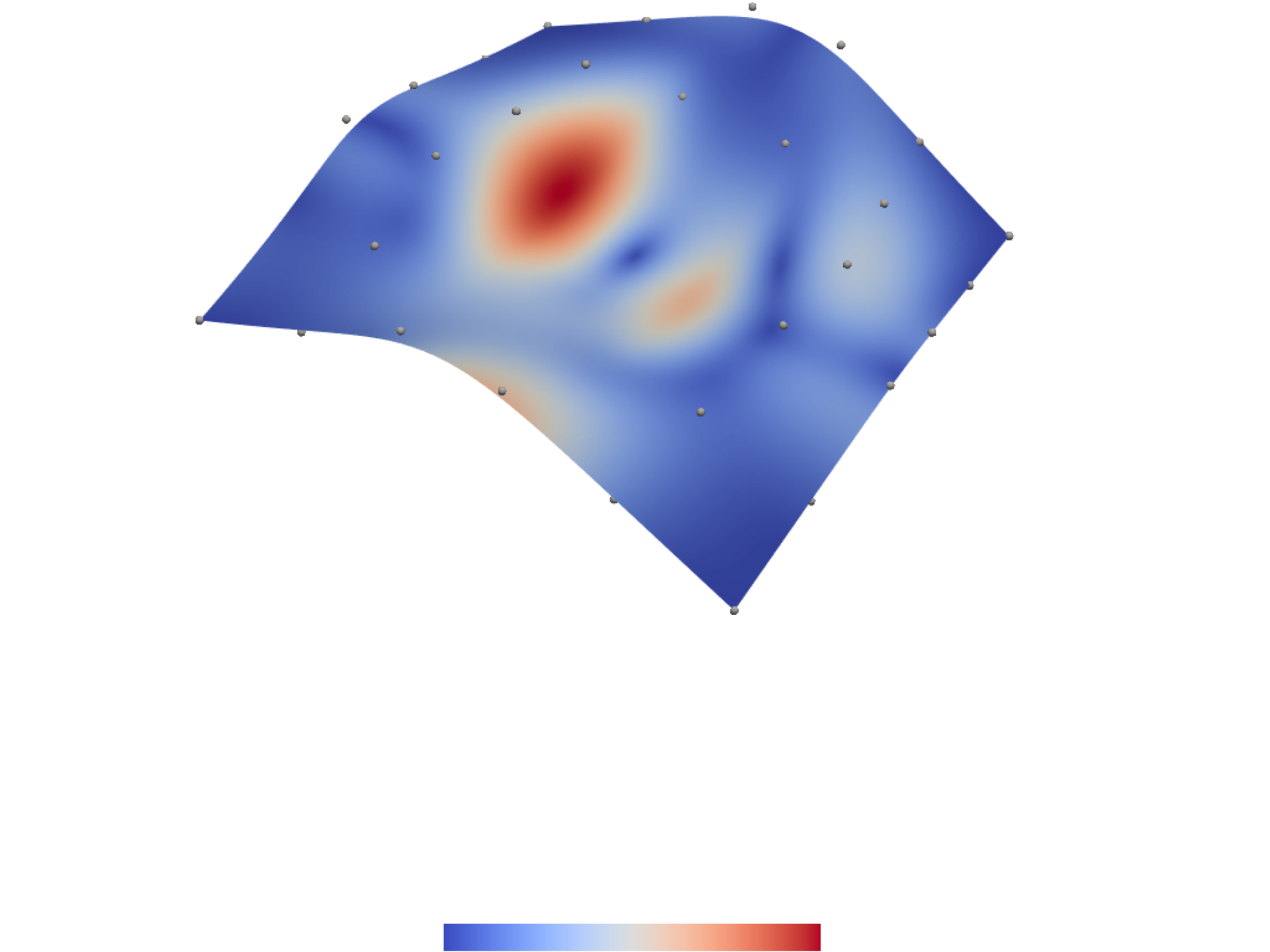}};
    \begin{axis}[
    scale only axis,
    scaled x ticks=false,
    scaled y ticks=false,
    at={(1.66cm,0.01cm)}, 
    tick label style={font=\footnotesize},
    title=$e_{relative}$,
    xticklabel=$\pgfmathprintnumber{\tick}$,
    yticklabel=$\pgfmathprintnumber{\tick}$,
    ymin=-0.009999999999761209,
    ymax=-0.00999809242784977,
    xmin=0.0,
    xmax=0.058846574284955655,
    ytick=\empty,
    height=0.1cm,
    width=1.43cm,  
    xtick={0.0,0.058846574284955655},
    xtick pos=right,
    xtick align=outside,
    title style={yshift=10pt,},
    ]
    \end{axis}
    \end{tikzpicture}%

%% file: fig/discretization_workflow/curved_surf_solve_system.tex
\begin{tikzpicture}
    \node[anchor=south west,inner sep=-0.2pt] (image) at (0,0) {\includegraphics[width=\textwidth]{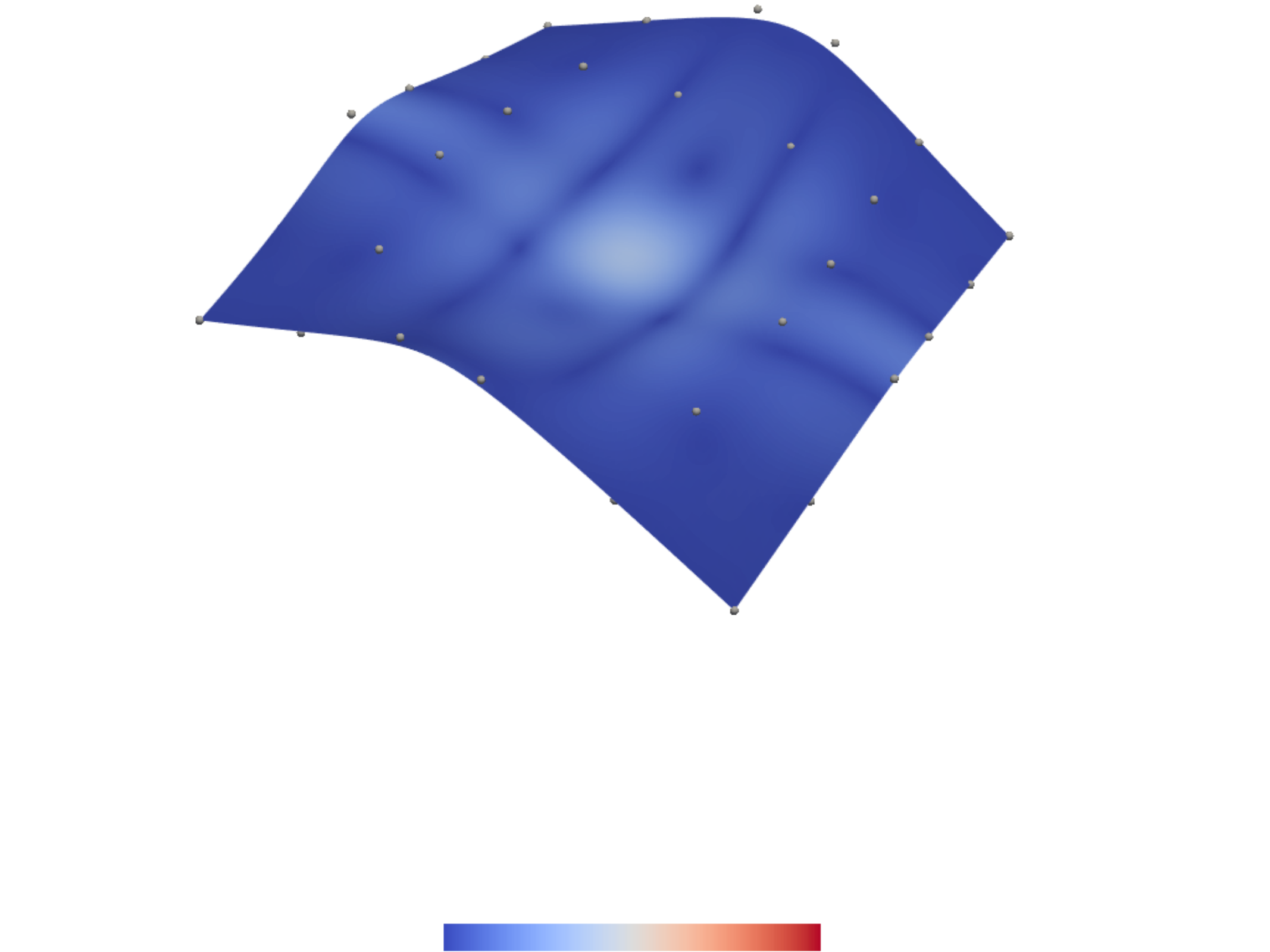}};
    \begin{axis}[
        scale only axis,
        scaled x ticks=false,
        scaled y ticks=false,
        at={(1.66cm,0.01cm)}, 
        tick label style={font=\footnotesize},
        title=$e_{relative}$,
        xticklabel=$\pgfmathprintnumber{\tick}$,
        yticklabel=$\pgfmathprintnumber{\tick}$,
        ymin=-0.009999999999761209,
        ymax=-0.00999809242784977,
        xmin=0.0,
        xmax=0.058846574284955655,
        ytick=\empty,
        height=0.1cm,
        width=1.43cm,  
        xtick={0.0,0.058846574284955655},
        xtick pos=right,
        xtick align=outside,
        title style={yshift=10pt,},
        ]
        \end{axis}
        \end{tikzpicture}%
    

%% file: fig/discretization_workflow/curved_surf_gradient_descent_method.tex
\begin{tikzpicture}
    \node[anchor=south west,inner sep=-0.2pt] (image) at (0,0) {\includegraphics[width=\textwidth]{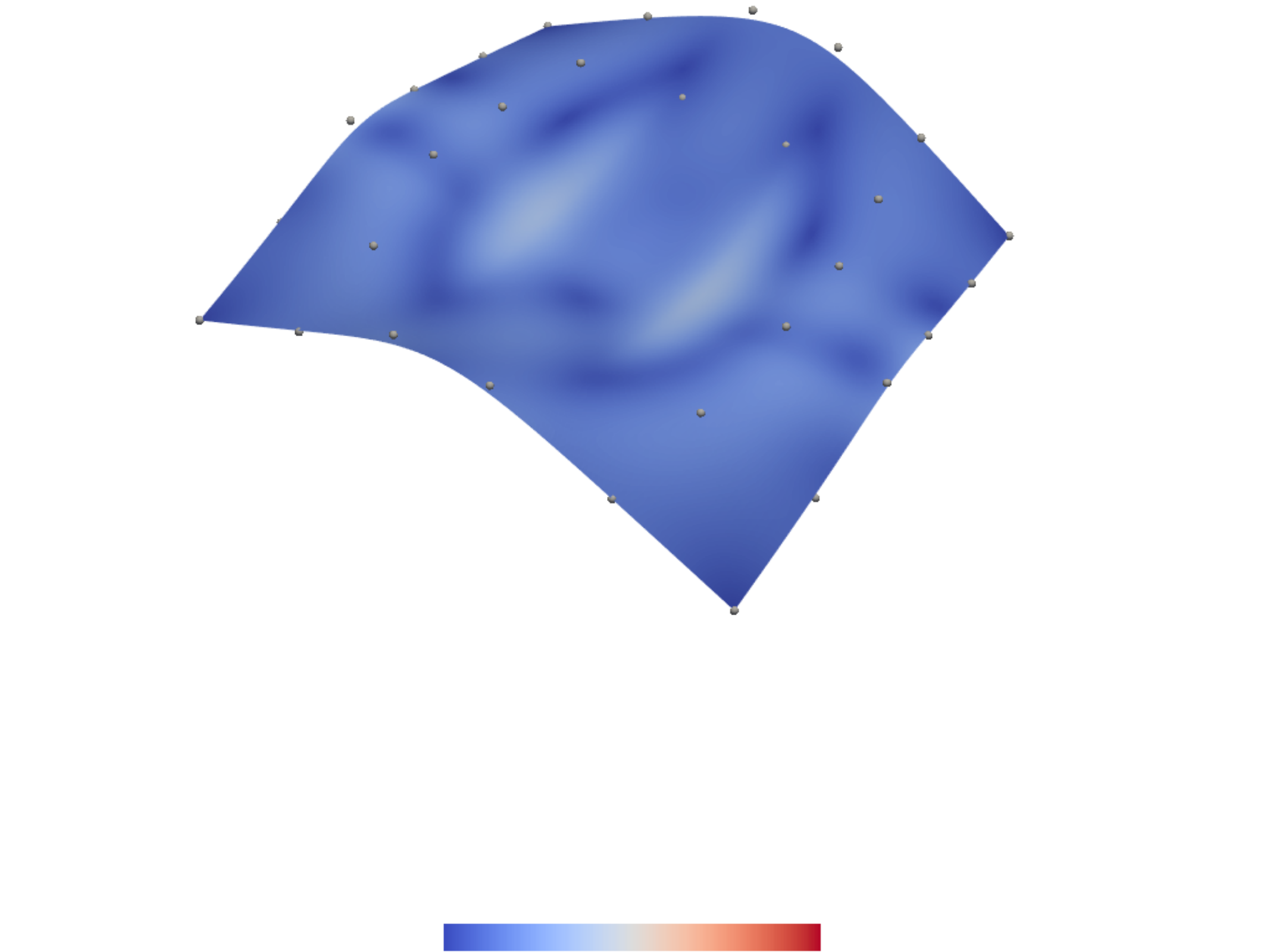}};
    \begin{axis}[
        scale only axis,
        scaled x ticks=false,
        scaled y ticks=false,
        at={(1.67cm,0.01cm)}, 
        tick label style={font=\footnotesize},
        title=$e_{relative}$,
        xticklabel=$\pgfmathprintnumber{\tick}$,
        yticklabel=$\pgfmathprintnumber{\tick}$,
        ymin=-0.009999999999761209,
        ymax=-0.00999809242784977,
        xmin=0.0,
        xmax=0.058846574284955655,
        ytick=\empty,
        height=0.1cm,
        width=1.43cm,  
        xtick={0.0,0.058846574284955655},
        xtick pos=right,
        xtick align=outside,
        title style={yshift=10pt,},
        ]
        \end{axis}
        \end{tikzpicture}%
    

%% file: fig/discretization_workflow/average_multipatches_kink_1.eps_tex
\begingroup%
  \makeatletter%
  \providecommand\color[2][]{%
    \errmessage{(Inkscape) Color is used for the text in Inkscape, but the package 'color.sty' is not loaded}%
    \renewcommand\color[2][]{}%
  }%
  \providecommand\transparent[1]{%
    \errmessage{(Inkscape) Transparency is used (non-zero) for the text in Inkscape, but the package 'transparent.sty' is not loaded}%
    \renewcommand\transparent[1]{}%
  }%
  \providecommand\rotatebox[2]{#2}%
  \newcommand*\fsize{\dimexpr\f@size pt\relax}%
  \newcommand*\lineheight[1]{\fontsize{\fsize}{#1\fsize}\selectfont}%
  \ifx\svgwidth\undefined%
    \setlength{\unitlength}{412.76205709bp}%
    \ifx\svgscale\undefined%
      \relax%
    \else%
      \setlength{\unitlength}{\unitlength * \real{\svgscale}}%
    \fi%
  \else%
    \setlength{\unitlength}{\svgwidth}%
  \fi%
  \global\let\svgwidth\undefined%
  \global\let\svgscale\undefined%
  \makeatother%
  \begin{picture}(1,0.5686607)%
    \lineheight{1}%
    \setlength\tabcolsep{0pt}%
    \put(0,0){\includegraphics[width=\unitlength]{average_multipatches_kink_1.eps}}%
    \put(0.76,0.11967928){\color[rgb]{0,0,0}\makebox(0,0)[lt]{\lineheight{0}\smash{\begin{tabular}[t]{l}Patch 2\end{tabular}}}}%
    \put(0.59590048,0.30370764){\color[rgb]{0,0,0}\makebox(0,0)[lt]{\lineheight{0}\smash{\begin{tabular}[t]{l}$\offsetdistance{} \aninwardsnormalvecwithindex{P1}$\end{tabular}}}}%
    \put(0.3,0.3091642){\color[rgb]{0,0,0}\makebox(0,0)[lt]{\lineheight{0}\smash{\begin{tabular}[t]{l}$\offsetdistance{} \aninwardsnormalvecwithindex{P2}$\end{tabular}}}}%
    \put(0.07,0.11967739){\color[rgb]{0,0,0}\makebox(0,0)[lt]{\lineheight{0}\smash{\begin{tabular}[t]{l}Patch 1\end{tabular}}}}%
  \end{picture}%
\endgroup%

%% file: fig/discretization_workflow/average_multipatches_kink_2.eps_tex
\begingroup%
  \makeatletter%
  \providecommand\color[2][]{%
    \errmessage{(Inkscape) Color is used for the text in Inkscape, but the package 'color.sty' is not loaded}%
    \renewcommand\color[2][]{}%
  }%
  \providecommand\transparent[1]{%
    \errmessage{(Inkscape) Transparency is used (non-zero) for the text in Inkscape, but the package 'transparent.sty' is not loaded}%
    \renewcommand\transparent[1]{}%
  }%
  \providecommand\rotatebox[2]{#2}%
  \newcommand*\fsize{\dimexpr\f@size pt\relax}%
  \newcommand*\lineheight[1]{\fontsize{\fsize}{#1\fsize}\selectfont}%
  \ifx\svgwidth\undefined%
    \setlength{\unitlength}{412.76205709bp}%
    \ifx\svgscale\undefined%
      \relax%
    \else%
      \setlength{\unitlength}{\unitlength * \real{\svgscale}}%
    \fi%
  \else%
    \setlength{\unitlength}{\svgwidth}%
  \fi%
  \global\let\svgwidth\undefined%
  \global\let\svgscale\undefined%
  \makeatother%
  \begin{picture}(1,0.5686607)%
    \lineheight{1}%
    \setlength\tabcolsep{0pt}%
    \put(0,0){\includegraphics[width=\unitlength]{average_multipatches_kink_2.eps}}%
    \put(0.76,0.11967928){\color[rgb]{0,0,0}\makebox(0,0)[lt]{\lineheight{0}\smash{\begin{tabular}[t]{l}Patch 2\end{tabular}}}}%
    \put(0.55335487,0.42308937){\color[rgb]{0,0,0}\makebox(0,0)[lt]{\lineheight{0}\smash{\begin{tabular}[t]{l}$\aninwardsnormalvecwithindex{P1}$\end{tabular}}}}%
    \put(0.46991978,0.23094638){\color[rgb]{0,0,0}\makebox(0,0)[lt]{\lineheight{0}\smash{\begin{tabular}[t]{l}$\offsetdistance{} \tilde{\aninwardsnormalvecwithindex{}}$\end{tabular}}}}%
    \put(0.35773269,0.42309242){\color[rgb]{0,0,0}\makebox(0,0)[lt]{\lineheight{0}\smash{\begin{tabular}[t]{l}$\aninwardsnormalvecwithindex{P2}$\end{tabular}}}}%
    \put(0.07,0.11967739){\color[rgb]{0,0,0}\makebox(0,0)[lt]{\lineheight{0}\smash{\begin{tabular}[t]{l}Patch 1\end{tabular}}}}%
  \end{picture}%
\endgroup%

%% file: section/problem_formulation.tex
\section{Problem formulation} \label{section:problem_formulation}

\subsection{Contact problem statement}

\begin{figure}
	\centering
	\def\svgwidth{0.64\linewidth}
	\rotatebox{90}
	{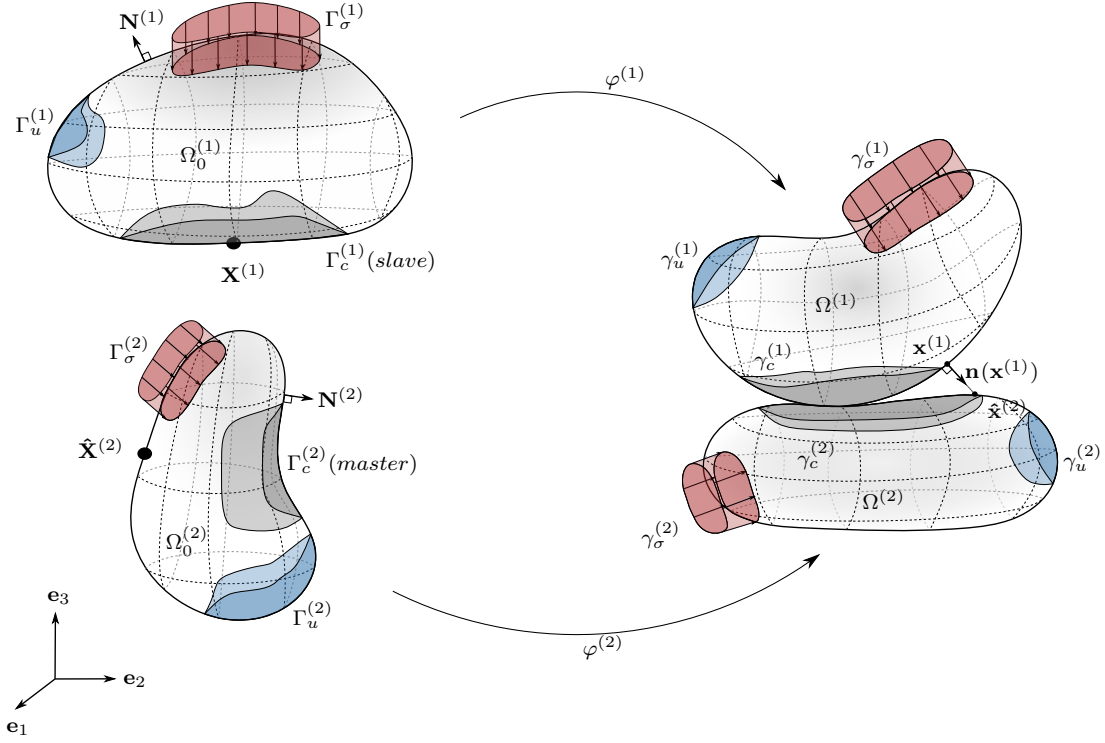}
	\caption{Setup of a three-dimensional contact problem.}
	\label{fig:two_bodies_contact}
\end{figure}

We consider a three-dimensional finite deformation contact problem involving two elastic bodies, 
as illustrated in Figure~\ref{fig:two_bodies_contact}. 
The bodies $i=1,2$ are represented by the open sets $\bodyref{i} \subset \realspace^3$ in the 
reference configuration and by $\bodycurrent{i}$ in the current configuration. 
For the sake of readability, and unless stated otherwise, all equations are assumed to hold for $i=1,2$. 
The finite deformation of each body is described by the deformation mapping \defmapping{i}.

The boundary \bodysurfref{i} can 
be subdivided into the Dirichlet boundary \interfacereference{u}{\abodyi}, the Neumann boundary 
\interfacereference{\sigma}{\abodyi}, and the potential contact boundary \interfacereference{c}{\abodyi}, 
on which displacements, tractions, and contact constraints are prescribed, respectively. 
These boundaries satisfy

\begin{equation}
	\centering
	\begin{aligned}
		\bodysurfref{i} = \interfacereference{u}{\abodyi} \cup \interfacereference{\sigma}{\abodyi} \cup \interfacereference{c}{\abodyi}, \\
		\interfacereference{u}{\abodyi} \cap \interfacereference{\sigma}{\abodyi} = \interfacereference{u}{\abodyi} \cap \interfacereference{c}{\abodyi} = \interfacereference{\sigma}{\abodyi} \cap \interfacereference{c}{\abodyi} = \emptyset.
	\end{aligned}
\end{equation}

In the current configuration, the corresponding boundaries are denoted by 
\interfacespatial{u}{\abodyi}, \interfacespatial{\sigma}{\abodyi}, and 
\interfacespatial{c}{\abodyi}, respectively. 
Following the standard nomenclature in contact mechanics, 
\interfacereference{c}{\abodynum{1}} is referred to as the \textit{slave} surface, 
while \interfacereference{c}{\abodynum{2}} is referred to as the \textit{master} surface.
Let \pospointref{(i)} and \pospointcurrent{(i)} denote the position of a material point in the 
reference and current configurations, respectively. 
The displacement field~\dispvector{i} is then defined as

\begin{equation}
	\dispvector{i} = \pospointcurrent{(i)} - \pospointref{(i)}.
\end{equation}

The strong form of the boundary value problem (BVP) of nonlinear elasticity reads

\begin{equation}
	\begin{aligned}
		\divergence (\defgrad{i} \cdot  \secondPK{\abodyi}) + \bodyforce{i} &= \mathbf{0} &\text{ in } \bodyref{i},\\ 
		\dispvector{i} &= \prescribeddisp{i} &\text{ on } \interfacereference{u}{\abodyi},\\ 
		\firstPK{i} \anoutwardsnormalvecrefconf{\abodyi} &= \prescribedtraction{i} &\text{ on } \interfacereference{\sigma}{\abodyi}.\\ 
	\end{aligned}
	\label{eq:strong_form_1}
\end{equation}

Here, $\defgrad{i}$ denotes the deformation gradient, while $\firstPK{i}$ and $\secondPK{\abodyi}$ represent 
the first and second Piola--Kirchhoff stress tensors, respectively. 
Furthermore, \anoutwardsnormalvecrefconf{\abodyi} denotes the outward unit normal 
vector on~\interfacereference{\sigma}{\abodyi}, and the vector $\bodyforce{i}$ represents the body 
forces. The quantities \prescribeddisp{i} and \prescribedtraction{i} denote the prescribed 
displacement and traction, respectively. Without loss of generality, 
all formulations are presented for the quasi-static case. The extension to dynamic problems is straightforward.

Assuming homogeneous bodies with hyperelastic material behavior, the constitutive relation in 
the reference configuration is given by

\begin{equation}
	\secondPK{} = \frac{\partial \strainenergyfunction{}}{\partial \greenlagstrain{}},
\end{equation}
where \strainenergyfunction{} is the strain energy density function and 
\greenlagstrain{} denotes the Green--Lagrange strain tensor. 
In the numerical examples, the St.\ Venant--Kirchhoff material model is employed, with

\begin{equation}
	\strainenergyfunction{}_{\mathrm{SVK}} = \frac{\lameparamlambda{}}{2}(\mathrm{tr}\mathbf{E})^2 + \lameparammu{} \mathbf{E} : \mathbf{E}, 
\end{equation}
where \lameparamlambda{} and \lameparammu{} are the Lam\'{e} parameters \cite{Basar2000}. 

For the solution of the contact problem, the gap function $\gapfunction(\mathbf{X}, t)$ 
is introduced to describe the non-penetration condition~\cite{Wriggers2006}. It is defined as

\begin{equation}
	\gapfunction(\pospointref{}, t) = -\anoutwardsnormalvec{} \cdot [\pospointcurrent{\abodynum{1}}(\pospointref{\abodynum{1}}, t) - 
	\hat{\mathbf{x}}^{\abodynum{2}}(\hat{\mathbf{X}}^{\abodynum{2}}(\pospointref{\abodynum{1}}, t),t)], \label{eq:gap_function}
\end{equation}
where \anoutwardsnormalvec{} denotes the outward unit normal vector at 
$\pospointcurrent{1} \in \interfacespatial{c}{\abodynum{1}}$ in the current configuration. 
The point~$\hat{\mathbf{x}}^{\abodynum{2}}$ is obtained from the closest-point projection of 
$\pospointcurrent{1}$ onto \interfacespatial{c}{\abodynum{2}}.
The contact traction \contacttraction{1} acting on the slave surface is 
decomposed into a normal component \normalcontactforce{} and a tangential 
component~\tangentialcontactforce{}, such that

\begin{equation}
	\contacttraction{1} = \normalcontactforce{} + \tangentialcontactforce{}.
	\label{eg:contact-traction}
\end{equation}

The normal component is given by 
$\normalcontactforce{} = \normalcontactpressure \anoutwardsnormalvec{}$, 
where \normalcontactpressure{} denotes the normal contact pressure. 
Although the proposed method can be extended to frictional contact, 
we consider frictionless sliding for simplicity, \ie{}~$\tangentialcontactforce{} = \zerovector{}$.
The contact conditions are expressed by the Karush--Kuhn--Tucker (KKT) conditions,

\begin{equation}
	\gapfunction{} \geq 0, \quad \normalcontactpressure{} \leq 0, \quad \gapfunction{} \: \normalcontactpressure{} = 0 \quad \text{ on } \interfacespatial{c}{\abodynum{1}}.
	\label{eq:strong_form_2}
\end{equation}

Equations~\eqref{eq:strong_form_1} and~\eqref{eq:strong_form_2} constitute the strong form 
of the unilateral frictionless contact problem. To derive the weak formulation 
of~\eqref{eq:strong_form_1}, we introduce the solution space \solutionspace{i} 
and the test space~\testingspace{i} for the displacement field,

\begin{equation}
	\solutionspace{i} = \{ \dispvector{i} \in \sobolevspace{i} \quad | \quad \dispvector{i} = \prescribeddisp{i} \text{ on } \interfacereference{u}{\abodyi}\},
\end{equation}
and 
\begin{equation}
	\testingspace{i} = \{ \delta\dispvector{i} \in \sobolevspace{i} \quad | \quad \delta\dispvector{i} = \mathbf{0} \text{ on } \interfacereference{u}{\abodyi}\},
\end{equation}
where \sobolevspace{i} represents a Sobolev function space \cite{Brenner2017}. 
To enforce the contact constraints given in~\eqref{eq:strong_form_2}, a Lagrange 
multiplier field \lagmultvector{} is introduced, belonging to the space 
\lagmultsolutionspace{}~\cite{Popp2009, Wohlmuth2000}. 
This Lagrange multiplier can be interpreted as the contact traction, \ie{} 
$\lagmultvector{} = -\contacttraction{1}$.
The solution space~\lagmultsolutionspace{} is defined as the dual space of the trace space 
\tracetestingspace{\abodynum{1}} associated with \testingspace{1}, restricted to 
\interfacespatial{c}{\abodynum{1}}. 
For frictionless contact, a simplification can be done in the solution space \lagmultsolutionspace{}. 
This simplification results in a convex cone of Lagrange multipliers defined as

\begin{equation}
	\lagmultsolutionspace{}^{+} 
	:= 
	\left\{
	\testlagmultvector{} \in \lagmultsolutionspace{} 
	\ \middle| \
	\testlagmultvector{\tau} = \zerovector, \,
	\dualitypairing{\testlagmultvectornormal{}}{\scalarfunction{}}_{\interfacespatial{c}{\abodynum{1}}} \ge 0, \,
	\scalarfunction{} \in \mathcal{W}^{+}
	\right\},
\end{equation}
where $\dualitypairing{\cdot}{\cdot}_{\interfacespatial{c}{\abodynum{1}}}$ 
denotes the duality pairing between the scalar components of the spaces 
\lagmultsolutionspace{} and~\tracetestingspace{} \cite{Hueber2005, Popp2012}. 
Furthermore, $\mathcal{W}^{+}$ is the closed, non-empty convex cone defined by
$\mathcal{W}^{+} := \left\{\scalarfunction{} \in \mathcal{W} \ \middle| \ \scalarfunction{} \ge 0\right\}$.

With these definitions, the weak formulation of the frictionless contact problem reads: Find 
$\dispvector{i} \in \solutionspace{i}$ and $\lagmultvector{} \in \lagmultsolutionspace{}^{+}$ 
such that 

\begin{align}
	-\intextenergy{}\left( \dispvector{i}, \delta \dispvector{i} \right) -
	 \contactenergy{}\left( \lagmultvector{}, \delta \dispvector{i} \right) &= 0 
	 &\forall \delta \dispvector{i} \in \testingspace{i}, 
	 \label{eq:weak_form_1}\\
	 \constraintenergy{}\left( \dispvector{i}, \delta \lagmultvector{} \right) &\geq 0 
	 &\forall \delta \lagmultvector{} \in \lagmultsolutionspace{}^{+}.
	 \label{eq:weak_form_2}
\end{align}

The internal and external virtual work contribution \intextenergy{}, 
the contact contribution \contactenergy{}, and the weak constraint contribution \constraintenergy{} 
are given by~\cite{Popp2012}

\begin{align}
	-\intextenergy{} 
	&= \sum_{i=1}^{2} 
	\Bigg[
	\int_{\bodyref{i}} 
	\secondPK{\abodyi} : \delta \greenlagstrain^{\abodyi} \, \mathrm{d}V_0
	- \int_{\bodyref{i}} 
	\bodyforce{i} \cdot \delta \dispvector{i} \, \mathrm{d}V_0
	- \int_{\interfacereference{\sigma}{\abodyi}} 
	\prescribedtraction{i} \cdot \delta \dispvector{i} \, \mathrm{d}A_0
	\Bigg], \\
	-\contactenergy{} 
	&= \int_{\interfacespatial{c}{\abodynum{1}}} 
	\lagmultvector{} \cdot 
	\left(
	\delta \dispvector{\abodynum{1}} 
	- \delta \dispvector{\abodynum{2}} \circ \chi
	\right) 
	\, \mathrm{d}A, 
	\label{eq:contact_energy} \\
	\constraintenergy{} 
	&= \int_{\interfacespatial{c}{\abodynum{1}}} 
	\left(
	\delta \lagmultscalarnormal{} 
	- \lagmultscalarnormal{}
	\right)
	\gapfunction 
	\, \mathrm{d}A.
	\label{eq:contact_weak_constraint_energy}
\end{align}

Here, $\chi : \interfacespatial{c}{\abodynum{1}} \rightarrow \interfacespatial{c}{\abodynum{2}}$ 
denotes a suitable projection mapping from a point on the slave side to its corresponding point on the master side. 
For frictionless contact, it is sufficient to enforce the non-penetration condition in the normal direction only. 
Accordingly, in equation~\eqref{eq:contact_weak_constraint_energy}, the constraint is imposed using the 
scalar normal component \lagmultscalarnormal{} of the Lagrange multiplier field.

\subsection{Embedded mesh problem statement} \label{subsection:em_problem_statement}

\begin{figure}
	\centering
	\def\svgwidth{0.59\linewidth}
	\rotatebox{90}
	{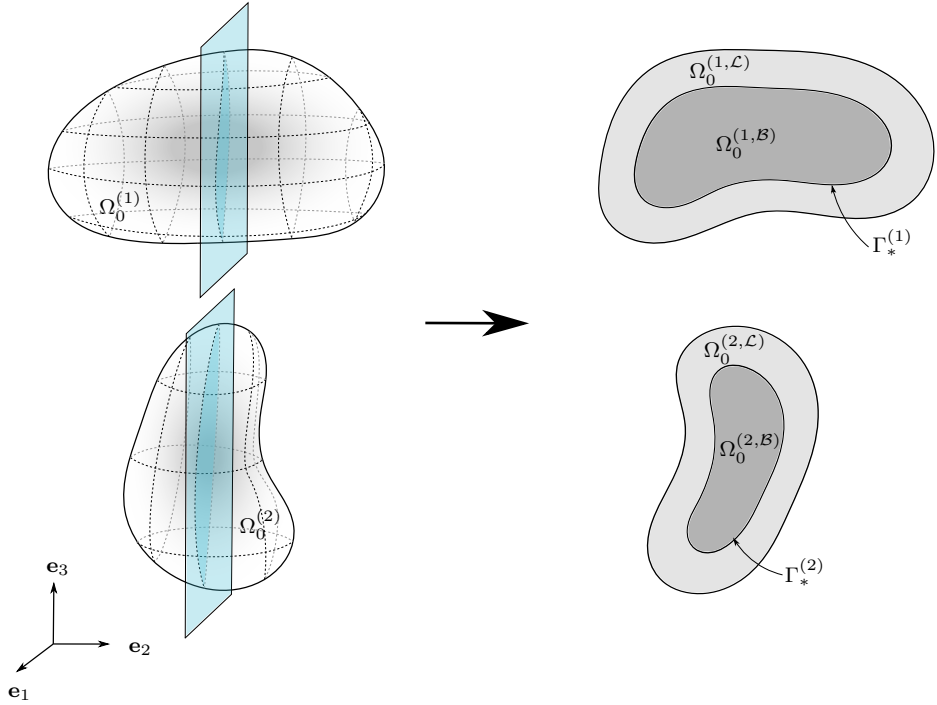}
	\caption{Subdivision of the solid domain into a boundary layer domain \boundarylayerdomain{i, } 
			 and a bulk domain \bulkdomain{i, }, separated by the interface \interfacestar{\abodyi}{}. 
			 For the three-dimensional case, the subdivision is illustrated using a cross-section defined by the blue plane.}
	\label{fig:subdivision_domains}
\end{figure}

So far, each solid body has been treated as a single domain. 
In the proposed approach, however, each contact body is subdivided into two subdomains: 
a boundary layer domain \boundarylayerdomain{i, } and a bulk domain~\bulkdomain{i, }. 
Furthermore, the interface \interfacestar{\abodyi}{} separates the two subdomains. 
This subdivision is illustrated in Figure~\ref{fig:subdivision_domains} for the three-dimensional case.
The body surface (and thus the contact boundary) \bodysurfref{i}, as well as a portion of the adjacent volume, 
is contained in the boundary layer domain \boundarylayerdomain{i, }, 
whereas the bulk domain \bulkdomain{i, } contains the remaining part of the body volume. 
Accordingly, the reference configuration satisfies 
\begin{equation}
\bodyref{i} = \boundarylayerdomain{i, } \cup \bulkdomain{i, }, 
\qquad
\boundarylayerdomain{i, } \cap \bulkdomain{i, } = \emptyset.
\end{equation}

In this work, homogeneous material properties are assumed. 
Consequently, both subdomains inherit the same material parameters as the original domain \bodyref{i}.

To couple the two subdomains, a coupling constraint is defined at the interface 
\interfacestar{\abodyi}{} and reads

\begin{equation}
	\dispboundarylayer{i} - \dispbulk{i} = \mathbf{0} \quad \text{ on } \quad \interfacestar{\abodyi}{}.
\end{equation}

As the proposed discretization approach described in Section~\ref{section:discretization_workflow} 
leads to overlapping meshes of~\boundarylayerdomain{i, } and \bulkdomain{i, }, 
a coupling constraint must be imposed between the two discretizations. 
This constraint is enforced using a Lagrange multiplier formulation. 
Analogous to the contact formulation, a Lagrange multiplier field 
\lambdastar{\abodyi}{} is introduced on the interface \interfacestar{\abodyi}{}.
The resulting contributions to the total virtual work consist of the coupling interface term

\begin{equation}
	-\couplinginterenergy{\abodyi}{} = \int_{\interfacestar{\abodyi}{}} \lambdastar{\abodyi}{} \left( \delta \dispboundarylayer{i} - 
	\delta \dispbulk{i}  \right) \text{d} A_0,
	\label{eq:coupling_interface_energy}
\end{equation}
and the variational form of the coupling constraint
\begin{equation}
	\couplingconstraintenergy{\abodyi}{} = \int_{\interfacestar{\abodyi}{}} \delta \lambdastar{\abodyi}{} \left( \dispboundarylayer{i} -
	 \dispbulk{i}  \right) \text{d} A_0.
	\label{eq:coupling_constraint_energy}
\end{equation} 

With this, the mixed formulation of the embedded mesh problem reads: Find 
$\dispvector{i} \in \solutionspace{i}$ and~$\lambdastar{\abodyi}{} \in \lagmultembeddedmeshsolutionspace{}$ 
such that 
\begin{align}
	 \intextenergy{}\left( \dispvector{i}, \delta \dispvector{i} \right) - 
	 \couplinginterenergy{\abodyi}{}\left( \lambdastar{\abodyi}{}, \delta \dispvector{i} \right) &= 0 
	 &\forall \delta \dispvector{i} \in \testingspace{i},\label{eq:weak_formulation_only_embedded_coupling} \\
	 \couplingconstraintenergy{\abodyi}{} \left( \dispvector{i}, \delta \lambdastar{\abodyi}{} \right) &= 0 
	 &\forall \delta \lambdastar{\abodyi}{} \in \lagmultembeddedmeshsolutionspace{}.
\end{align}
where \lagmultembeddedmeshsolutionspace{} is the dual space of the trace space of \testingspace{i} 
restricted to \interfacestar{\abodyi}{}~\cite{Bechet2009}. 

\subsection{Combined contact-embedded mesh formulation} \label{subsection:combined_statement}

The contact and embedded mesh coupling formulations are now combined 
to obtain a mixed finite element formulation in which the displacement field $\dispvector{i}$, 
the contact tractions \lagmultvector{}, and the interface tractions 
\lambdastar{\abodyi}{} are solved simultaneously. Incorporating the embedded mesh coupling contributions 
\eqref{eq:coupling_interface_energy} and \eqref{eq:coupling_constraint_energy} 
into the contact problem defined by \eqref{eq:weak_form_1} and \eqref{eq:weak_form_2} 
yields the following weak formulation

\begin{align}
	-\intextenergy{}\left( \dispvector{i}, \delta \dispvector{i} \right) - 
	 \contactenergy{}\left( \lagmultvector{}, \delta \dispvector{i} \right) - 
	 \couplinginterenergy{\abodyi}{}\left( \lambdastar{\abodyi}{}, \delta \dispvector{i} \right) &= 0 
	 &\forall \delta \dispvector{i} \in \testingspace{i}, \label{eq:global_formulation}\\
	 \constraintenergy{}\left( \dispvector{i}, \delta \lagmultvector{} \right) &\geq 0 
	 &\forall \delta \lagmultvector{} \in \lagmultsolutionspace{}^{+}, \label{eq:contact_weak_formulation} \\
	 \couplingconstraintenergy{\abodyi}{} \left( \dispvector{i}, \delta \lambdastar{\abodyi}{} \right) &= 0 
	 &\forall \delta \lambdastar{\abodyi}{} \in \lagmultembeddedmeshsolutionspace{}.\label{eq:coupling_weak_formulation}
\end{align}

Several key differences can be identified between the contact contributions 
\eqref{eq:contact_energy} and \eqref{eq:contact_weak_constraint_energy} 
and the embedded mesh contributions 
\eqref{eq:coupling_interface_energy} and \eqref{eq:coupling_constraint_energy}.
Due to the nature of contact problems, the contact terms 
depend on the current deformation of the bodies.
Thus, the contact 
contributions are formulated in the current configuration, whereas the embedded 
mesh coupling terms are formulated in the reference configuration. 
Consequently, the projection $\chi$ in 
\eqref{eq:contact_energy} must be evaluated continuously to capture the relative motion 
between slave and master sides in the current configuration. 
In contrast, although a projection is also required for the 
embedded mesh contribution \eqref{eq:coupling_interface_energy}, 
it is evaluated only once in the reference configuration.
Finally, the weak contact formulation~\eqref{eq:contact_weak_formulation} 
contains an inequality constraint, whereas the embedded mesh formulation~\eqref{eq:coupling_weak_formulation} 
leads to an equality constraint. 
The inequality requires specialized numerical treatment, typically based on active-set 
strategies~\cite{Hartmann2007, Hueber2005, Popp2009, Popp2010}.

%% file: fig/problem_formulation/contact_3d.pdf_tex
\begingroup%
  \makeatletter%
  \providecommand\color[2][]{%
    \errmessage{(Inkscape) Color is used for the text in Inkscape, but the package 'color.sty' is not loaded}%
    \renewcommand\color[2][]{}%
  }%
  \providecommand\transparent[1]{%
    \errmessage{(Inkscape) Transparency is used (non-zero) for the text in Inkscape, but the package 'transparent.sty' is not loaded}%
    \renewcommand\transparent[1]{}%
  }%
  \providecommand\rotatebox[2]{#2}%
  \newcommand*\fsize{\dimexpr\f@size pt\relax}%
  \newcommand*\lineheight[1]{\fontsize{\fsize}{#1\fsize}\selectfont}%
  \ifx\svgwidth\undefined%
    \setlength{\unitlength}{419.52755906bp}%
    \ifx\svgscale\undefined%
      \relax%
    \else%
      \setlength{\unitlength}{\unitlength * \real{\svgscale}}%
    \fi%
  \else%
    \setlength{\unitlength}{\svgwidth}%
  \fi%
  \global\let\svgwidth\undefined%
  \global\let\svgscale\undefined%
  \makeatother%
  \begin{picture}(1,1.41891892)%
    \lineheight{1}%
    \setlength\tabcolsep{0pt}%
    \put(0,0){\includegraphics[width=\unitlength,page=1]{contact_3d.pdf}}%
    \put(0.09401693,1.25){\rotatebox{-89.876308}{\makebox(0,0)[lt]{\lineheight{1.25}\smash{\begin{tabular}[t]{l}\footnotesize{$\yvector$}\end{tabular}}}}}%
    \put(0.03124356,1.40108026){\rotatebox{-89.876308}{\makebox(0,0)[lt]{\lineheight{1.25}\smash{\begin{tabular}[t]{l}\footnotesize{$\xvector$}\end{tabular}}}}}%
    \put(0,0){\includegraphics[width=\unitlength,page=2]{contact_3d.pdf}}%
    \put(0.37247346,1.03839716){\rotatebox{-89.87630774}{\makebox(0,0)[lt]{\lineheight{1.25}\smash{\begin{tabular}[t]{l}\footnotesize{$\interfacereference{c}{(2)}(master)$}\end{tabular}}}}}%
    \put(0.17096348,1.03135481){\rotatebox{-89.5952753}{\makebox(0,0)[lt]{\lineheight{1.25}\smash{\begin{tabular}[t]{l}\footnotesize{$\interfacereference{u}{(2)}$}\end{tabular}}}}}%
    \put(0.63916842,0.55569725){\rotatebox{-89.87630744}{\makebox(0,0)[lt]{\lineheight{1.25}\smash{\begin{tabular}[t]{l}\footnotesize{$\interfacespatial{u}{(1)}$}\end{tabular}}}}}%
    \put(0.3757023,0.0359627){\rotatebox{-89.22887484}{\makebox(0,0)[lt]{\lineheight{1.25}\smash{\begin{tabular}[t]{l}\footnotesize{$\interfacespatial{u}{(2)}$}\end{tabular}}}}}%
    \put(0.27534897,0.58090938){\rotatebox{-89.87630801}{\makebox(0,0)[lt]{\lineheight{1.25}\smash{\begin{tabular}[t]{l}\footnotesize{$\interfacespatial{\sigma}{(2)}$}\end{tabular}}}}}%
    \put(0,0){\includegraphics[width=\unitlength,page=3]{contact_3d.pdf}}%
    \put(0.20127816,1.34797033){\rotatebox{-89.87630835}{\makebox(0,0)[lt]{\lineheight{1.25}\smash{\begin{tabular}[t]{l}\footnotesize{$\zvector$}\end{tabular}}}}}%
    \put(0.26927758,1.19512691){\rotatebox{-89.876308}{\makebox(0,0)[lt]{\lineheight{1.25}\smash{\begin{tabular}[t]{l}\footnotesize{$\Omega_0^{(2)}$}\end{tabular}}}}}%
    \put(0.51608829,1.2679384){\rotatebox{-89.876308}{\makebox(0,0)[lt]{\lineheight{1.25}\smash{\begin{tabular}[t]{l}\footnotesize{$\interfacereference{\sigma}{(2)}$}\end{tabular}}}}}%
    \put(0.38815131,1.30864128){\rotatebox{-89.876308}{\makebox(0,0)[lt]{\lineheight{1.25}\smash{\begin{tabular}[t]{l}\footnotesize{$\mathbf{\hat{X}}^{(2)}$}\end{tabular}}}}}%
    \put(0,0){\includegraphics[width=\unitlength,page=4]{contact_3d.pdf}}%
    \put(0.63673556,0.98873706){\rotatebox{-89.876308}{\makebox(0,0)[lt]{\lineheight{1.25}\smash{\begin{tabular}[t]{l}\footnotesize{$\interfacereference{c}{(1)}(slave)$}\end{tabular}}}}}%
    \put(0.81261814,1.38921185){\rotatebox{-89.876308}{\makebox(0,0)[lt]{\lineheight{1.25}\smash{\begin{tabular}[t]{l}\footnotesize{$\interfacereference{u}{(1)}$}\end{tabular}}}}}%
    \put(0,0){\includegraphics[width=\unitlength,page=5]{contact_3d.pdf}}%
    \put(0.61636152,1.12439599){\rotatebox{-89.876308}{\makebox(0,0)[lt]{\lineheight{1.25}\smash{\begin{tabular}[t]{l}\footnotesize{$\mathbf{X}^{(1)}$}\end{tabular}}}}}%
    \put(0,0){\includegraphics[width=\unitlength,page=6]{contact_3d.pdf}}%
    \put(0.94976719,0.98944129){\rotatebox{-89.55099706}{\makebox(0,0)[lt]{\lineheight{1.25}\smash{\begin{tabular}[t]{l}\footnotesize{$\interfacereference{\sigma}{(1)}$}\end{tabular}}}}}%
    \put(0.77264468,1.17680087){\rotatebox{-89.876308}{\makebox(0,0)[lt]{\lineheight{1.25}\smash{\begin{tabular}[t]{l}\footnotesize{$\Omega_0^{(1)}$}\end{tabular}}}}}%
    \put(0,0){\includegraphics[width=\unitlength,page=7]{contact_3d.pdf}}%
    \put(0.94601573,1.25552074){\rotatebox{-89.876308}{\makebox(0,0)[lt]{\lineheight{1.25}\smash{\begin{tabular}[t]{l}\footnotesize{$\mathbf{N}^{(1)}$}\end{tabular}}}}}%
    \put(0,0){\includegraphics[width=\unitlength,page=8]{contact_3d.pdf}}%
    \put(0.87277623,0.62790673){\rotatebox{-89.876308}{\makebox(0,0)[lt]{\lineheight{1.25}\smash{\begin{tabular}[t]{l}\footnotesize{\defmapping{1}}\end{tabular}}}}}%
    \put(0,0){\includegraphics[width=\unitlength,page=9]{contact_3d.pdf}}%
    \put(0.45452758,0.9989025){\rotatebox{-89.876308}{\makebox(0,0)[lt]{\lineheight{1.25}\smash{\begin{tabular}[t]{l}\footnotesize{$\mathbf{N}^{(2)}$}\end{tabular}}}}}%
    \put(0,0){\includegraphics[width=\unitlength,page=10]{contact_3d.pdf}}%
    \put(0.4395202,0.13365836){\rotatebox{-89.65832144}{\makebox(0,0)[lt]{\lineheight{1.25}\smash{\begin{tabular}[t]{l}\footnotesize{$\mathbf{\hat{x}}^{(2)}$}\end{tabular}}}}}%
    \put(0.37898846,0.37967627){\rotatebox{-89.8393801}{\makebox(0,0)[lt]{\lineheight{1.25}\smash{\begin{tabular}[t]{l}\footnotesize{$\interfacespatial{c}{(2)}$}\end{tabular}}}}}%
    \put(0.32189576,0.295491){\rotatebox{-89.61512003}{\makebox(0,0)[lt]{\lineheight{1.25}\smash{\begin{tabular}[t]{l}\footnotesize{$\Omega^{(2)}$}\end{tabular}}}}}%
    \put(0.5227213,0.23044187){\rotatebox{-89.61821906}{\makebox(0,0)[lt]{\lineheight{1.25}\smash{\begin{tabular}[t]{l}\footnotesize{$\mathbf{x}^{(1)}$}\end{tabular}}}}}%
    \put(0.76984562,0.30957228){\rotatebox{-89.8924659}{\makebox(0,0)[lt]{\lineheight{1.25}\smash{\begin{tabular}[t]{l}\footnotesize{$\interfacespatial{\sigma}{(1)}$}\end{tabular}}}}}%
    \put(0.57724017,0.35647593){\rotatebox{-89.876308}{\makebox(0,0)[lt]{\lineheight{1.25}\smash{\begin{tabular}[t]{l}\footnotesize{$\Omega^{(1)}$}\end{tabular}}}}}%
    \put(0.50726545,0.4356613){\rotatebox{-89.87630705}{\makebox(0,0)[lt]{\lineheight{1.25}\smash{\begin{tabular}[t]{l}\footnotesize{$\interfacespatial{c}{(1)}$}\end{tabular}}}}}%
    \put(0.49156092,0.16317949){\rotatebox{-89.80262439}{\makebox(0,0)[lt]{\lineheight{1.25}\smash{\begin{tabular}[t]{l}\footnotesize{$\anoutwardsnormalvec{}(\mathbf{x}^{(1)})$}\end{tabular}}}}}%
    \put(0.13145947,0.65677529){\rotatebox{-89.876308}{\makebox(0,0)[lt]{\lineheight{1.25}\smash{\begin{tabular}[t]{l}\footnotesize{\defmapping{2}}\end{tabular}}}}}%
    \put(0,0){\includegraphics[width=\unitlength,page=11]{contact_3d.pdf}}%
  \end{picture}%
\endgroup%

%% file: fig/problem_formulation/subdomains.pdf_tex
\begingroup%
  \makeatletter%
  \providecommand\color[2][]{%
    \errmessage{(Inkscape) Color is used for the text in Inkscape, but the package 'color.sty' is not loaded}%
    \renewcommand\color[2][]{}%
  }%
  \providecommand\transparent[1]{%
    \errmessage{(Inkscape) Transparency is used (non-zero) for the text in Inkscape, but the package 'transparent.sty' is not loaded}%
    \renewcommand\transparent[1]{}%
  }%
  \providecommand\rotatebox[2]{#2}%
  \newcommand*\fsize{\dimexpr\f@size pt\relax}%
  \newcommand*\lineheight[1]{\fontsize{\fsize}{#1\fsize}\selectfont}%
  \ifx\svgwidth\undefined%
    \setlength{\unitlength}{419.52755906bp}%
    \ifx\svgscale\undefined%
      \relax%
    \else%
      \setlength{\unitlength}{\unitlength * \real{\svgscale}}%
    \fi%
  \else%
    \setlength{\unitlength}{\svgwidth}%
  \fi%
  \global\let\svgwidth\undefined%
  \global\let\svgscale\undefined%
  \makeatother%
  \begin{picture}(1,1.41891892)%
    \lineheight{1}%
    \setlength\tabcolsep{0pt}%
    \put(0,0){\includegraphics[width=\unitlength,page=1]{subdomains.pdf}}%
    \put(0.07446074,1.18737388){\rotatebox{-90.091722}{\makebox(0,0)[lt]{\lineheight{1.25}\smash{\begin{tabular}[t]{l}\footnotesize{$\yvector$}\end{tabular}}}}}%
    \put(0.01089734,1.35628211){\rotatebox{-90.091722}{\makebox(0,0)[lt]{\lineheight{1.25}\smash{\begin{tabular}[t]{l}\footnotesize{$\xvector$}\end{tabular}}}}}%
    \put(0,0){\includegraphics[width=\unitlength,page=2]{subdomains.pdf}}%
    \put(0.18715559,1.30250931){\rotatebox{-90.09172308}{\makebox(0,0)[lt]{\lineheight{1.25}\smash{\begin{tabular}[t]{l}\footnotesize{$\zvector$}\end{tabular}}}}}%
    \put(0.24325724,1.03234006){\rotatebox{-90.091722}{\makebox(0,0)[lt]{\lineheight{1.25}\smash{\begin{tabular}[t]{l}\footnotesize{$\Omega_0^{(2)}$}\end{tabular}}}}}%
    \put(0,0){\includegraphics[width=\unitlength,page=3]{subdomains.pdf}}%
    \put(0.69271912,1.22888853){\rotatebox{-90.091722}{\makebox(0,0)[lt]{\lineheight{1.25}\smash{\begin{tabular}[t]{l}\footnotesize{$\Omega_0^{(1)}$}\end{tabular}}}}}%
    \put(0,0){\includegraphics[width=\unitlength,page=4]{subdomains.pdf}}%
    \put(0.17423492,0.2684053){\rotatebox{-90.21840383}{\makebox(0,0)[lt]{\lineheight{1.25}\smash{\begin{tabular}[t]{l}\footnotesize{\interfacestar{(2)}{}}\end{tabular}}}}}%
    \put(0,0){\includegraphics[width=\unitlength,page=5]{subdomains.pdf}}%
    \put(0.88441,0.40182225){\rotatebox{-90.21840385}{\makebox(0,0)[lt]{\lineheight{1.25}\smash{\begin{tabular}[t]{l}\footnotesize{\boundarylayerdomain{1, }}\end{tabular}}}}}%
    \put(0,0){\includegraphics[width=\unitlength,page=6]{subdomains.pdf}}%
    \put(0.78215706,0.36510392){\rotatebox{-90.21840354}{\makebox(0,0)[lt]{\lineheight{1.25}\smash{\begin{tabular}[t]{l}\footnotesize{\bulkdomain{1, }}\end{tabular}}}}}%
    \put(0.49111578,0.38171977){\rotatebox{-90.2184037}{\makebox(0,0)[lt]{\lineheight{1.25}\smash{\begin{tabular}[t]{l}\footnotesize{\boundarylayerdomain{2, }}\end{tabular}}}}}%
    \put(0.35165721,0.36006465){\rotatebox{-90.21840397}{\makebox(0,0)[lt]{\lineheight{1.25}\smash{\begin{tabular}[t]{l}\footnotesize{\bulkdomain{2, }}\end{tabular}}}}}%
    \put(0,0){\includegraphics[width=\unitlength,page=7]{subdomains.pdf}}%
    \put(0.63954274,0.14561718){\rotatebox{-90.21840375}{\makebox(0,0)[lt]{\lineheight{1.25}\smash{\begin{tabular}[t]{l}\footnotesize{\interfacestar{(1)}{}}\end{tabular}}}}}%
  \end{picture}%
\endgroup%

%% file: section/spatial_discretization.tex
\section{Spatial discretization and numerical integration} \label{section:spatial_discretization}

In Section~\ref{section:discretization_workflow}, the mesh generation workflow for 
the bodies involved in the contact problem was described, resulting in discretizations of the subdomains 
\boundarylayerdomain{i, } and \bulkdomain{i, }. 
As introduced previously, the mesh associated with \boundarylayerdomain{i, } is referred to as the \textit{boundary layer mesh}, 
whereas the mesh representing \bulkdomain{i, } is referred to as the \textit{background mesh}. 
Furthermore, in Section~\ref{section:problem_formulation}, a mixed finite element formulation 
was derived involving three unknown fields: the displacement \dispvector{i}, 
the contact tractions \lagmultvector{}, and the embedded mesh interface tractions 
\lambdastar{\abodyi}{}.

In the following, the spatial discretization of these fields is presented. 
We begin with the discretization of the subdomains \boundarylayerdomain{i, } 
and \bulkdomain{i, }. Subsequently, a mortar finite element discretization 
is introduced for the Lagrange multiplier fields. 
First, the mortar discretization of the contact tractions \lagmultvector{} is discussed, 
followed by that of the embedded mesh interface tractions \lambdastar{\abodyi}{}. 
For the latter, particular attention is paid to the selection of suitable discrete Lagrange multiplier spaces.

\begin{figure}
	\centering
	\def\svgwidth{5.0cm}
	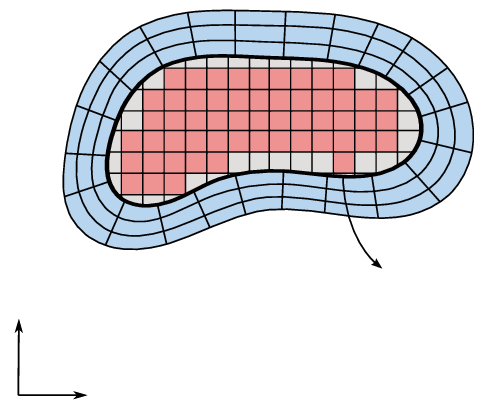
	\caption{Spatial discretization approach. The boundary layer mesh and the background mesh are shown in blue 
	and red, respectively. Background elements intersected by the coupling interface \interfacestar{}{} are highlighted in gray.}
	\label{fig:spatial_discretization}
\end{figure}

The proposed discretization approach can be fully automated, 
as outlined in Section~\ref{section:discretization_workflow}, 
where the process originates from an arbitrary NURBS-based surface representation defined in a CAD/CAM system. 
Following the meshing pipeline illustrated in Figure~\ref{fig:discretization-pipeline} 
leads to the mesh configuration shown in Figure~\ref{fig:spatial_discretization}, 
in which the boundary layer mesh overlaps the Cartesian background mesh. 
As a result, some elements of the background mesh are intersected by the coupling interface 
$\interfacestar{}{}$ of the boundary layer mesh. 
These intersected elements, referred to as \textit{cut elements}, require special treatment 
in the numerical integration procedure. 
The corresponding treatment, together with additional details on the contact integration, 
is discussed at the end of this section.

\begin{remark}
	For the discretization of the boundary layer and background meshes, 
	both isogeometric and classical Lagrangian elements may be employed and combined arbitrarily. 
	In isogeometric elements, the degrees of freedom are associated with control points, 
	whereas in Lagrangian elements they are associated with nodes. 
	For simplicity of terminology, both control points and nodes are hereafter collectively referred to as nodes.
\end{remark}

\subsection{Structure discretization}

The solid subdomains \boundarylayerdomain{i, } and \bulkdomain{i, } are discretized using an 
isoparametric approach. For the subdomains $j = \layerindex, \backgroundindex$, the 
discrete element position 
\elepos{i}{j}, displacement \eledis{i}{j}, and virtual displacement \elevirtualdis{i}{j} 
are approximated as

\begin{equation}
	\elepos{i}{j} = \sum_{k = 1}^{\numnodesdiscret{i}{j}} \ashapefunction{k} \ \nodalpos{i,j}{k},
\end{equation}

\begin{equation}
	\eledis{i}{j} = \sum_{k = 1}^{\numnodesdiscret{i}{j}} \ashapefunction{k} \ \nodaldis{(i,j)}{k},
\end{equation}
and 
\begin{equation}
	\elevirtualdis{i}{j} = \sum_{k = 1}^{\numnodesdiscret{i}{j}} \ashapefunction{k} \ \nodalvirtualdis{i,j}{k}.
	\label{eq:discrete_virtual_dis}
\end{equation}

Here, $\ashapefunction{k} \in \realspace$ denotes the interpolation function associated with 
element node~$k$, while $\nodalpos{i,j}{k}$, $\nodaldis{(i,j)}{k}$, and $\nodalvirtualdis{i,j}{k}$ 
represent the nodal positions, displacements, and virtual displacements, respectively. 
For body~$i$, each subdomain~$j$ contains $\numnodesdiscret{i}{j}$ nodes. 
The definition of the interpolation functions $\ashapefunction{k}$ depends on the chosen 
discretization technology. Following the discretization workflow described 
in Section~\ref{section:discretization_workflow}, 
the boundary layer domain \boundarylayerdomain{i, } is discretized using NURBS basis 
functions in order to ensure a $C^1$-continuous surface description, 
which is advantageous for the contact formulation. 
The background mesh~\bulkdomain{i, } may be discretized either 
by classical Lagrangian polynomials within the \fem{} framework or by NURBS basis 
functions within \iga{}.

\subsection{Discretization of the contact Lagrange multiplier}

A mortar finite element discretization is applied to the contact traction 
\lagmultvector{}, see~\cite{Popp2009, Wohlmuth2000}. 
The resulting discrete Lagrange multiplier field \lagmultvector{h} is given by

\begin{equation}
	\lagmultvector{h} = \sum_{q = 1}^{\numnodeswithlagrange{}} \alagmultshapefunction{q} \ \nodallagmult{q}{}.
	\label{eq:contact_discrete_lagrange_mult}
\end{equation}

Here, \alagmultshapefunction{q} is a matrix of shape functions, 
and \nodallagmult{q}{} represents the corresponding nodal Lagrange multipliers. 
In mortar discretizations, it is common that each node on the slave 
side carries a Lagrange multiplier, \ie{} 
$\numnodeswithlagrange{} = \numnodesslavebl{}$, where \numnodesslavebl{} denotes the 
number of nodes on the slave side. However, this is not a strict requirement. 
For second-order Lagrangian elements, it is often preferable that not all 
slave-side nodes carry a Lagrange multiplier, \ie{} $\numnodeswithlagrange{} < \numnodesslavebl{}$, 
see~\cite{Wohlmuth2012,Popp2012}. 

In~\cite{Seitz2016}, a mortar framework based on dual shape functions (which will be briefly discussed 
in the following lines) was proposed for 
contact mechanics and classical domain decomposition for NURBS discretizations of 
second and third order. Optimal convergence rates under uniform 
mesh refinement were demonstrated. 
In the present approach, the contact surfaces are  
is discretized using a boundary layer based on NURBS shape functions. 
Consequently, we follow the mortar discretization strategy of~\cite{Seitz2016} 
and assign one scalar Lagrange multiplier to each slave-side node. 

To complete the discretization, suitable shape functions for the 
Lagrange multiplier field \lagmultvector{} 
must be defined. 
We therefore introduce a discrete Lagrange multiplier space 
$\lagmultsolutionspace{}_{,h}^{+}$ 
that approximates the continuous space $\lagmultsolutionspace{}^{+}$.
In mortar finite element discretizations, the selection of an appropriate discrete space
for $\lagmultsolutionspace{}_{,h}^{+}$ is crucial to ensure 
stability of the method~\cite{Hueber2005}. 
Two well-known types of Lagrange multiplier interpolation are \textit{standard} 
and \textit{dual} shape functions. For standard shape functions, \alagmultshapefunction{q} 
are chosen identical to the shape functions used to interpolate the primal variable on 
the slave side~$\interfacereference{c}{\abodynum{1}}$, \ie{} $\alagmultshapefunction{q} = \ashapefunction{a}$.
On the other hand, dual shape functions are constructed such that they satisfy 
the biorthogonality condition

\begin{equation}
	\int_{\interfacespatial{c, h}{\abodynum{1}}} 
	\alagmultshapefunction{q} \, \ashapefunction{a} \, \mathrm{d}A
	=
	\kronecker{q a}
	\int_{\interfacespatial{c, h}{\abodynum{1}}} 
	\ashapefunction{a} \, \mathrm{d}A,
	\quad 
	q, a = 1, \ldots, \numnodeswithlagrange{}.
\end{equation}

Here, \kronecker{q a} denotes the Kronecker delta. 
Satisfaction of the biorthogonality condition ensures that the dual shape functions \alagmultshapefunction{q} are 
well defined and possess the required approximation properties. 
Dual shape functions simplify the enforcement of mortar coupling conditions while 
preserving the mathematical properties of the method. 
Furthermore, they allow for efficient condensation of the discrete Lagrange multiplier degrees 
of freedom, thereby eliminating the saddle-point structure of the resulting system. 
For further details on the construction of dual shape functions, we refer 
to~\cite{Scott1990, Wohlmuth2000}.

Substituting \eqref{eq:discrete_virtual_dis} and \eqref{eq:contact_discrete_lagrange_mult} 
into the contact virtual work expression \eqref{eq:contact_energy} yields 
the discretized contact virtual work
\begin{align}
	\contactenergy{,h} &= \sum_{q = 1}^{\numnodeswithlagrange{}} \sum_{a = 1}^{\numnodesslavebl{}}  \nodallagmult{q}{T}
						 \left( \int_{\interfacespatial{c}{\abodynum{1}}} \alagmultshapefunction{q} \ \ashapefunction{a} \ \mathrm{d} A  \right) \delta \nodaldisslave{}{a} \nonumber \\
					   &- \sum_{q = 1}^{\numnodeswithlagrange{}} \sum_{b = 1}^{\numnodesmasterbl{}} \nodallagmult{q}{T} 
					     \left( \int_{\interfacespatial{c}{\abodynum{1}}} \alagmultshapefunction{q} \left(\ashapefunction{b} \circ \projectionsurfacetosurface{} \right) \mathrm{d} A  \right) \delta \nodaldismaster{}{b},
    \label{eq:discretized_contact_energy}
\end{align}
where, $\delta\nodaldisslave{}{a}$ and $\delta\nodaldismaster{}{b}$ denote 
the virtual displacements on the slave and master side, respectively.
In the present formulation, $\delta\nodaldisslave{}{a} \in \delta \nodaldis{(1,\layerindex)}{a}$ and 
$\delta\nodaldismaster{}{b} \in \delta \nodaldis{(2,\layerindex)}{b}$.
The operator $\projectionsurfacetosurface{} : \interfacespatial{c,h}{\abodynum{2}} \rightarrow \interfacespatial{c,h}{\abodynum{1}}$ 
defines a suitable projection from the discretized master side onto the 
discretized slave side. The master side is described by 
\numnodesmasterbl{} number of nodes. From \eqref{eq:discretized_contact_energy}, the 
mortar matrices \localmortaroperatorD{} 
and \localmortaroperatorM{} are defined as
\begin{align}
	\localmortaroperatorD{[q, a]} &= \int_{\interfacespatial{c}{\abodynum{1}}} \alagmultshapefunction{q} \ \ashapefunction{a} \ \mathrm{d} A \ \identitymatrix{}, &q = 1, ..., \numnodeswithlagrange{}, a = 1, ..., \numnodesslavebl{}  \label{eq:mortar_matrix_D_contact}\\
	\localmortaroperatorM{[q, b]} &= \int_{\interfacespatial{c}{\abodynum{1}}} \alagmultshapefunction{q} \left(\ashapefunction{b} \circ \projectionsurfacetosurface{} \right) \mathrm{d} A \ \identitymatrix{}, &q = 1, ..., \numnodeswithlagrange{}, b = 1, ..., \numnodesmasterbl{},  \label{eq:mortar_matrix_M_contact}
\end{align}
where $\identitymatrix{} \in \realspace^{3 \times 3}$ denotes the identity matrix. 
In general, both \localmortaroperatorD{} and \localmortaroperatorM{} are rectangular matrices.
However, if $\numnodeswithlagrange{} = \numnodesslavebl{}$, the matrix \localmortaroperatorD{} 
becomes square. To simplify the notation, 
we can define a discrete displacement vector \contactdiscretedisplacementvector{},
which groups the displacements related to the contact interface in 
$\contactdiscretedisplacementvector{} = (\nodaldismaster{}{}, \nodaldisslave{}{})$. 
Substituting the mortar matrices~\eqref{eq:mortar_matrix_D_contact} 
and \eqref{eq:mortar_matrix_M_contact} into 
\eqref{eq:discretized_contact_energy} yields
\begin{align}
	\contactenergy{,h} 
	&= 
	\delta \nodaldisslave{^\transpose}{} 
	\localmortaroperatorD{}^{\transpose} 
	\nodallagmult{}{}
	-
	\delta \nodaldismaster{^\transpose}{} 
	\localmortaroperatorM{}^{\transpose} 
	\nodallagmult{}{} \nonumber \\
	&= 
	\delta \contactdiscretedisplacementvector{}^{\transpose}
	\underbrace{
	\begin{bmatrix}
	-\localmortaroperatorM{}^{\transpose} \\
	\localmortaroperatorD{}^{\transpose}
	\end{bmatrix}
	\nodallagmult{}{}
	}_{\contactforce(\contactdiscretedisplacementvector{}, \nodallagmult{}{})}
	=
	\delta \contactdiscretedisplacementvector{}^{\transpose}
	\contactforce(\contactdiscretedisplacementvector{}, \nodallagmult{}{}).
	\label{eq:discretized_contact_energy_matrix}
\end{align}

Finally, the weak constraint contribution for frictionless contact, 
defined in \eqref{eq:weak_form_2} and \eqref{eq:contact_weak_constraint_energy}, 
is discretized. In~\cite{Hueeber2008}, 
the discrete counterpart of these expressions is derived and shown to be equivalent to pointwise conditions. 
The corresponding discrete conditions read

\begin{equation}
	\left(\gapfunctiondiscrete\right)_q \geq 0, \quad \left(\lagmultscalarnormal\right)_q \geq 0, \quad \left(\gapfunctiondiscrete\right)_q \left(\lagmultscalarnormal\right)_q = 0, \quad q=1, ..., \numnodeswithlagrange{},
	\label{eq:discretized-pointwise-conditions}
\end{equation}
where $\left(\gapfunctiondiscrete\right)_q$ denotes the discrete weighted gap function at the slave node $q$, defined by
\begin{equation}
	\left(\gapfunctiondiscrete\right)_q
	=
	\int_{\interfacespatial{c}{\abodynum{1}}} 
	\alagmultshapefunction{q} \, \gapfunction{}_{,h} 
	\, \mathrm{d}A,
\end{equation}
with $\gapfunction{}_{,h}$ representing the discretized gap function introduced in \eqref{eq:gap_function}.

\subsection{Discretization of the embedded mesh Lagrange multiplier}\label{subsection:em_coupling_discretization}

Analogously to the contact traction, the embedded mesh coupling traction $\lambdastar{\abodyi}{}$ 
is discretized using a mortar-based approach. Since this discretization is performed for each body~$i$, 
the superscript~$\abodyi$ is omitted in the following for notational simplicity. 
For an arbitrary body, the discrete embedded mesh coupling traction is given by

\begin{equation}
	\lambdastar{}{,h} = \sum_{r = 1}^{\numinterfacenodes{}} \alagmultshapefunctionstar{r} \, \nodallagmultstar{r}{}, 
	\label{eq:embedded_discrete_lagrange_mult}
\end{equation}
where \alagmultshapefunctionstar{r} denote the Lagrange multiplier shape functions, and 
\nodallagmultstar{r}{} are the associated nodal Lagrange multipliers defined on the 
coupling interface \interfacestar{}{}. 

The selection of suitable Lagrange multiplier shape functions 
in~\eqref{eq:embedded_discrete_lagrange_mult} 
is more involved than in classical mesh-tying configurations. 
In standard mesh tying, two boundary-conforming meshes are coupled along a common interface. 
In contrast, in an embedded mesh configuration only one mesh--here, the 
boundary layer mesh--remains boundary conforming and defines the coupling interface, 
which is embedded into the second mesh, namely the Cartesian background mesh.
As a consequence, stability properties that hold for commonly used Lagrange multiplier spaces in classical mortar methods 
do not automatically transfer to the embedded mesh setting. 
In particular, an inappropriate choice of the discrete Lagrange multiplier space 
may lead to an unstable multiplier field due to a violation of the discrete inf–sup condition~\cite{Lew2008}.

In this work, the Lagrange multipliers are defined at the nodes on \interfacestar{}{}, 
following the approach proposed in~\cite{Sanders2012}. 
For an interface \interfacestar{}{} described by the inner surface of the boundary layer mesh 
with $\numnodesinterfacestar{}$ nodes, this yields 
$\numinterfacenodes{} = \numnodesinterfacestar{}$. 
Furthermore, a standard mortar interpolation is employed for the Lagrange multiplier field, 
\ie{} $\alagmultshapefunctionstar{r} = \ashapefunction{c}$, 
where $\ashapefunction{c}$ denote the shape functions defining \interfacestar{}{}.

As shown in~\cite{Sanders2012}, the use of Lagrange multiplier fields in embedded mesh 
configurations may lead to mesh locking if the overlapping mesh is finer and 
stiffer than the underlying mesh. It should be noted that alternative coupling 
strategies exist, such as Nitsche’s method~\cite{Nitsche1971, Sanders2012, Wei2021} 
and discontinuous Galerkin approaches~\cite{Lew2008, Sanders2012a}, which do not 
exhibit mesh locking and could be employed instead of a mortar-based formulation. 
Nevertheless, for the class of problems considered in this work, stable and convergent 
results can be achieved using a mortar-based approach. This is demonstrated by 
the numerical experiments presented in Subsections~\ref{subsection:mesh_locking} 
and~\ref{subsection:hertzian-contact-problem}.

By inserting the approximations \eqref{eq:discrete_virtual_dis} and \eqref{eq:embedded_discrete_lagrange_mult} 
into \eqref{eq:coupling_interface_energy} and \eqref{eq:coupling_constraint_energy}, 
the discretized coupling contributions are obtained as
\begin{equation}
	-\couplinginterenergy{}{,h} = \sum_{c = 1}^{\numnodesinterfacestar{}} \sum_{r = 1}^{\numinterfacenodes{}} \nodallagmultstar{r}{\transpose} 
	                              \left( \int_{\interfacestar{}{}}  \alagmultshapefunctionstar{r} \ \ashapefunction{c} \ \mathrm{d} A  \right) \delta \nodaldisinterface{}{c}  
								- \sum_{d = 1}^{\numnodescutelements{}} \sum_{r = 1}^{\numinterfacenodes{}} \nodallagmultstar{r}{\transpose} 
								  \left( \int_{\interfacestar{}{}} \alagmultshapefunctionstar{r} \left( \ashapefunction{d} \circ \projectionsurfcetovolume{} \right) \mathrm{d} A \right) \delta \nodaldiscut{}{d},
	\label{eq:discretized_coupling_interface_energy}
\end{equation}

and 
\begin{equation}
	\couplingconstraintenergy{}{,h} = \sum_{c = 1}^{\numnodesinterfacestar{}} \sum_{r = 1}^{\numinterfacenodes{}} \delta \nodallagmultstar{r}{\transpose} 
										\left( \int_{\interfacestar{}{}}  \alagmultshapefunctionstar{r} \ \ashapefunction{c} \ \mathrm{d} A  \right)  \nodaldisinterface{}{c}
										- \sum_{d = 1}^{\numnodescutelements{}} \sum_{r = 1}^{\numinterfacenodes{}} \delta \nodallagmultstar{r}{\transpose} 
								         \left( \int_{\interfacestar{}{}} \alagmultshapefunctionstar{r} \left( \ashapefunction{d} \circ \projectionsurfcetovolume{} \right) \mathrm{d} A \right) \nodaldiscut{}{d},
	\label{eq:discretized_coupling_constraint_energy}
\end{equation}
where $\delta \nodaldisinterface{}{c}$ and $\delta \nodaldiscut{}{d}$ 
denote the virtual displacements on the interface and on
the corresponding cut elements of the background mesh, respectively, such that 
$\delta \nodaldisinterface{}{c} \in \delta \nodaldis{\layerindex}{c}$ 
and $\delta \nodaldiscut{}{d} \in \delta \nodaldis{\backgroundindex}{d}$.
Furthermore,~$\numnodescutelements{}$ is the number of nodes associated with the corresponding cut elements. 
The operator~$\projectionsurfcetovolume{} : \interfacestar{\interfaceindex}{,h} \rightarrow \interfacestar{\cuteleindex}{,h}$ 
defines a projection mapping a point on the boundary layer interface to a corresponding point 
in the background mesh. It is important to emphasize that the projection operator $\projectionsurfcetovolume{}$
differs fundamentally from the contact projection~$\projectionsurfacetosurface{}$ introduced in~\eqref{eq:discretized_contact_energy}. 
While~$\projectionsurfcetovolume{}$ defines a \textit{surface-to-volume} projection, $\projectionsurfacetosurface{}$ 
corresponds to a \textit{surface-to-surface} projection. This distinction reflects a fundamental 
difference between mortar formulations for embedded mesh coupling and those used in contact problems, 
with significant implications for the evaluation of discrete projections, mesh intersection procedures, 
and the required geometric operations.
Based on \eqref{eq:discretized_coupling_interface_energy} and \eqref{eq:discretized_coupling_constraint_energy}, the mortar matrices $\localmortaroperatorDstar{}{}$ and $\localmortaroperatorMstar{}{}$ are defined by
\begin{align}
	\localmortaroperatorDstar{}{[r, c]} &= \int_{\interfacestar{}{}} \alagmultshapefunctionstar{r} \ \ashapefunction{c}  \ \mathrm{d} A \ \identitymatrix{}, &r = 1, \ldots, \numinterfacenodes{}, c = 1, \ldots, \numnodesinterfacestar{}, \label{eq:mortar_matrix_D_embeddedmesh}\\
	\localmortaroperatorMstar{}{[r, d]} &= \int_{\interfacestar{}{}} \alagmultshapefunctionstar{r} \left( \ashapefunction{d} \circ \projectionsurfcetovolume{} \right) \mathrm{d} A \ \identitymatrix{},  &r = 1, \ldots, \numinterfacenodes{}, d = 1, \ldots, \numnodescutelements{} \label{eq:mortar_matrix_M_embeddedmesh}.  
\end{align}	

Similar to the previous subsection, the notation can be simplified by defining a discrete displacement vector
$\emdiscretedisplacementvector{} = (\nodaldiscut{}{}, \nodaldisinterface{}{})$, which collects the 
displacements associated with the embedded mesh coupling.
Using the mortar matrices \localmortaroperatorDstar{}{} and \localmortaroperatorMstar{}{}, 
we can reformulate~\eqref{eq:discretized_coupling_interface_energy} and~\eqref{eq:discretized_coupling_constraint_energy}, which yields 

\begin{equation}
	-\couplinginterenergy{}{,h} =   \delta \nodaldisinterface{^\transpose}{} \localmortaroperatorDstar{\transpose} \nodallagmultstar{}{} - \delta \nodaldiscut{^\transpose}{} \localmortaroperatorMstar{\transpose} \nodallagmultstar{}{} = \delta \emdiscretedisplacementvector{}^{\transpose} \underbrace{ \left[\begin{array}{c}
		-\localmortaroperatorMstar{\transpose} \\
		\localmortaroperatorDstar{\transpose}
		\end{array}\right] \nodallagmultstar{}{}}_{\couplingforce{}(\nodallagmultstar{}{})} = \delta \emdiscretedisplacementvector{}^{\transpose} \couplingforce{}(\nodallagmultstar{}{}),
    \label{eq:discretized_coupling_interface_energy_reform}
\end{equation}
and 
\begin{equation}
	\couplingconstraintenergy{}{,h} = \delta \nodallagmultstar{}{\transpose} \localmortaroperatorDstar{} \nodaldisinterface{}{} -\delta \nodallagmultstar{}{\transpose} \localmortaroperatorMstar{} \nodaldiscut{}{} = \delta \nodallagmultstar{}{\transpose} \underbrace{\left[\begin{array}{ll}
	-\localmortaroperatorMstar{} & \localmortaroperatorDstar{}
	\end{array}\right]\emdiscretedisplacementvector{}}_{\couplingconstraint{}\left(\emdiscretedisplacementvector{}\right)} .
	\label{eq:discretized_coupling_constraint_energy_reform}
\end{equation}
Here, \couplingforce{} represents the vector of discretized forces 
acting on the degrees of freedom (\dofs{}) involved in the embedded mesh coupling, 
including both interface and cut element nodes. 
The vector \couplingconstraint{} denotes the discrete coupling constraint.

\subsection{Complete discretized contact-embedded mesh formulation}\label{subsection:complete_discretization}

\newcommand{\blockMatrixKco}{%
\colorbox{red!15}{\parbox[c][1.2cm][c]{1.2cm}{\centering $\stiffnessmatrixcontact$}}}

\newcommand{\blockzeroupperright}{%
\colorbox{gray!15}{\parbox[c][1.2cm][c]{0.9cm}{\centering $\zeromatrix$}}}

\newcommand{\blockzerolowerleft}{%
\colorbox{gray!15}{\parbox[c][0.7cm][c]{1.2cm}{\centering $\zeromatrix$}}}

\newcommand{\blockMatrixKstar}{%
\colorbox{cyan!15}{\parbox[c][0.7cm][c]{0.9cm}{\centering $\stiffnessmatrixembedded$}}}

Finally, all discretized residual contributions from the internal and external forces, 
as well as from the contact and embedded mesh coupling formulations,  
are assembled into the global weak formulation introduced in 
\eqref{eq:global_formulation}–\eqref{eq:coupling_weak_formulation}. 
To this end, we define the global displacement vector  
$\globaldiscretedisplacementvector = (\nodaldisnocontactnoembedded, \contactdiscretedisplacementvector{}, \emdiscretedisplacementvector{})$, 
where \nodaldisnocontactnoembedded{} groups the remaining displacements 
that are not associated with the contact or embedded mesh coupling formulations.
The discretized problem formulation reads
\begin{align}
	\internalforce{}\left(\globaldiscretedisplacementvector\right) + \contactforce\left(\globaldiscretedisplacementvector{}, \nodallagmult{}{}\right) + \couplingforce{}(\nodallagmultstar{}{}) - \externalforce{} &= \zerovector, \label{eq:final_residual}\\
	\left(\gapfunctiondiscrete\right)_q \geq 0, \quad \left(\lagmultscalarnormal\right)_q \geq 0, \quad \left(\gapfunctiondiscrete\right)_q \left(\lagmultscalarnormal\right)_q &= 0, \quad q=1, \ldots, \numnodeswithlagrange{}, \label{eq:final_contact_global_inequality} \\
	\couplingconstraint{}\left(\globaldiscretedisplacementvector\right) &= \zerovector. \label{eq:final_em_global_quality}
\end{align}
The resulting nonlinear system is solved using the Newton-Raphson method.
The system exhibits a double saddle-point structure due to 
the mixed formulation for both the contact and embedded mesh coupling. 
In the following, we briefly describe the solution strategies to solve 
the nonlinear and linear systems.

The inequality constraint in \eqref{eq:final_contact_global_inequality} introduces 
an additional nonlinearity, since the active set of slave nodes in contact 
with the master side is not known a priori.
To address this, a primal-dual active set strategy (PDASS) is employed 
within the contact solution algorithm. In short, the~PDASS can be interpreted as a 
semismooth Newton method~\cite{Hintermueller2002}, which treats the nonlinearities arising from 
finite deformations, nonlinear material behaviour, and contact 
within a unified iterative scheme. This requires a reformulation 
of the KKT conditions using a nonlinear complementarity function \complementarityfunction{j} 
for each slave node $j$. Results from the mathematical literature show that the resulting 
formulation is semismooth, such that a Newton method can still be applied. 
Furthermore, the algorithm requires a consistent 
linearization of all deformation-dependent contact quantities, 
such as the mortar matrices \localmortaroperatorD{} and \localmortaroperatorM{}.  
The algorithmic details of the PDASS and the consistent linearization 
of deformation-dependent contact terms are 
described in~\cite{Gitterle2010, Popp2009, Popp2010, Hueber2005}. 

Although the saddle-point system in \eqref{eq:final_residual}-\eqref{eq:final_em_global_quality} 
can be solved using either direct solvers or tailored iterative solvers, 
some techniques may be employed to simplify the system. 
For the contact block, the use of dual shape functions for the interpolation 
of the Lagrange multipliers is particularly advantageous. 
The biorthogonality property reduces the mortar matrix \localmortaroperatorD{} 
to a diagonal matrix, which enables straightforward elimination of the discrete Lagrange multipliers 
by static condensation~\cite{Wohlmuth2000, Popp2009, Popp2010, Gitterle2010, Popp2012}.

To eliminate the Lagrange multipliers in the embedded mesh coupling problem, 
a node-wise penalty regularization is introduced~\cite{Yang2004,Steinbrecher2020,Steinbrecher2022,Steinbrecher2025}. 
The regularization reads
\begin{equation}
	\nodallagmultstar{}{} \approx \penaltyparam{} \scalingmatrix{-1} \couplingconstraint{}\left(\emdiscretedisplacementvector{}\right),
	\label{eq:penalty_regularization}
\end{equation}
thus, the Lagrange multiplier \nodallagmultstar{}{} becomes a pure function 
of the unknown displacements \emdiscretedisplacementvector{}. 
Here,~$\penaltyparam~\in~\realspace^{+}$ denotes the penalty parameter, 
and \scalingmatrix{} is a diagonal nodal scaling matrix. 
An exact fulfillment of~\eqref{eq:penalty_regularization} 
is achieved for $\penaltyparam{} \rightarrow \infty$, which, however, leads to poor conditioning of the system. 
For practical purposes, a penalty parameter of the order $\penaltyparam{} \sim 100\,\Youngsmodulus{}$ has proven 
sufficiently accurate, where~\Youngsmodulus{} denotes the Young's modulus of the material. \responsetoview{Subsection~\ref{subsection:hertzian-contact-problem} 
illustrates the sensitivity of the numerical solution to the penalty parameter.} 
The contribution of the discrete coupling constraint \couplingconstraint{} is proportional to 
the support of the associated Lagrange multiplier shape functions; \ie{} it depends on the 
interface element area. As described in~\cite{Steinbrecher2020}, to ensure the fulfillment 
of the patch tests presented in Subsection~\ref{subsection:patch_test}, the vector \couplingconstraint{} 
is scaled by the inverse of the diagonal matrix~\scalingmatrix{}, following the approach 
presented in~\cite{Yang2004}.
For each Lagrange multiplier node $r$, the local scaling matrix is defined as

\begin{equation}
	\scalingmatrix{}[r,r] = \int_{\interfacestar{}{,h}} \alagmultshapefunctionstar{r} \ \mathrm{d} A \ \identitymatrix^{3 \times 3},
\end{equation}
which is assembled into the global scaling matrix \scalingmatrix{}.

\subsection{Numerical integration}\label{subsection:numerical_integration_cutelements}

Special attention is required for the numerical integration of the individual blocks arising 
from the contact and embedded mesh formulations. Focusing first on the contact contributions, 
the mortar matrices \localmortaroperatorD{} and \localmortaroperatorM{}, introduced 
in~\eqref{eq:mortar_matrix_D_contact} and~\eqref{eq:mortar_matrix_M_contact}, respectively, 
require integration over the current slave side. To this end, a projection between the discretized 
master and slave sides is performed in order to determine suitable integration points. 
Established algorithms, such as integration schemes based on segmentation procedures, 
can be employed for this purpose. Such approaches have been successfully applied 
in contact formulations based on classical finite elements~\cite{Puso2004, Puso2008, Popp2010}, as well 
as in contact problems using \iga{} discretizations~\cite{Hesch2012, Dittmann2014, Seitz2016}. 

The embedded mesh discretization illustrated in Figure~\ref{fig:spatial_discretization} requires 
modifications to the element quadrature routines for cut elements that are overlapped by 
the boundary layer mesh. Moreover, within the embedded mesh formulation, the mortar 
matrices defined in~\eqref{eq:mortar_matrix_D_embeddedmesh} and~\eqref{eq:mortar_matrix_M_embeddedmesh} 
require integration over the coupling interface $\interfacestar{}{,h}$. 
Such modifications are well established in the context of extended finite element 
methods (XFEM)~\cite{Mayer2010, Moes1999} and CutFEM approaches~\cite{Burman2015, Winter2018}. 
In this work, we adopt the well-tested routines presented in~\cite{Mayer2009, Sudhakar2014}, 
which are briefly summarized in the following.

\begin{figure}
	\centering

	\mbox{}\hfill 
	\begin{subfigure}[t]{4.5cm}
		\centering
		\def\svgwidth{4.5cm}
		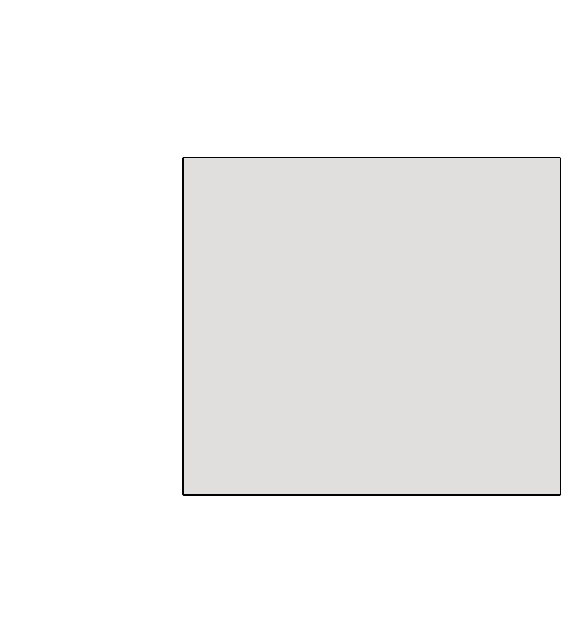
		\caption{}
		\label{fig:numerical_integration_1}
	\end{subfigure} \hfill
	\begin{subfigure}[t]{4.5cm}
		\centering
		\def\svgwidth{4.5cm}
		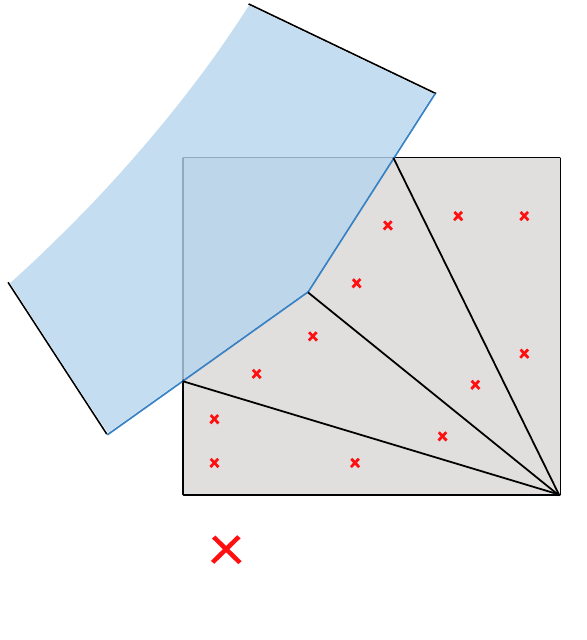
		\caption{}
		\label{fig:numerical_integration_2}
	\end{subfigure} \hfill
	\begin{subfigure}[t]{4.5cm}
		\centering
		\def\svgwidth{4.5cm}
		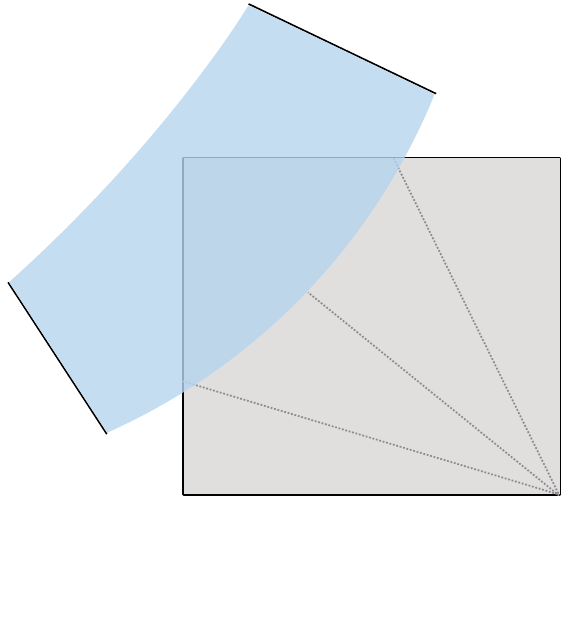
		\caption{}
		\label{fig:numerical_integration_3}
	\end{subfigure} 
	\hfill\mbox{}

	\caption{(a) Numerical integration of a cut element (gray) intersected by 
	\anyinterfacestar{,h} (blue).	(b) Resulting triangulation and Gauss points with the 
	linearized interface~\lininterfacestar{}{} (figure adapted from~\cite{Sanders2012}). (c) Mapping of boundary segments and their Gauss points 
	onto \interfacestar{}{,h} and into the cut element.}
	\label{fig:numerical_integration}
\end{figure}

In Figure~\ref{fig:numerical_integration_1}, a two-dimensional cut element 
is intersected by the interface $\interfacestar{}{,h}$. 
\responsetoview{Two subdomains can be distinguished: an inactive region, which does not contribute to the element stiffness, 
and an active region, which does contribute to the stiffness.} 
This decomposition requires computing the geometric intersection between the cut element and the interface.
Since the interface~$\interfacestar{}{,h}$ is represented by a nonlinear NURBS geometry, 
it is first subdivided into linear segments, yielding a linearized interface 
denoted by $\lininterfacestar{}{}$. 
The cut element is then intersected with these linear segments and partitioned into integration cells 
using a constrained Delaunay triangulation. 
\responsetoview{For a two-dimensional element, the active part of the cut element is decomposed 
into triangular subcells.}
Each subcell is integrated using standard Gauss quadrature with $\numgpstetrahedra{}$ Gauss points 
and corresponding weights, as illustrated in Figure~\ref{fig:numerical_integration_2}.



For the evaluation of the surface terms in the embedded mesh coupling, 
an integration rule must be defined on each interface element and its corresponding underlying cut element. 
This is achieved using the boundary segments obtained from the Delaunay triangulation. 
On these boundary segments, Gauss integration points are defined. 
The boundary segments and their associated Gauss points are then mapped onto the interface $\interfacestar{}{,h}$, 
as illustrated in Figure~\ref{fig:numerical_integration_3}. In a final step, the integration points on 
$\interfacestar{}{,h}$ are then mapped to the background element.

\begin{remark}
\responsetoview{A well-known challenge in immersed boundary and embedded mesh approaches arises when the intersection of the coupling interface with the background mesh gives rise to cut elements with very small active volume fractions.
Such small integration domains may severely deteriorate the conditioning of the discrete system and, depending on the formulation, may also affect its stability. Within CutFEM frameworks, these difficulties are commonly addressed by means of ghost-penalty stabilization~\cite{Burman2010b,dePrenter2023}.
Although ghost-penalty stabilization has predominantly been employed in formulations in which the coupling conditions are imposed weakly using Nitsche's method~\cite{Schott2016,Burman2012,Burman2014}, it can also be incorporated into formulations based on Lagrange multipliers~\cite{Burman2010a,Burman2024}.
In the present work, no additional stabilization technique is introduced to address cut elements with arbitrarily small active volume fractions, as the primary focus lies on the proposed contact discretization and embedded mesh coupling framework rather than on the treatment of small-cut instabilities.
In the numerical examples considered in this work, no severe difficulties associated with small cut elements were encountered.
In the final three-dimensional numerical example presented in Section~\ref{subsection:tori_contact}, cut elements whose active volume fraction falls below a prescribed threshold are removed in order to prevent severe ill-conditioning of the resulting system.
The development and investigation of dedicated stabilization techniques for the proposed framework constitute an important topic for future research.}
\end{remark}

%% file: fig/spatial_discretization/spatial_discretization.eps_tex
\begingroup%
  \makeatletter%
  \providecommand\color[2][]{%
    \errmessage{(Inkscape) Color is used for the text in Inkscape, but the package 'color.sty' is not loaded}%
    \renewcommand\color[2][]{}%
  }%
  \providecommand\transparent[1]{%
    \errmessage{(Inkscape) Transparency is used (non-zero) for the text in Inkscape, but the package 'transparent.sty' is not loaded}%
    \renewcommand\transparent[1]{}%
  }%
  \providecommand\rotatebox[2]{#2}%
  \newcommand*\fsize{\dimexpr\f@size pt\relax}%
  \newcommand*\lineheight[1]{\fontsize{\fsize}{#1\fsize}\selectfont}%
  \ifx\svgwidth\undefined%
    \setlength{\unitlength}{232.00000763bp}%
    \ifx\svgscale\undefined%
      \relax%
    \else%
      \setlength{\unitlength}{\unitlength * \real{\svgscale}}%
    \fi%
  \else%
    \setlength{\unitlength}{\svgwidth}%
  \fi%
  \global\let\svgwidth\undefined%
  \global\let\svgscale\undefined%
  \makeatother%
  \begin{picture}(1,0.8620689)%
    \lineheight{1}%
    \setlength\tabcolsep{0pt}%
    \put(0,0){\includegraphics[width=\unitlength]{spatial_discretization.eps}}%
    \put(0.81025484,0.26316163){\color[rgb]{0,0.00784314,0.01176471}\makebox(0,0)[lt]{\lineheight{0}\smash{\begin{tabular}[t]{l}$\interfacestar{}{}$\end{tabular}}}}%
    \put(0.01601474,0.2357308){\color[rgb]{0,0.00784314,0.01176471}\makebox(0,0)[lt]{\lineheight{0}\smash{\begin{tabular}[t]{l}$\yvector$\end{tabular}}}}%
    \put(0.18948588,0.03656821){\color[rgb]{0,0.00784314,0.01176471}\makebox(0,0)[lt]{\lineheight{0}\smash{\begin{tabular}[t]{l}$\xvector$\end{tabular}}}}%
  \end{picture}%
\endgroup%

%% file: fig/spatial_discretization/numerical_integration_1.pdf_tex
\begingroup%
  \makeatletter%
  \providecommand\color[2][]{%
    \errmessage{(Inkscape) Color is used for the text in Inkscape, but the package 'color.sty' is not loaded}%
    \renewcommand\color[2][]{}%
  }%
  \providecommand\transparent[1]{%
    \errmessage{(Inkscape) Transparency is used (non-zero) for the text in Inkscape, but the package 'transparent.sty' is not loaded}%
    \renewcommand\transparent[1]{}%
  }%
  \providecommand\rotatebox[2]{#2}%
  \newcommand*\fsize{\dimexpr\f@size pt\relax}%
  \newcommand*\lineheight[1]{\fontsize{\fsize}{#1\fsize}\selectfont}%
  \ifx\svgwidth\undefined%
    \setlength{\unitlength}{281.50830847bp}%
    \ifx\svgscale\undefined%
      \relax%
    \else%
      \setlength{\unitlength}{\unitlength * \real{\svgscale}}%
    \fi%
  \else%
    \setlength{\unitlength}{\svgwidth}%
  \fi%
  \global\let\svgwidth\undefined%
  \global\let\svgscale\undefined%
  \makeatother%
  \begin{picture}(1,1.09051665)%
    \lineheight{1}%
    \setlength\tabcolsep{0pt}%
    \put(0,0){\includegraphics[width=\unitlength,page=1]{numerical_integration_1.pdf}}%
    \put(0.32612718,0.70167027){\makebox(0,0)[lt]{\lineheight{1.25}\smash{\begin{tabular}[t]{l}\responsetoview{Inactive}\end{tabular}}}}%
    \put(0,0){\includegraphics[width=\unitlength,page=2]{numerical_integration_1.pdf}}%
    \put(0.76206792,0.92286452){\makebox(0,0)[lt]{\lineheight{1.25}\smash{\begin{tabular}[t]{l}\anyinterfacestar{,h}\end{tabular}}}}%
    \put(0.60458218,0.28637955){\makebox(0,0)[lt]{\lineheight{1.25}\smash{\begin{tabular}[t]{l}\responsetoview{Active}\end{tabular}}}}%
    \put(0,0){\includegraphics[width=\unitlength,page=3]{numerical_integration_1.pdf}}%
  \end{picture}%
\endgroup%

%% file: fig/spatial_discretization/numerical_integration_2.pdf_tex
\begingroup%
  \makeatletter%
  \providecommand\color[2][]{%
    \errmessage{(Inkscape) Color is used for the text in Inkscape, but the package 'color.sty' is not loaded}%
    \renewcommand\color[2][]{}%
  }%
  \providecommand\transparent[1]{%
    \errmessage{(Inkscape) Transparency is used (non-zero) for the text in Inkscape, but the package 'transparent.sty' is not loaded}%
    \renewcommand\transparent[1]{}%
  }%
  \providecommand\rotatebox[2]{#2}%
  \newcommand*\fsize{\dimexpr\f@size pt\relax}%
  \newcommand*\lineheight[1]{\fontsize{\fsize}{#1\fsize}\selectfont}%
  \ifx\svgwidth\undefined%
    \setlength{\unitlength}{281.50830847bp}%
    \ifx\svgscale\undefined%
      \relax%
    \else%
      \setlength{\unitlength}{\unitlength * \real{\svgscale}}%
    \fi%
  \else%
    \setlength{\unitlength}{\svgwidth}%
  \fi%
  \global\let\svgwidth\undefined%
  \global\let\svgscale\undefined%
  \makeatother%
  \begin{picture}(1,1.09051665)%
    \lineheight{1}%
    \setlength\tabcolsep{0pt}%
    \put(0,0){\includegraphics[width=\unitlength,page=1]{numerical_integration_2.pdf}}%
    \put(0.75636793,0.92496105){\makebox(0,0)[lt]{\lineheight{1.25}\smash{\begin{tabular}[t]{l}\lininterfacestar{}\end{tabular}}}}%
    \put(0.44506073,0.13788709){\makebox(0,0)[lt]{\lineheight{1.25}\smash{\begin{tabular}[t]{l}Gauss point\end{tabular}}}}%
  \end{picture}%
\endgroup%

%% file: fig/spatial_discretization/numerical_integration_3.pdf_tex
\begingroup%
  \makeatletter%
  \providecommand\color[2][]{%
    \errmessage{(Inkscape) Color is used for the text in Inkscape, but the package 'color.sty' is not loaded}%
    \renewcommand\color[2][]{}%
  }%
  \providecommand\transparent[1]{%
    \errmessage{(Inkscape) Transparency is used (non-zero) for the text in Inkscape, but the package 'transparent.sty' is not loaded}%
    \renewcommand\transparent[1]{}%
  }%
  \providecommand\rotatebox[2]{#2}%
  \newcommand*\fsize{\dimexpr\f@size pt\relax}%
  \newcommand*\lineheight[1]{\fontsize{\fsize}{#1\fsize}\selectfont}%
  \ifx\svgwidth\undefined%
    \setlength{\unitlength}{281.50830847bp}%
    \ifx\svgscale\undefined%
      \relax%
    \else%
      \setlength{\unitlength}{\unitlength * \real{\svgscale}}%
    \fi%
  \else%
    \setlength{\unitlength}{\svgwidth}%
  \fi%
  \global\let\svgwidth\undefined%
  \global\let\svgscale\undefined%
  \makeatother%
  \begin{picture}(1,1.09051665)%
    \lineheight{1}%
    \setlength\tabcolsep{0pt}%
    \put(0,0){\includegraphics[width=\unitlength,page=1]{numerical_integration_3.pdf}}%
    \put(0.27898494,0.12734001){\makebox(0,0)[lt]{\lineheight{1.25}\smash{\begin{tabular}[t]{l}Boundary segment\end{tabular}}}}%
    \put(0,0){\includegraphics[width=\unitlength,page=2]{numerical_integration_3.pdf}}%
    \put(0.27956258,0.01591898){\makebox(0,0)[lt]{\lineheight{1.25}\smash{\begin{tabular}[t]{l}Mapped segment\end{tabular}}}}%
    \put(0,0){\includegraphics[width=\unitlength,page=3]{numerical_integration_3.pdf}}%
  \end{picture}%
\endgroup%

%% file: section/numerical_examples.tex
\section{Numerical examples} \label{section:numerical_examples}

In this section, we present several numerical examples to validate the proposed discretization approach. 
Table~\ref{table:validation_building_blocks} illustrates how the algorithmic building blocks are validated 
through these examples. Section~\ref{subsection:patch_test} presents patch tests that 
verify the consistency of the embedded mesh coupling between the boundary layer and the background meshes. 
In Section~\ref{subsection:mesh_locking}, the applicability of the mortar-based method, described in 
Section~\ref{subsection:em_coupling_discretization}, is demonstrated using the numerical example 
from~\cite[Section~7]{Sanders2012}. Section~\ref{subsection:spatial_convergence} investigates 
the accuracy of the embedded mesh discretization in a contact problem through a spatial convergence study. 
Section~\ref{subsection:hertzian-contact-problem} then presents a two-dimensional Hertzian contact problem. 
Finally, Section~\ref{subsection:tori_contact} 
demonstrates the approach in a three-dimensional setting, showcasing dynamic impact between two tori. 
For the numerical examples related to contact, the mortar finite element approach based on dual shape 
functions for NURBS discretizations presented in~\cite{Seitz2016} was used.

The isogeometric B-reps are defined using the open-source library NURBS-Python~\cite{bingol2019geomdl}. 
Offset generation and preprocessing for all numerical examples are carried out with the open-source finite element 
pre-processor BeamMe~\cite{BeamMe}. The simulations are performed using the open-source multi-physics 
research code 4C~\cite{4C}.

\begin{table}[h]
	\centering
	\caption{Validation of the algorithmic building blocks.}
	\begin{tabular}{p{3.2cm} p{5cm}}
	\toprule
	\textbf{Embedded mesh} &
	\ref{subsection:patch_test} Patch test\\
	& \ref{subsection:mesh_locking} Mesh locking \\
	\midrule
	\textbf{Embedded mesh with contact} &
	\ref{subsection:spatial_convergence} Convergence study\\
	& \ref{subsection:hertzian-contact-problem} Hertzian contact problem\\
	& \ref{subsection:tori_contact} Two tori impact \\
	\bottomrule
	\end{tabular}
	\label{table:validation_building_blocks}
\end{table}

	





\subsection{Patch test} \label{subsection:patch_test}

\begin{figure}[H]
	\centering
	
	\def\svgwidth{0.3\linewidth}
	    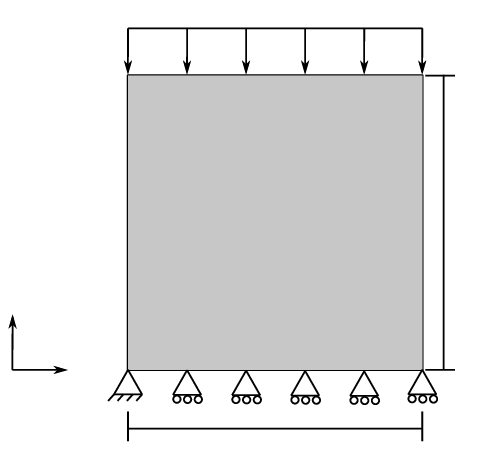

	\caption{Problem setup of the patch test.}
	\label{fig:patch_bcs}
\end{figure}

\begin{figure}[H]
	\centering
	
	\begin{subfigure}{0.2\textwidth}
		\includegraphics[width=\textwidth]{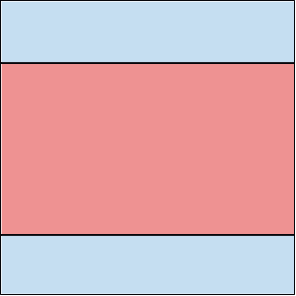}
		\caption{}
		\label{subfig:patch_bcs_case_1}
	\end{subfigure}
	\hspace{1cm}
	\begin{subfigure}{0.2\textwidth}
		\includegraphics[width=\textwidth]{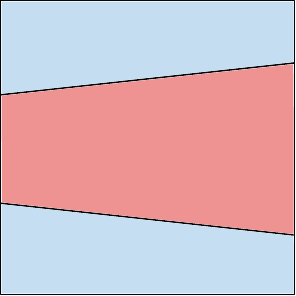}
		\caption{}
		\label{subfig:patch_bcs_case_2}
	\end{subfigure}
	\hspace{1cm}
	\begin{subfigure}{0.2\textwidth}
		\includegraphics[width=\textwidth]{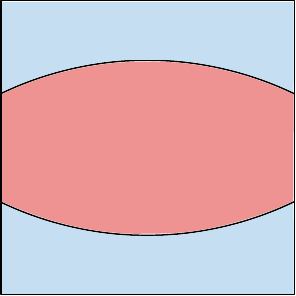}
		\caption{}
		\label{subfig:patch_bcs_case_3}
	\end{subfigure}
	
	\caption{Mesh configurations used for the patch test: 
	(a) straight interfaces, (b) inclined interfaces and (c) curved interfaces.}
	\label{fig:patch_configs}
\end{figure}

Patch tests are widely used to assess 
the consistency of finite element discretizations. 
A patch test is considered successful if an arbitrary patch is able to 
reproduce a state of constant stress~\cite{Irons1972, Elabbasi2001}. 
In this work, we consider the patch test shown in Figure~\ref{fig:patch_bcs}, which 
consists of an elastic block~($\Youngsmodulus{}~=~1,~\Poissonratio{}~=~0.3$) subjected to a uniform pressure load $p = -0.01$ 
in $\yaxis$-direction.
The continuum is described using nonlinear kinematics. 
The penalty parameter is set to $\epsilon = 1000$, and the width and height of the block are both~$a = 3$. 
Although the problem is presented in a two-dimensional setting, it can be straightforwardly extended to three dimensions, 
as repeatedly demonstrated by the authors for similar contact and mesh coupling scenarios~\cite{Popp2012, Popp2013}. 
To assess the consistency of the proposed methodology, the following representative mesh configurations are analyzed (see Figure~\ref{fig:patch_configs}):

\begin{enumerate}[label=(\alph*)]
	\item boundary layer with straight interfaces,
	\item boundary layer with inclined interfaces,
	\item boundary layer with curved interfaces.
\end{enumerate}

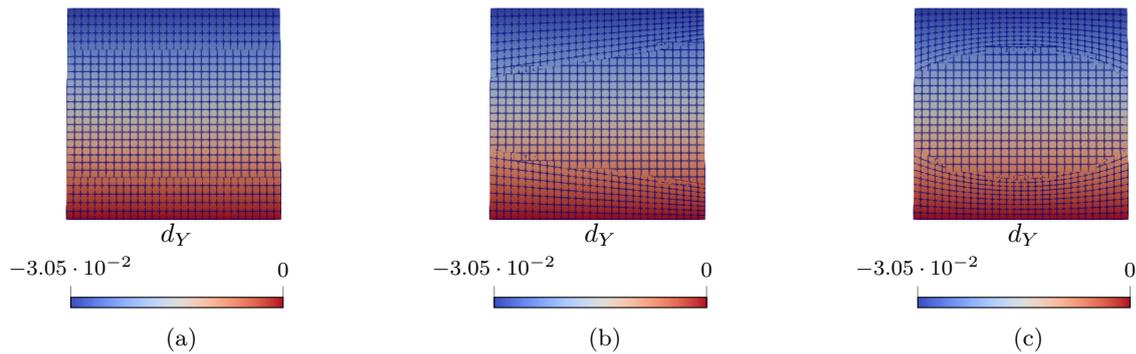
\begin{figure}[H]
	\centering
	\begin{subfigure}{0.3\textwidth}
		\import{fig/numerical_examples/patch_tests/}{straight_fine_2_nonlin_kinem_displacement_Y.tex}
		\caption{}
	\end{subfigure}
	\hfill
	\begin{subfigure}{0.3\textwidth}
		\import{fig/numerical_examples/patch_tests/}{inclined_fine_2_nonlin_kinem_displacement_Y.tex}
		\caption{}
	\end{subfigure}
	\hfill
	\begin{subfigure}{0.3\textwidth}
		\import{fig/numerical_examples/patch_tests/}{curved_fine_2_nonlin_kinem_displacement_Y.tex}
		\caption{}
	\end{subfigure}
	\caption{Displacement field $u_Y$ of the patch tests: 
	(a) straight interfaces, (b) inclined interfaces and (c) curved  
	interfaces.}
	\label{fig:patch_test_displacement}
\end{figure}

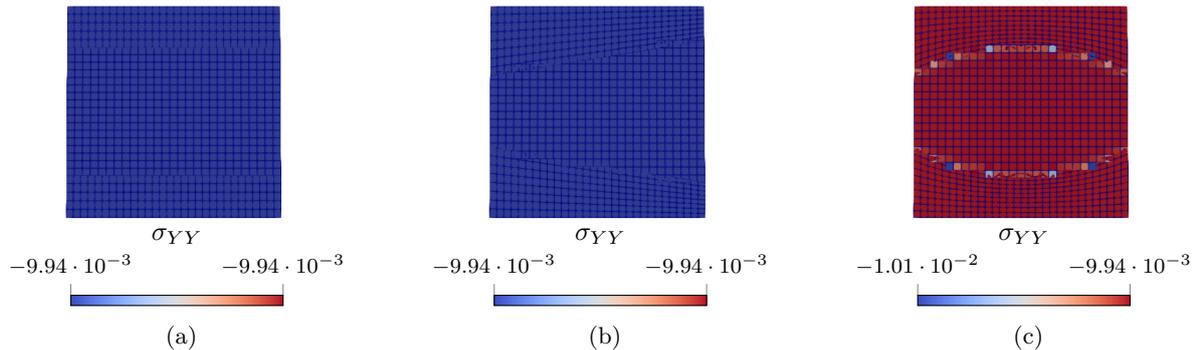
\begin{figure}[h]
	\centering
	\begin{subfigure}{0.3\textwidth}
		\import{fig/numerical_examples/patch_tests/}{straight_fine_2_nonlin_kinem_stress_scale_element_cauchy_stresses_xyz_YY.tex}
		\caption{}
		\label{subfig:patch_test_stress_case_1}
	\end{subfigure}
	\hfill
	\begin{subfigure}{0.3\textwidth}
		\import{fig/numerical_examples/patch_tests/}{inclined_fine_2_nonlin_kinem_stress_scale_element_cauchy_stresses_xyz_YY.tex}
		\caption{}
		\label{subfig:patch_test_stress_case_2}
	\end{subfigure}
	\hfill
	\begin{subfigure}{0.3\textwidth}
		\import{fig/numerical_examples/patch_tests/}{curved_fine_2_nonlin_kinem_element_cauchy_stresses_xyz_YY.tex}
		\caption{}
		\label{subfig:patch_test_stress_case_3}
	\end{subfigure}
		
	\caption{\responsetoview{Cauchy stress $\sigma_{YY}$ of the patch tests: (a)
	straight interfaces, (b) inclined interfaces and (c) curved 
	interfaces. The color scale used in (c) differs from that used in (a) and (b). }}
	\label{fig:patch_test_stress}
\end{figure}

For all three configurations, a linear displacement field $u_Y$ is obtained, as illustrated in Figure~\ref{fig:patch_test_displacement}. 
Figure~\ref{fig:patch_test_stress} 
shows the results of the Cauchy stress field $\sigma_{YY}$ for the three configurations. 
The exact fulfillment of the patch test is observed for the first and second configurations 
(Figures~\ref{subfig:patch_test_stress_case_1} and~\ref{subfig:patch_test_stress_case_2}, respectively), 
where a constant stress field is recovered up to machine precision. 
These results confirm the consistency of the proposed approach for coupling isogeometric and Cartesian meshes. 
The third configuration exhibits deviations of less than $1.6\%$ in the element-wise stress field, 
which are localized to background elements intersected by the curved interfaces, as shown in 
Figure~\ref{subfig:patch_test_stress_case_3}. 
This behavior is expected, since the numerical integration described in 
Subsection~\ref{subsection:numerical_integration_cutelements} produces 
geometrically inexact integrations cells of the actual material domain 
for curved cutting interfaces~\cite{Popp2013, Sudhakar2014, Steinbrecher2020, Steinbrecher2025}. 

\subsection{Mesh locking} \label{subsection:mesh_locking}

The mortar method in embedded mesh applications is known to 
potentially exhibit mesh-locking effects under specific conditions. 
The most critical configuration occurs when a fine mesh 
overlays a coarse mesh and the stiffness of the fine mesh is 
significantly higher than that of the coarse mesh~\cite{Sanders2012}. 
In the present application, however, this issue is not expected, since the 
discretized bodies are assumed to possess homogeneous material properties. 
Consequently, the overlapping meshes share identical material parameters. 
Nevertheless, the potential occurrence of mesh locking is investigated 
in the following example.

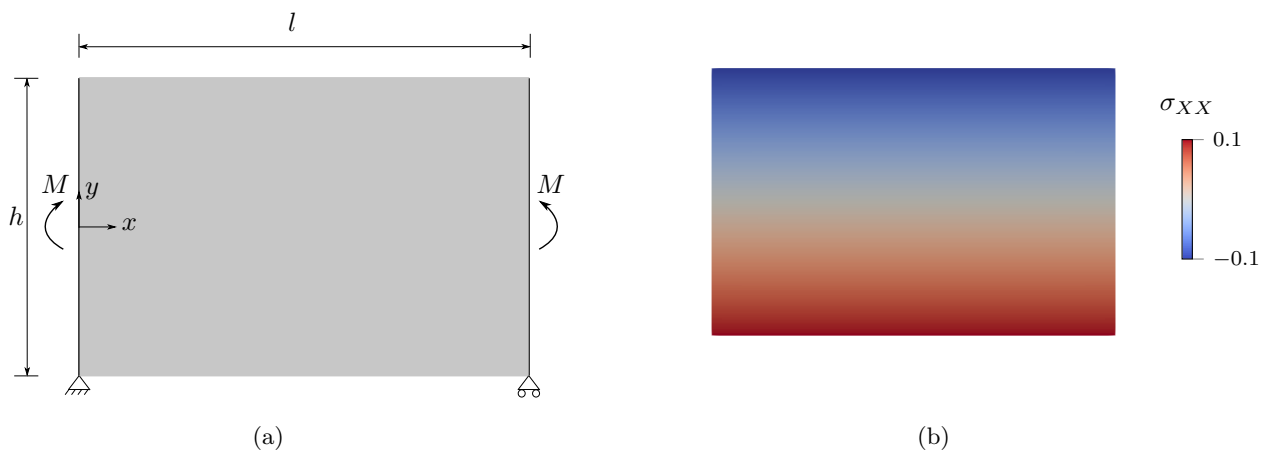
\begin{figure}[h]
	\centering
	\begin{subfigure}[t]{0.45\textwidth}
		\begin{tikzpicture}[
			description/.style={rectangle, text width=5cm}
		]
			\node (bcs) at (0, 0) 
				{\def\svgwidth{\linewidth}
				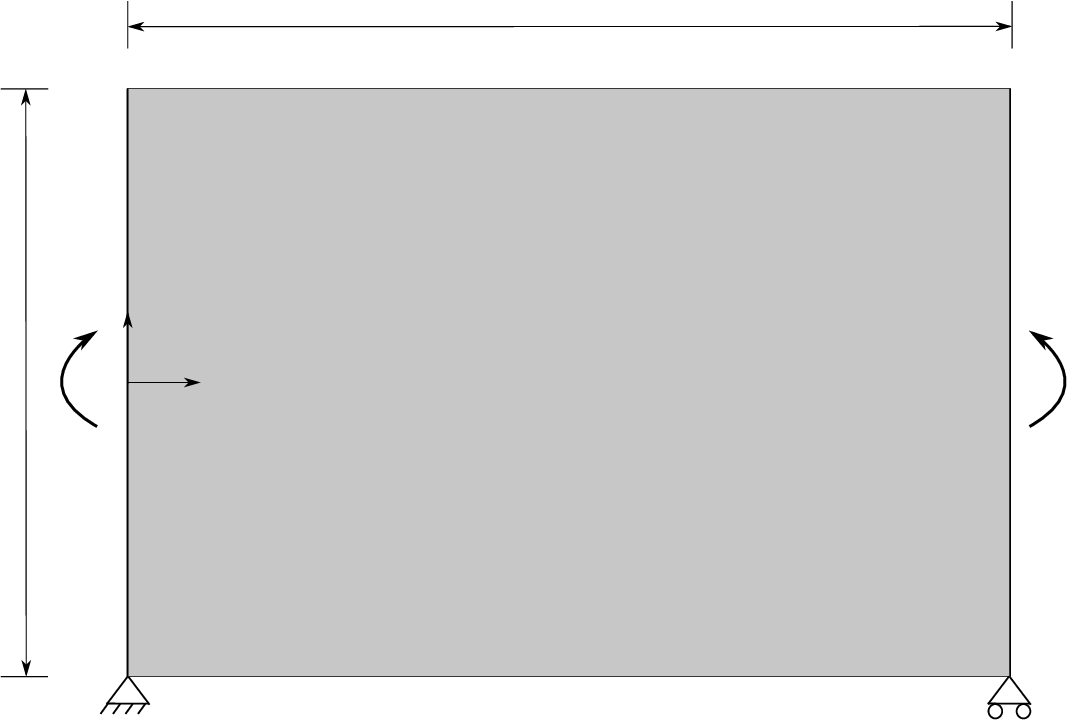};
			
		\end{tikzpicture}
		\caption{}
		\label{subfig:bending_beam_bcs}
	\end{subfigure}
	\hfill
	\begin{subfigure}[t]{0.45\textwidth}
		\begin{tikzpicture}[
			description/.style={rectangle, text width=5cm}
		]
		\node (bcs) at (0, 0) 
			{\import{fig/numerical_examples/bending_beam/}{reference_linear_kinem_element_cauchy_stresses_xyz_ZZ.tex}};
			
		\end{tikzpicture}
		\caption{}
		\label{subfig:bending_beam_refsol}
	\end{subfigure}
	
	\caption{(a) Boundary conditions for bending beam and (b) its analytical reference solution. }
	\label{fig:bending_beam}
\end{figure}

We consider a problem setup similar to that presented in 
\cite[Section~7]{Sanders2012}; see Figure~\ref{subfig:bending_beam_bcs}. 
The configuration consists of a two-dimensional, linear-elastic beam 
with dimensions $l = 1.5$ and $h = 1.0$. 
A bending moment $M$ is applied at both ends of the beam, induced by 
a linearly varying distribution of normal forces 
$p(\yaxis{}) = -0.2\,\yaxis{}$. Since the loading results in pure bending, 
the Cauchy stress component $\sigma_{XX}$ is expected to vary linearly 
in the $y$-direction and to be independent of the beam stiffness. 
Figure~\ref{subfig:bending_beam_refsol} shows the analytical reference 
solution obtained using a highly refined mesh.

\begin{figure}[h]
	\centering
	\begin{tikzpicture}[
		description/.style={rectangle, text width=5cm}
	]
		\node (bcs) at (0, 0) 
			{\def\svgwidth{0.45\textwidth}
			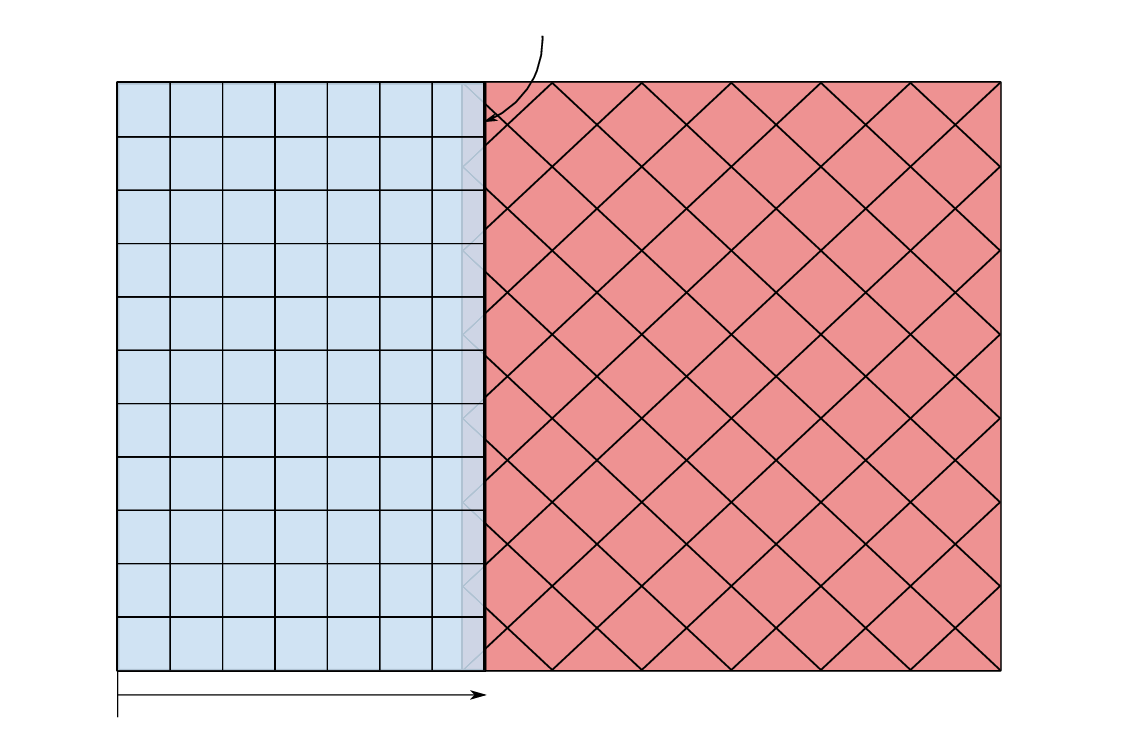};
		
	\end{tikzpicture}
	
	\caption{Discretization of the bending beam with overlapping NURBS (blue) and Lagrange (red) meshes.}
	\label{fig:bending_beam_emconfig}
\end{figure}

For the mesh-locking study, the beam is discretized using two overlapping 
meshes, denoted as~\boundarylayerdomainnumexample{} and \bulkdomainnumexample{}. 
The interface $\Gamma^{*}$ of \boundarylayerdomainnumexample{} is located 
at $x = 0.625$ (Figure~\ref{fig:bending_beam_emconfig}). The foreground 
mesh~\boundarylayerdomainnumexample{} is discretized using NURBS elements of size 
\elesizeboundarylayer{}, while the background mesh \bulkdomainnumexample{} 
consists of linear triangular and quadrilateral Lagrangian elements of size 
\elesizebulkdomain{}. As suggested in~\cite[Section~7]{Sanders2012}, the 
background mesh is arranged in a cross-hatched pattern to represent a 
worst-case configuration, in which the interface $\Gamma^{*}$ diagonally 
intersects several background elements.
Both meshes share identical material parameters, 
$\Youngsmodulus{}_{1} = \Youngsmodulus{}_{2} = 50$ and 
$\Poissonratio{} = 0.0$. The coupling constraint, enforced by 
Lagrange multipliers, is applied along the interface $\Gamma^{*}$, 
and the penalty parameter is set to $\epsilon = 10^{6}$.

Two configurations are considered:
\begin{enumerate}
	\item the foreground and background mesh elements are of comparable size, with $\elesizebulkdomain{} / \elesizeboundarylayer{} \approx  1.2$,
	\item the foreground mesh elements are finer than the background mesh elements, with $\elesizebulkdomain{} / \elesizeboundarylayer{} \approx  4.2$.
\end{enumerate}

\begin{figure}[]
	\centering
	\begin{subfigure}[t]{0.45\textwidth}
		\import{fig/numerical_examples/bending_beam/}{case_2_linear_kinem_element_cauchy_stresses_xyz_ZZ.tex}
		\caption{}
		\label{subfig:bending_beam_config1}
	\end{subfigure}
	\hfill
	\begin{subfigure}[t]{0.45\textwidth}
		\begin{tikzpicture}[
			description/.style={rectangle, text width=5cm}
		]
		{\includegraphics[width=0.9\textwidth]{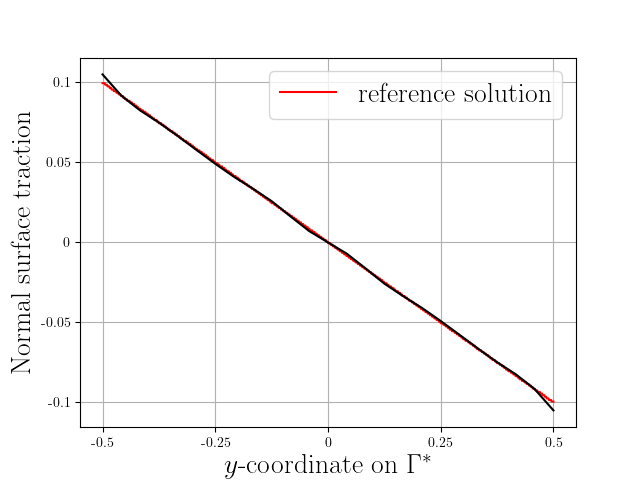}};
			
		\end{tikzpicture}
		\caption{}
		\label{subfig:bending_beam_config1_plot}
	\end{subfigure}
	
	\caption{Results for configuration 1: (a) Cauchy stresses $\sigma_{XX}$ on the beam. (b) Normal surface tractions on the interface $\Gamma^{*}$.}
	\label{fig:bending_beam_config1}
\end{figure}

\begin{figure}[]
	\centering
	\begin{subfigure}[]{0.45\textwidth}
		\import{fig/numerical_examples/bending_beam/}{case_1_linear_kinem_element_cauchy_stresses_xyz_ZZ.tex}
		\caption{}
		\label{subfig:bending_beam_config2}
	\end{subfigure}
	\hfill
	\begin{subfigure}[]{0.45\textwidth}
		\includegraphics[width=\textwidth]{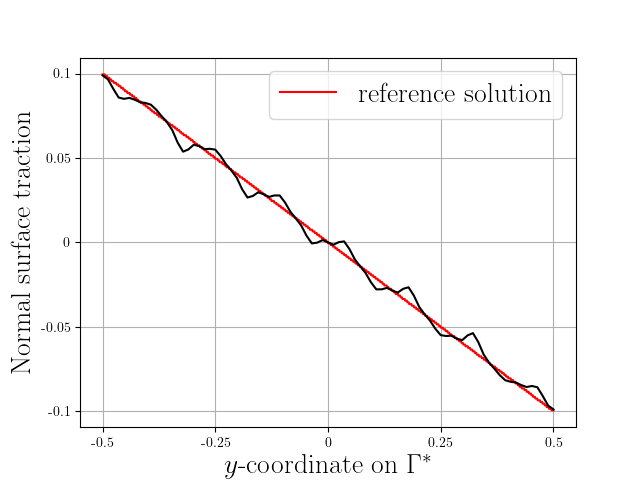}
		\caption{}
		\label{subfig:bending_beam_config2_plot}
	\end{subfigure}
	
	\caption{Results for configuration 2: (a) Cauchy stresses $\sigma_{XX}$ on the beam. (b) Normal surface tractions on the interface $\Gamma^{*}$.}
	\label{fig:bending_beam_config2}
\end{figure}

These configurations are motivated by contact applications, in which the 
isogeometric boundary layer mesh might be refined to accurately 
resolve contact interactions. 
For the first configuration, the computed stress $\sigma_{XX}$ and the
normal interface tractions exhibit 
very good agreement with the reference solution (Figure~\ref{fig:bending_beam_config1}).  
In the second configuration, oscillations in the normal interface tractions are observed 
near the intersections of the interface $\Gamma^{*}$ with the edges of the 
cut elements (see Figure~\ref{fig:bending_beam_config2}). These results are consistent with those reported 
in~\cite[Section~7]{Sanders2012}. The observed oscillations 
remain small and localized, showing that the adopted embedded mesh approach 
is suitable for the intended setting of overlapping meshes with identical material properties.

\begin{figure}[]
	\centering
	\begin{subfigure}[t]{0.45\textwidth}
		\import{fig/numerical_examples/bending_beam/}{case_3_linear_kinem_element_cauchy_stresses_xyz_ZZ.tex}
		\caption{}
		\label{subfig:bending_beam_config3}
	\end{subfigure}
	\hfill
	\begin{subfigure}[t]{0.45\textwidth}
		\begin{tikzpicture}[
			description/.style={rectangle, text width=5cm}
		]
		{\includegraphics[width=\textwidth]{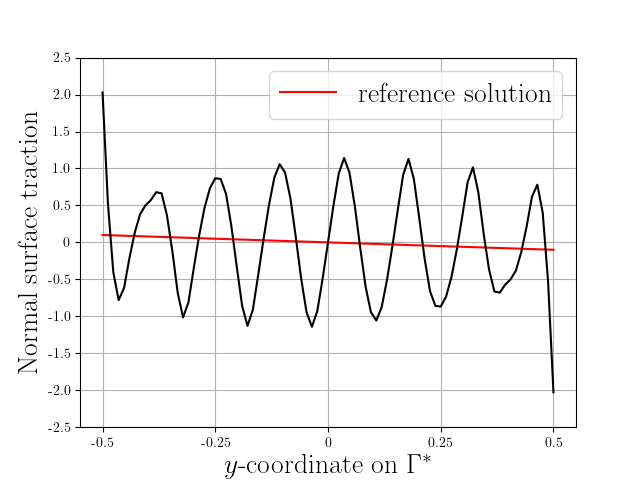}};
			
		\end{tikzpicture}
		\caption{}
		\label{subfig:bending_beam_config3_plot}
	\end{subfigure}
	
	\caption{Results for configuration 3: (a) Cauchy stresses $\sigma_{XX}$ on the beam. (b) Normal surface tractions on the interface $\Gamma^{*}$.}
	\label{fig:bending_beam_config3}
\end{figure}

Although not directly relevant to our application, a 
third configuration is included for completeness. This configuration uses 
the same mesh ratio as the second configuration, with the difference 
that the NURBS mesh presents a significantly higher stiffness 
than the background mesh, namely~$\Youngsmodulus_{1} = 50000$ and 
$\Youngsmodulus_{2} = 50$. As expected, mesh locking is observed (Figure~\ref{fig:bending_beam_config3}), 
characterized by significant stress fluctuations in the elements of 
the isogeometric boundary layer near the interface~$\Gamma^{*}$. 

\subsection{Convergence study in a contact problem}\label{subsection:spatial_convergence}

\begin{figure}[h]
	\centering
	
	\begin{subfigure}[t]{0.3\textwidth}
		\begin{tikzpicture}[scale=0.93]
			\node (bcs) at (0,0) {
				\def\svgwidth{\textwidth}
				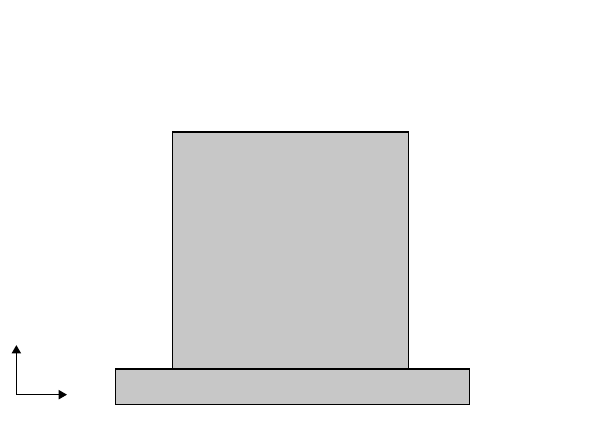};
	
			\draw[domain=-1:1, smooth, variable=\x, blue, thin, shift={(-0.03,0.7)}] 
			  plot ({\x}, {\x*\x*\x*\x});
		
		\end{tikzpicture}
		\caption{}
		\label{subfig:single_block_bcs}
	\end{subfigure}
	\hfill
	\begin{subfigure}[t]{0.3\textwidth}
		\begin{tikzpicture}[scale=1.398]
			\node[right=0.1cm of bcs] {
				\def\svgwidth{\textwidth}
				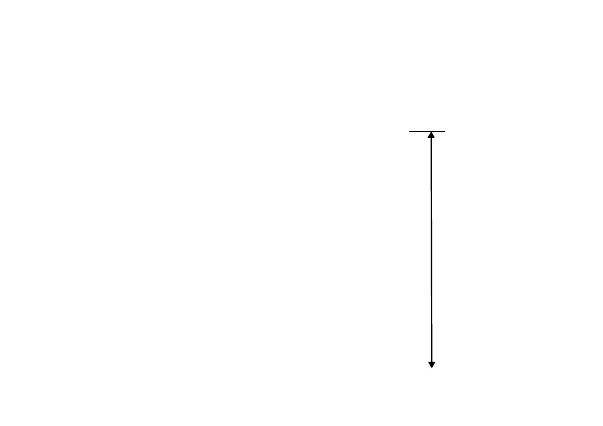};
		
		\end{tikzpicture}
		\caption{}
		\label{subfig:single_block_subdomains}
	\end{subfigure}
	\hfill
	\begin{subfigure}[t]{0.3\textwidth}
		\begin{tikzpicture}[scale=1.398]
			\node[right=0.1cm of bcs] {
				\def\svgwidth{\textwidth}
				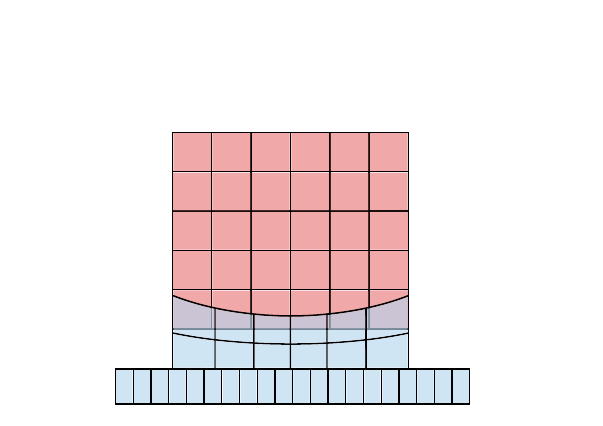};
		
		\end{tikzpicture}
		\caption{}
		\label{subfig:single_block_mesh}
	\end{subfigure}

	\caption{Contact problem between a block and a rigid surface. (a) Boundary conditions. (b) Subdivision of block's domain. (c) Discretization of the domains with NURBS (blue) and Lagrange elements~(red).}
	\label{fig:single_block}
\end{figure}

In this example, we investigate the spatial convergence of the proposed discretization approach 
applied to a contact problem. Figure~\ref{subfig:single_block_bcs} shows the boundary conditions of a 
simple two-dimensional contact problem between a linear-elastic block ($\Youngsmodulus{}=1$, $\Poissonratio{}=0$) and a rigid surface. 
The block has a width and height of $w=h=3$. A distributed force $f_y = 0.1x^4$ is applied to press the block 
against the rigid surface. Furthermore, the block is constrained to eliminate rigid body modes. 

The block’s domain is subdivided as shown in Figure~\ref{subfig:single_block_subdomains}. 
Although the proposed discretization method generates \interfacestar{}{} through an offset operation, 
in this example it is prescribed as an arbitrary curved interface. This choice is made to verify the 
method in a more complex cut scenario. The NURBS patch representing the boundary layer is then 
constructed to conform to the contact interface and the curved boundary layer. 
The rigid surface is discretized using a second-order NURBS patch, while the background domain 
is discretized with a Cartesian mesh. For the contact boundary conditions, the slave side 
corresponds to the base of the block, whereas the master side is defined on the rigid surface. 
The penalty parameter associated with the embedded mesh formulation is set to $\epsilon = 10^4$. 

Two mesh configurations are considered: 
\begin{enumerate}
	\item The boundary layer and background domains are discretized using second-order NURBS (\nurbssecondordertwodim{}) and first-order Lagrangian elements (\lagrangefirstordertwodim{}), respectively 
	\item The boundary layer and background domains are discretized using second-order NURBS (\nurbssecondordertwodim{}) and second-order serendipity Lagrangian elements (\lagrangeseconddordertwodim{}), respectively. 
\end{enumerate}

The reference solution is obtained from a numerical simulation using a fine mesh 
size ($h_{\text{ref}} = 2.344 \times 10^{-2}$), employing second-order NURBS for 
the boundary layer and second-order Lagrangian elements for the background mesh.

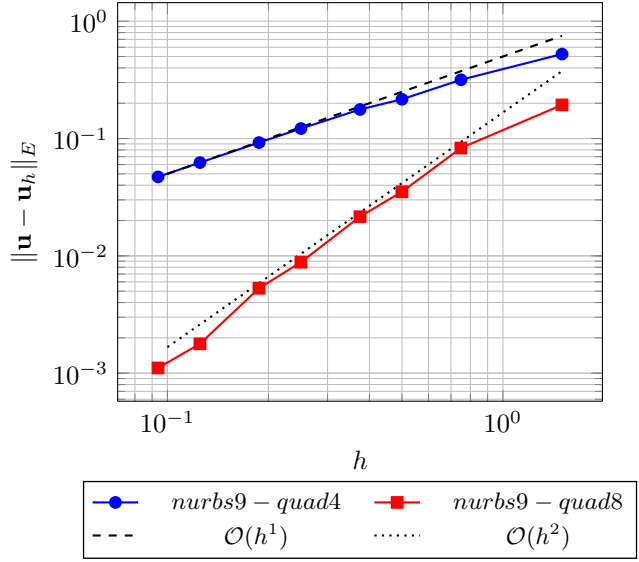
\begin{figure}[h]
	\centering
	\input{fig/numerical_examples/convergence_plot/convergence_plot_nurbs27-hex8.tex}
	\caption{Convergence rates of two mesh configurations for the contact problem with embedded mesh discretization.}
	\label{fig:single_block_convergence_plot}
\end{figure}

For this problem, the contact boundary does not evolve during deformation. As a result, 
optimal convergence rates in the energy norm are observed for both mesh configurations, 
as shown in Figure~\ref{fig:single_block_convergence_plot}. In the first configuration, 
although the overlapping mesh is discretized using second-order NURBS, the convergence 
rate is limited to $\mathcal{O}(h^1)$ due to the use of linear Lagrangian elements in the 
background mesh. In contrast, the second configuration achieves the optimal 
convergence rate of $\mathcal{O}(h^2)$ by employing second-order Lagrangian elements in the 
background mesh. 
These results demonstrate that the proposed discretization approach effectively 
exploits the properties of both discretizations: the high-order continuity of the 
isogeometric boundary layer for the contact formulation and the mesh flexibility of 
independent background meshes, without compromising optimal spatial convergence.

\subsection{Hertzian contact problem}\label{subsection:hertzian-contact-problem}

\begin{figure}
	\centering
	
	\begin{subfigure}[t]{0.45\textwidth}
		\begin{tikzpicture}[scale=1]
			\node (bcs) at (0,0) {
				\def\svgwidth{\textwidth}
				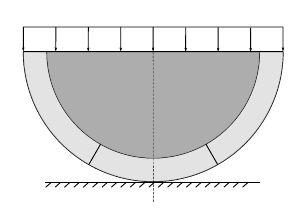};	
		\end{tikzpicture}
		\caption{}
		\label{subfig:2d_hertzian_contact_bcs}
	\end{subfigure}
	\hfill
	\begin{subfigure}[t]{0.45\textwidth}
		\includegraphics[width=\linewidth]{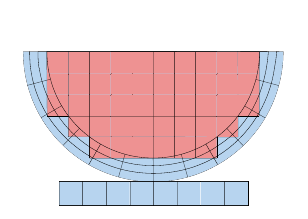}	
		\caption{}
		\label{subfig:2d_hertzian_contact_mesh}
	\end{subfigure}

	\caption{Hertzian contact problem. (a) Boundary conditions and subdivision of half-cylinder domain. (b) Discretization of the domains with NURBS (blue) and Lagrange elements (red).}
	\label{fig:2d_hertzian_contact}
\end{figure}

To further demonstrate the capabilities of the proposed workflow, 
we consider a classical benchmark in contact mechanics: the Hertzian 
contact problem. In the two-dimensional configuration studied here, 
a linear-elastic half-cylinder 
($\Youngsmodulus{}=250$, $\Poissonratio{}=0.0$) 
with radius $R=10$ is compressed against a rigid plate under a 
constant pressure $p$; see Figure~\ref{subfig:2d_hertzian_contact_bcs}. 
For this problem, an analytical expression for the normal contact traction 
\normalcontacttraction{} within the contact zone can be derived under the 
assumption of small deformations, \cf{}~\cite{Kikuchi1988}. The  
linear Hertzian solution is given by

\begin{equation}
	\normalcontacttraction{} = \frac{4 R p}{\pi b^2} \sqrt{b^2 - x^2} \quad \text{ with } \quad \widthcontactzone{}=2\sqrt{\frac{2 R^2 p (1-\Poissonratio{}^2)}{\Youngsmodulus{} \pi}},
\end{equation}
where \widthcontactzone{} denotes the width of the contact zone. 
The maximum normal contact traction~\maxnormalcontacttraction{} occurs at 
$x = 0$. 

It should be noted that the employed numerical contact formulation fully 
accounts for deformation-dependent nonlinearities. Therefore, an exact 
match between the numerical solution (which incorporate nonlinear effects) 
and the Hertzian solution (which is strictly based on 
linear kinematics) cannot be expected. 

Applying the proposed discretization approach, the half-cylinder 
domain is subdivided into a NURBS boundary layer with an offset 
thickness of $\offsetdistance{} = 0.1$ and a bulk domain, as shown in 
Figure~\ref{subfig:2d_hertzian_contact_bcs}. The boundary layer is 
discretized using second-order NURBS shape functions, while the 
background domain is discretized with a Cartesian mesh composed of 
Lagrangian elements (Figure~\ref{subfig:2d_hertzian_contact_mesh}). 
The surface of the half-cylinder is defined as the slave interface, 
whereas the rigid plate acts as the master interface. 
For the embedded mesh coupling, the penalty parameter is set to 
$\epsilon = 10^4$.

\begin{figure}[]
	\begin{subfigure}{0.23\textwidth}
		\includegraphics[width=\textwidth]{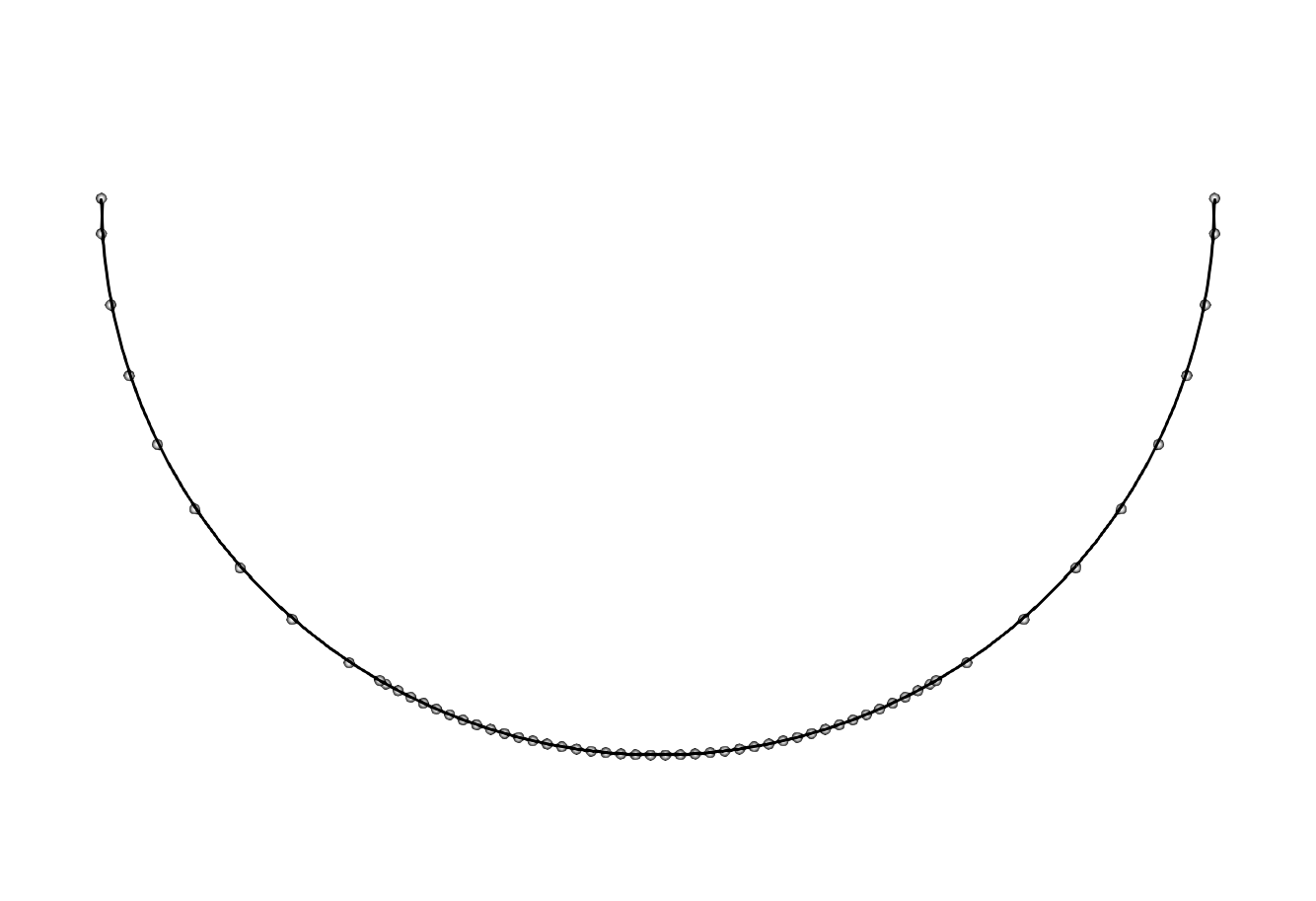}
		\caption{}
	\end{subfigure}
	\hfill
	\begin{subfigure}{0.23\textwidth}
		\includegraphics[width=\textwidth]{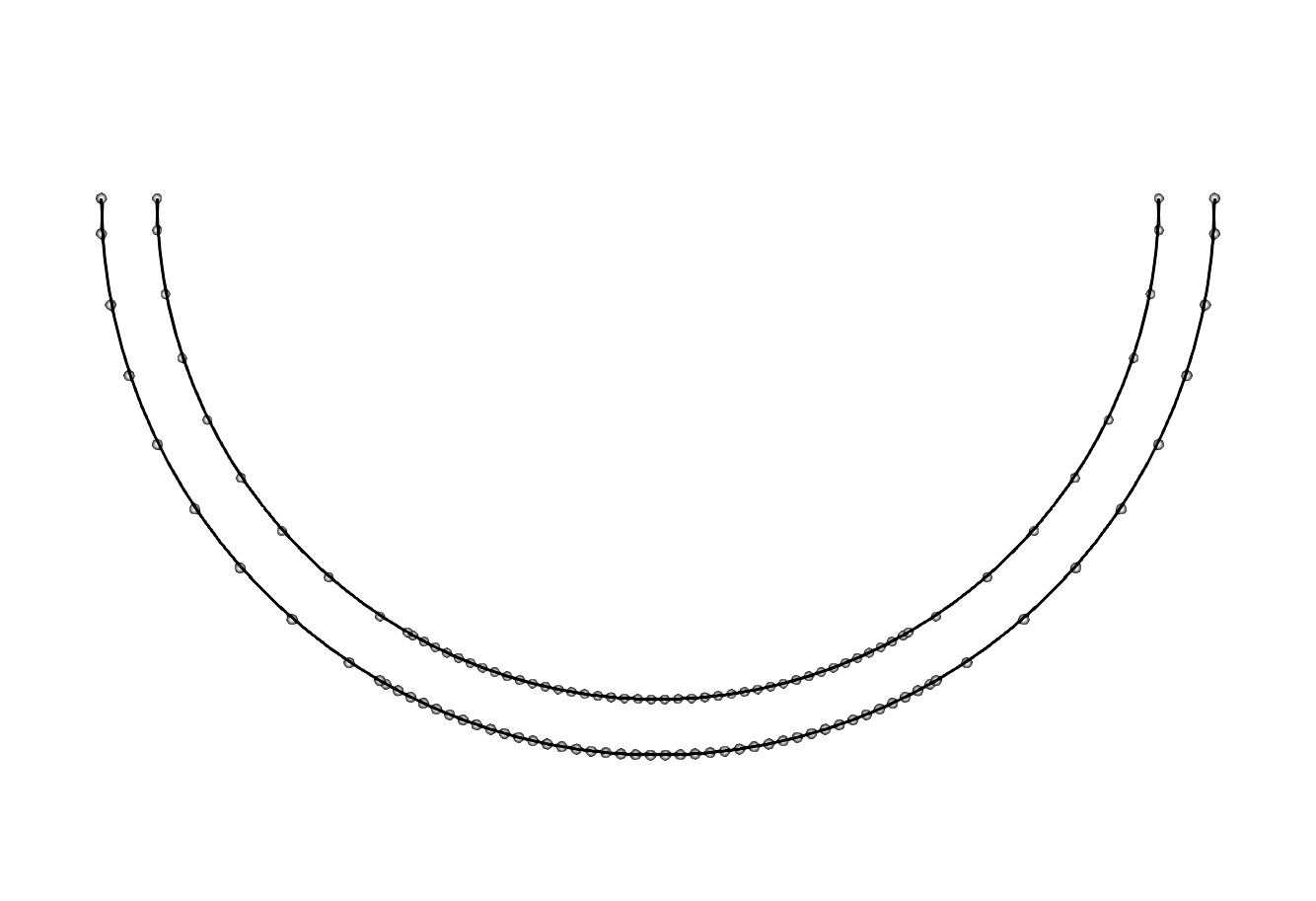}
		\caption{}
	\end{subfigure}
	\hfill
	\begin{subfigure}{0.23\textwidth}
		\includegraphics[width=\textwidth]{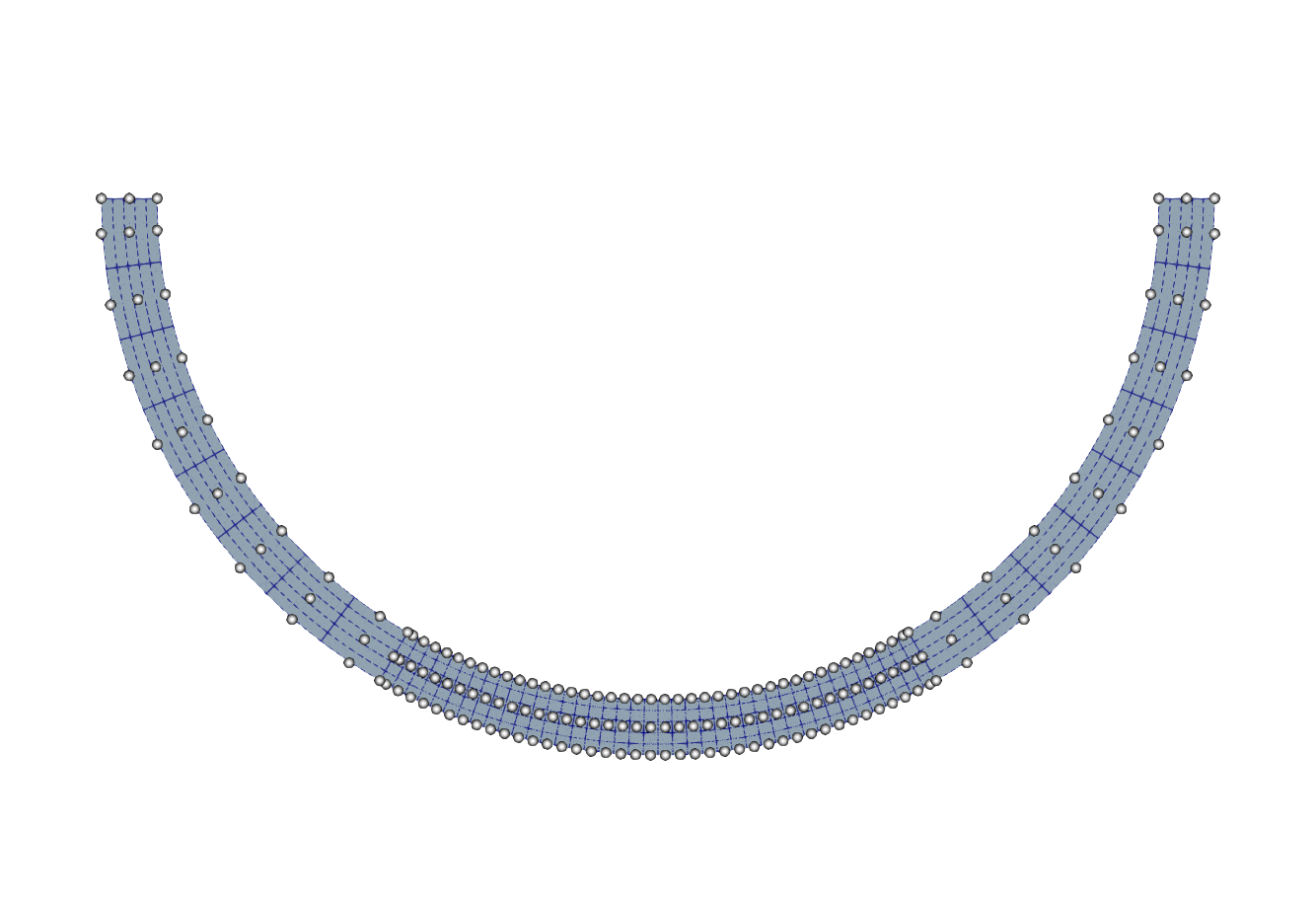}
		\caption{}
	\end{subfigure}
	\hfill
	\begin{subfigure}{0.23\textwidth}
		\includegraphics[width=\textwidth]{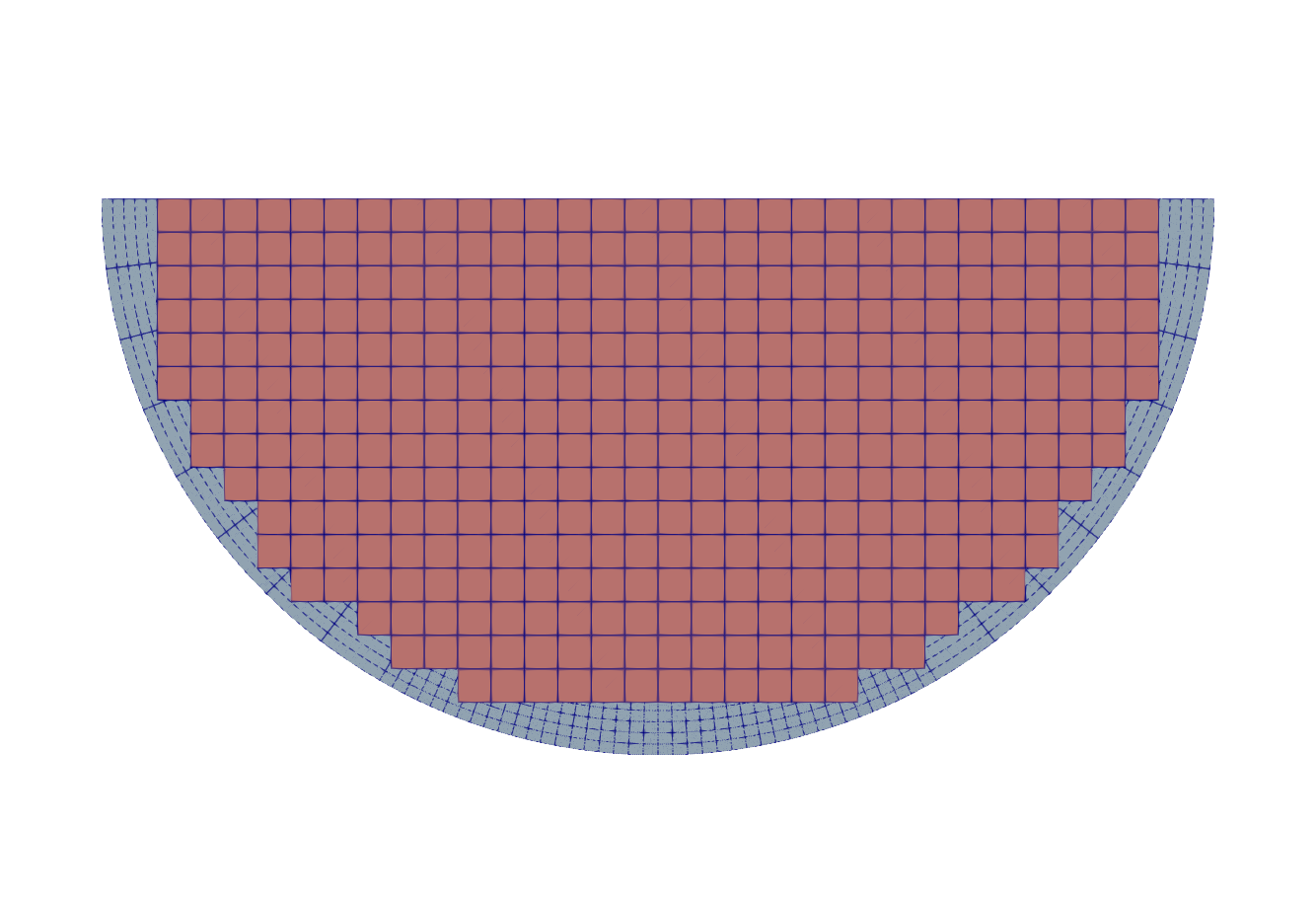}
		\caption{}
	\end{subfigure}

	\caption{Mesh generation of a Hertzian contact problem: (a) base \brep{} of half-cylinder, (b) offset of base \brep{}, (c) NURBS boundary layer mesh, (d) discretization of bulk domain with a Cartesian mesh.}
	\label{fig:hertzian_contact_mesh_gen}
\end{figure}

The mesh generation process is illustrated in 
Figure~\ref{fig:hertzian_contact_mesh_gen}. The procedure starts 
from the base \brep{}, which in this case consists of three NURBS 
curve segments representing the potential contact boundary of the 
half-cylinder. These curves are offset by the distance 
$\offsetdistance{}$, generating three NURBS surface patches. Finally, the 
remaining bulk domain of the half-cylinder is discretized using a 
Cartesian mesh.
	
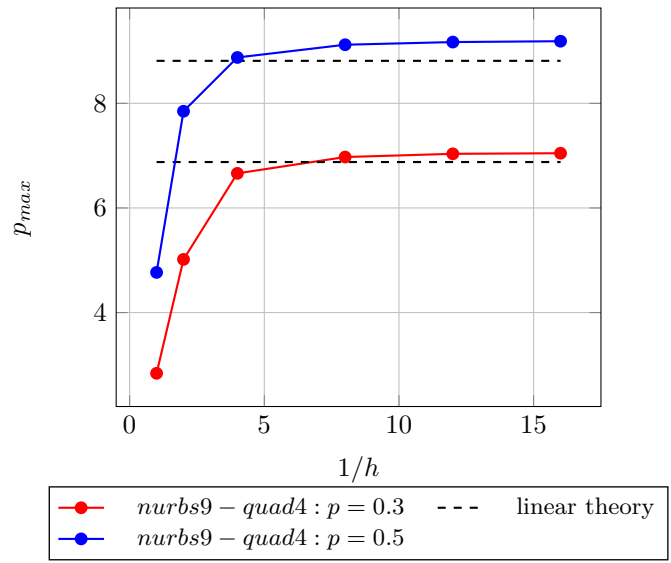
\begin{figure}
	\centering
	\input{fig/numerical_examples/2d_hertzian_contact/convergence_plot_hetzian_contact.tex}
	\caption{Spatial convergence of the Hertzian contact problem for the load cases $p = 0.3$ and $p = 0.5$.}
	\label{fig:spatial_convergence_hertzian_contact}
\end{figure}

\begin{figure}[]
	\centering
	\input{fig/numerical_examples/2d_hertzian_contact/pressure_plot_hertzian_contact_mesh_1_16.tex}
	\caption{Traction along the contact interface for $p=0.3$: comparison between the numerical solution on the finest mesh and the linear theory.}
	\label{fig:pressure_profile_hertzian_contact}
\end{figure}
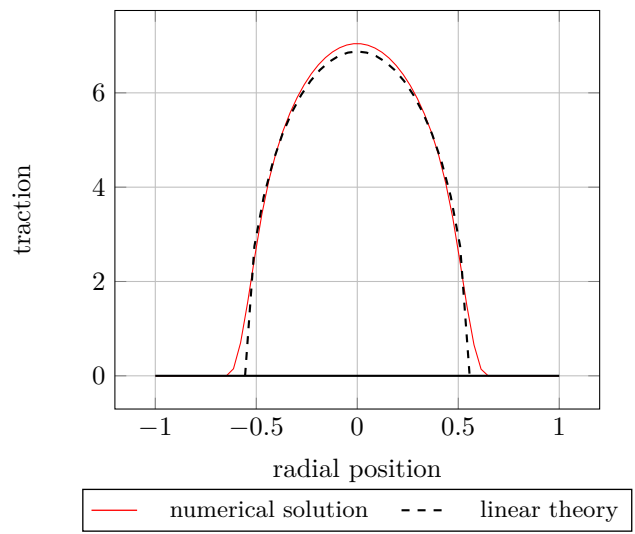

The spatial convergence of the numerical solution for 
\maxnormalcontacttraction{} is investigated using uniform mesh refinement 
for the load cases $p = 0.3$ and $p = 0.5$. The corresponding results 
are presented in Figure~\ref{fig:spatial_convergence_hertzian_contact}. 
For both load levels, the solution converges to a stable value, 
exhibiting a consistent convergence trend under mesh refinement. 
As discussed above, a deviation from the Hertzian solution, 
which is based on linear theory, is expected. This discrepancy becomes less pronounced at lower load levels.
Figure~\ref{fig:pressure_profile_hertzian_contact} depicts the contact 
traction \normalcontacttraction{} obtained on the finest mesh for~$p = 0.3$, 
plotted along the radial coordinate of the half-cylinder. The numerical 
results show very close agreement with the Hertzian solution 
in terms of the traction distribution.

\pgfmathsetmacro{\numdofs}{2}
\pgfmathtruncatemacro{\centralpatchdofs}{384*\numdofs} 
\pgfmathtruncatemacro{\sidedofs}{66*\numdofs}  
\pgfmathtruncatemacro{\cartesianmeshdofs}{166*\numdofs}  
\pgfmathtruncatemacro{\totaldisdofs}{\centralpatchdofs + \sidedofs*2 + \cartesianmeshdofs}

\begin{figure}[]
	\begin{subfigure}{0.5\textwidth}
		\centering
		\def\svgwidth{\linewidth}
		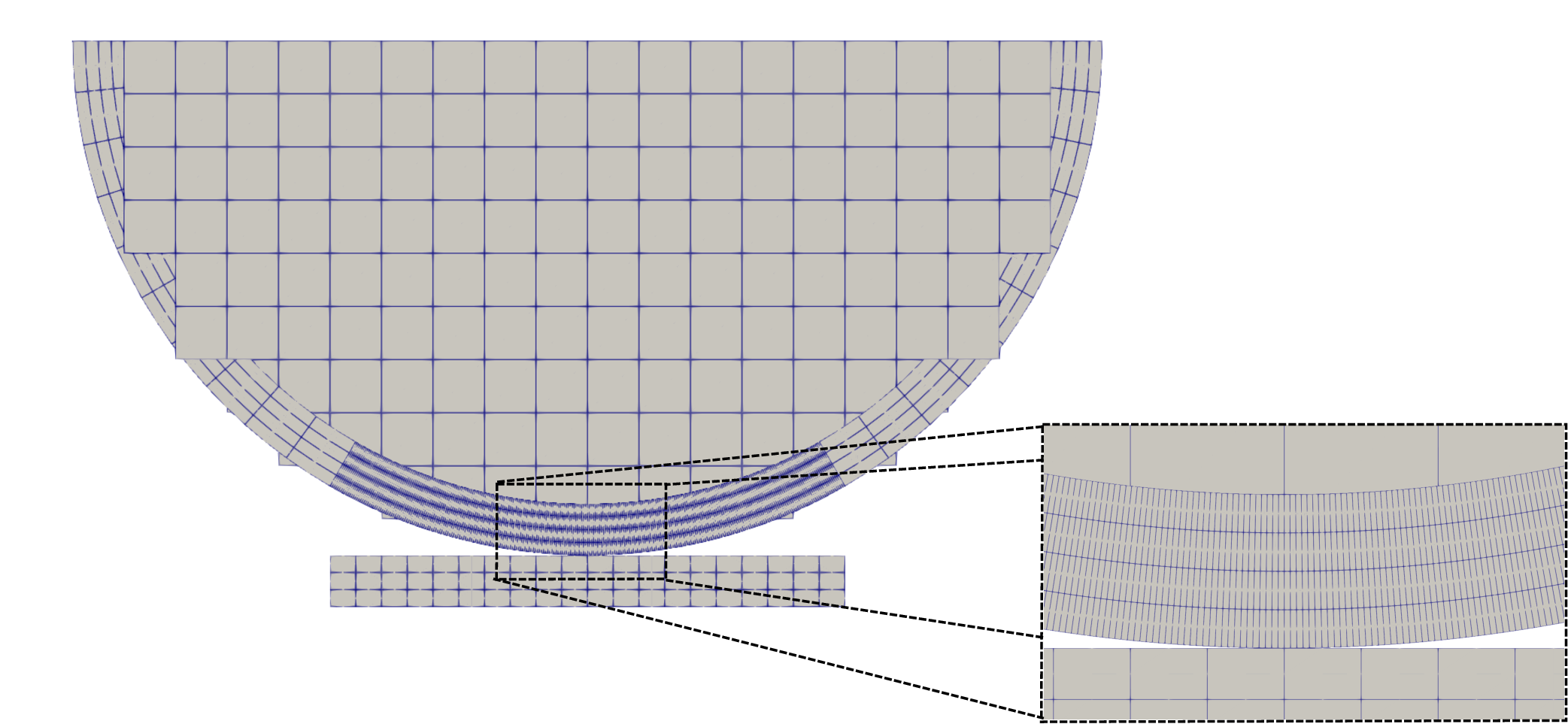
		\caption{}
		\label{subfig:flexibility_mesh_gen}
	\end{subfigure}
	\hfill
	\begin{subfigure}{0.45\textwidth}
		\centering
		\input{fig/numerical_examples/2d_hertzian_contact/pressure_plot_hertzian_contact_mesh_optimal_refinement.tex}
		\caption{}
		\label{subfig:flexibility_result}
	\end{subfigure}

	\caption{Demonstration of selective mesh refinement for the Hertzian contact problem: 
	(a) locally refined boundary layer mesh; 
	(b) resulting contact traction compared to a uniformly refined Lagrangian reference solution.}
	\label{fig:flexibility_method_hertzian_contact_mesh_gen}
\end{figure}

With this example, we further demonstrate the capabilities of the 
proposed discretization workflow, in particular its ability to 
selectively refine the boundary layer mesh independently of the 
bulk mesh while exploiting the high accuracy of NURBS discretizations 
in the contact region.  
To improve the contact stress solution, mesh refinement via knot insertion 
is applied exclusively to the central patch of the boundary layer. 
For the given boundary conditions and load $p = 0.3$, the side patches 
are not expected to come into contact with the rigid plate; refinement 
in these regions is therefore unnecessary. Since refinement of the bulk 
domain does not necessarily enhance the solution quality at the contact 
interface, the Cartesian mesh can remain relatively coarse. 
With this strategy, the final mesh consists of \centralpatchdofs{} \dofs{} 
in the central patch, \sidedofs{} \dofs{} in each side patch, and 
\cartesianmeshdofs{} \dofs{} in the Cartesian mesh, resulting in a total 
of \totaldisdofs{} \dofs{}. The resulting mesh configuration is shown in 
Figure~\ref{subfig:flexibility_mesh_gen}. 
The contact tractions obtained with the selectively refined mesh are 
compared to those computed using a uniformly refined Lagrangian mesh 
with more than~1.3 million \dofs{} (Figure~\ref{subfig:flexibility_result}), 
showing excellent agreement. These results highlight that the proposed 
discretization approach combines flexibility in mesh generation with 
high accuracy in the computed contact stresses.

\responsetoview{Finally, we investigate the sensitivity of the solution with respect to the embedded mesh penalty parameter~$\epsilon$.
We consider the Hertzian contact problem with~$E=250$,~$\nu=0$, load case~$p=0.3$, and a target element size of~$h=0.25$.
The boundary layer is discretized using quadratic NURBS elements, whereas linear Lagrange elements are employed for 
the Cartesian background mesh.
The resulting total energy~$\Pi$ is evaluated for penalty parameters ranging from $\epsilon=10^2$ to $\epsilon=10^8$.
Figure~\ref{fig:convergence_plot_penalty_energy_a} shows that the total energy approaches an asymptotic value as the penalty parameter increases.
In Figure~\ref{fig:convergence_plot_penalty_energy_b}, the error in the total energy is plotted against the penalty parameter, where a linear convergence rate is observed.
Of course, the error values shown in~\ref{fig:convergence_plot_penalty_energy_b} are problem dependent.
Nevertheless, a useful rule of thumb for practical applications is that a choice of $\epsilon \sim 100,\Youngsmodulus{}$ usually provides an adequate enforcement of the embedded mesh coupling constraint.
}

\begin{figure}[h]
	\centering
	\begin{subfigure}{0.45\textwidth}
		\input{fig/numerical_examples/2d_hertzian_contact/convergence_plot_penalty.tex}
		\caption{}
		\label{fig:convergence_plot_penalty_energy_a}
	\end{subfigure}
	\begin{subfigure}{0.45\textwidth}
		\input{fig/numerical_examples/2d_hertzian_contact/convergence_plot_penalty_error.tex}
		\caption{}
		\label{fig:convergence_plot_penalty_energy_b}
	\end{subfigure}
	\caption{\responsetoview{Sensitivity of the Hertzian contact problem with respect to the embedded mesh penalty parameter~$\epsilon$. In~(a), the total energy~$\Pi$ is plotted against the penalty parameter. In~(b), the error in the total energy,~$e_E = |\Pi - \Pi_{\text{ref}}|/\Pi_{\text{ref}}$, is plotted against the penalty parameter, where $\Pi_{\text{ref}}$ is the energy obtained with~$\epsilon=10^8$.}}
	\label{fig:convergence_plot_penalty_energy}
\end{figure}

\subsection{Two tori in contact}\label{subsection:tori_contact}

The final numerical example demonstrates the applicability of the proposed 
discretization workflow to a three-dimensional time-dependent contact problem. The configuration 
is inspired by the example presented in~\cite{Yang2008}, in which two hollow 
tori collide. In the present study, both tori are not hollow, but modeled as solid bodies.
Each torus is described by a Neo--Hookean material model 
($\Youngsmodulus{}=40$, $\Poissonratio{}=0.1$, $\Density{}=0.1$) and has 
major and minor radii of 1.5 and~0.5, respectively. The boundary layer 
thickness is $\offsetdistance{} = 0.1$. The centers of the first and second torus are 
located at $O_{T_1} = (0, 0, 0)$ and $O_{T_2} = (-0.5, 0, 0)$, respectively. The 
axis of rotation of the first torus is defined by the unit vector $\mathbf{a}_{T_1} = 1/\sqrt{2}(-1,1,0)$, 
while the axis of rotation of the second torus is the $z$-axis. This configuration is shown in 
Figure~\ref{subfig:tori_contact_initial_config}.  


For each body, the spatial discretization consists of a NURBS boundary 
layer with 49,248 \dofs{}, while the Cartesian background mesh is composed 
of linear Lagrangian elements with 22,221 \dofs{}, resulting in a total of 
142,932 \dofs{} for the two-body system. The penalty parameter is set 
to~$\epsilon = 1000$. Time integration is performed using the generalized-$\alpha$ 
method with a spectral radius of $\rho_{\infty} = 0.9$. The simulation is 
carried out up to $t = 5$ using a constant time step of $\Delta t = 0.01$. 


Both bodies move toward each other with an initial velocity of 
$v_0 = 0.162$. It is well known that cut elements with very small 
integration volumes may lead to ill-conditioned systems and, consequently, 
convergence difficulties. To prevent this, cut elements with an integration volume smaller 
than a carefully chosen threshold tolerance are removed.


Characteristic deformation stages during contact are shown in 
Figure~\ref{fig:tori_contact}. To improve visualization of the boundary 
layer and Cartesian meshes, the discretization is intersected with the 
$xy$-plane, providing a side view of the deformed configuration, \cf{} Figure~\ref{fig:tori_contact_side_view}. 
The results illustrate large deformations and significant sliding between 
the tori, benefiting from the $C^1$ continuity of the isogeometric boundary 
layer discretization. Overall, this example demonstrates the applicability of the proposed 
approach to complex three-dimensional contact problems involving large 
deformations, highlighting its potential for even more demanding 
realistic scenarios.


\newcommand{\imagerowwithlabel}[4]{%
  \begin{subfigure}[t]{.48\textwidth}
    \includegraphics[width=\linewidth]{#1}
	\subcaption{#2}
	\label{subfig:tori_contact_initial_config}
  \end{subfigure}\hfill
  \begin{subfigure}[t]{.48\textwidth}
    \includegraphics[width=\linewidth]{#3}
	\subcaption{#4}
  \end{subfigure}\par\smallskip
}

\newcommand{\imagerow}[4]{%
  \begin{subfigure}[t]{.48\textwidth}
    \includegraphics[width=\linewidth]{#1}
	\subcaption{#2}
  \end{subfigure}\hfill
  \begin{subfigure}[t]{.48\textwidth}
    \includegraphics[width=\linewidth]{#3}
	\subcaption{#4}
  \end{subfigure}\par\smallskip
}

\begin{figure} 
  \centering
  \imagerowwithlabel{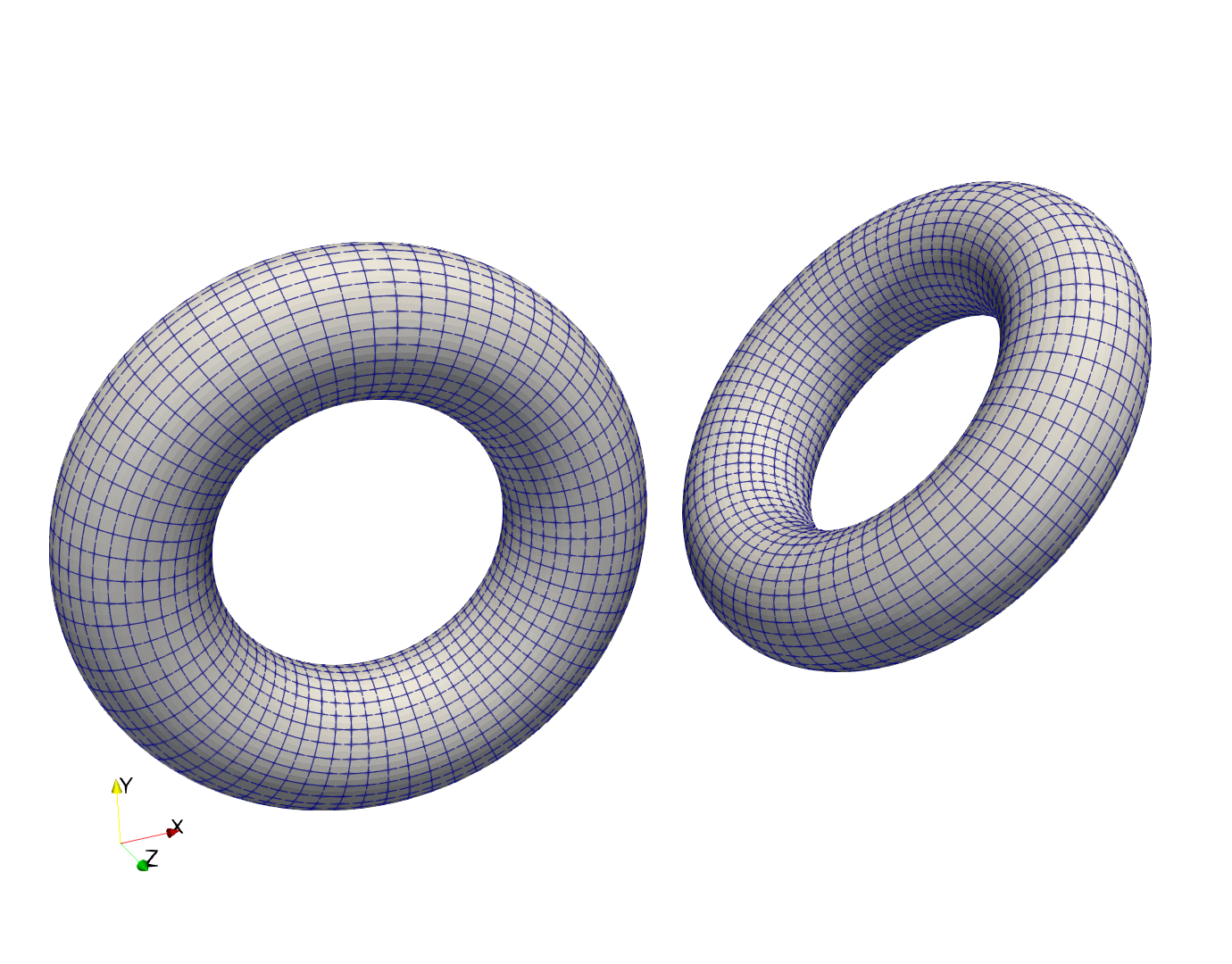}{$t=0.0$ }{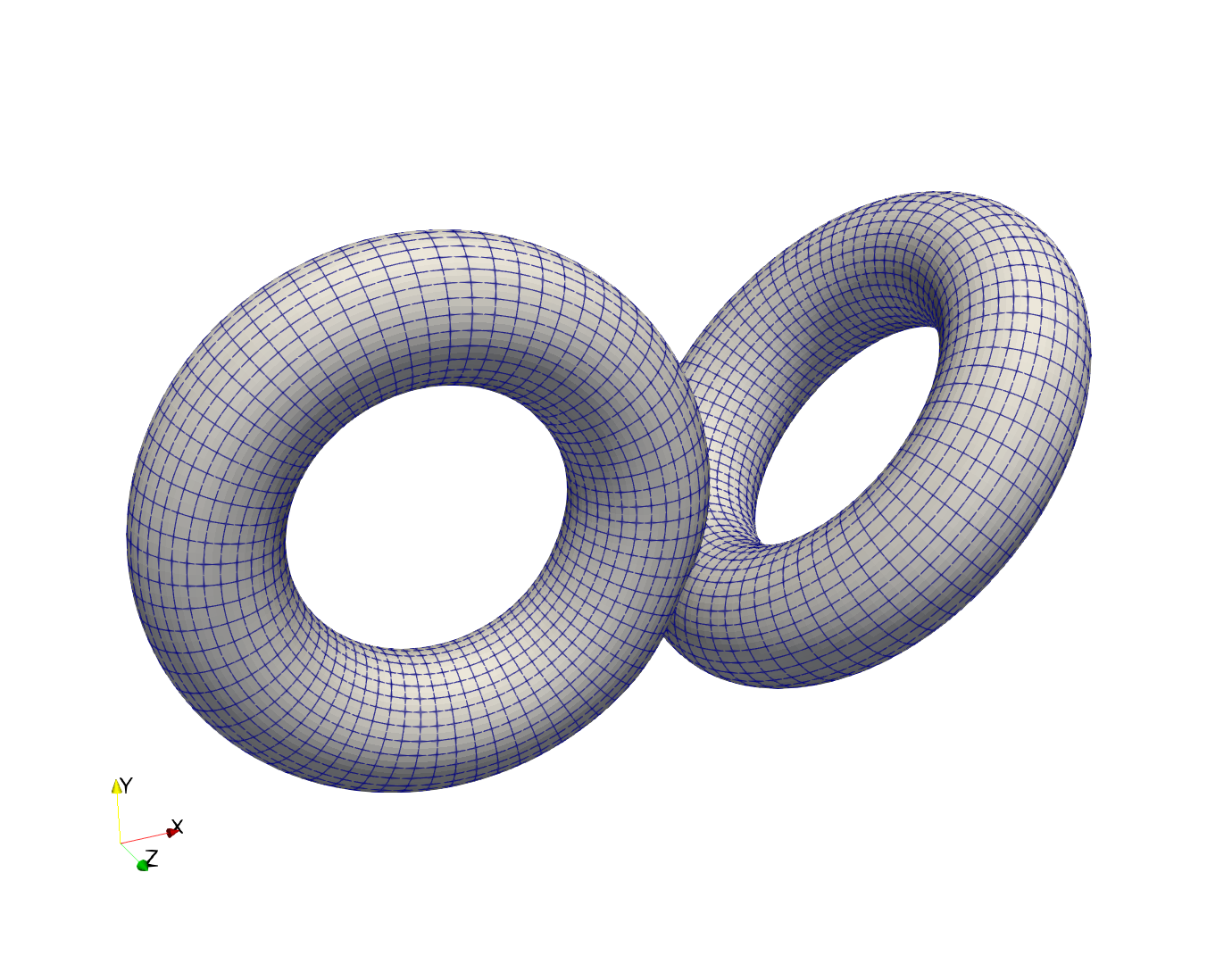}{$t=0.4$ }
  \imagerow{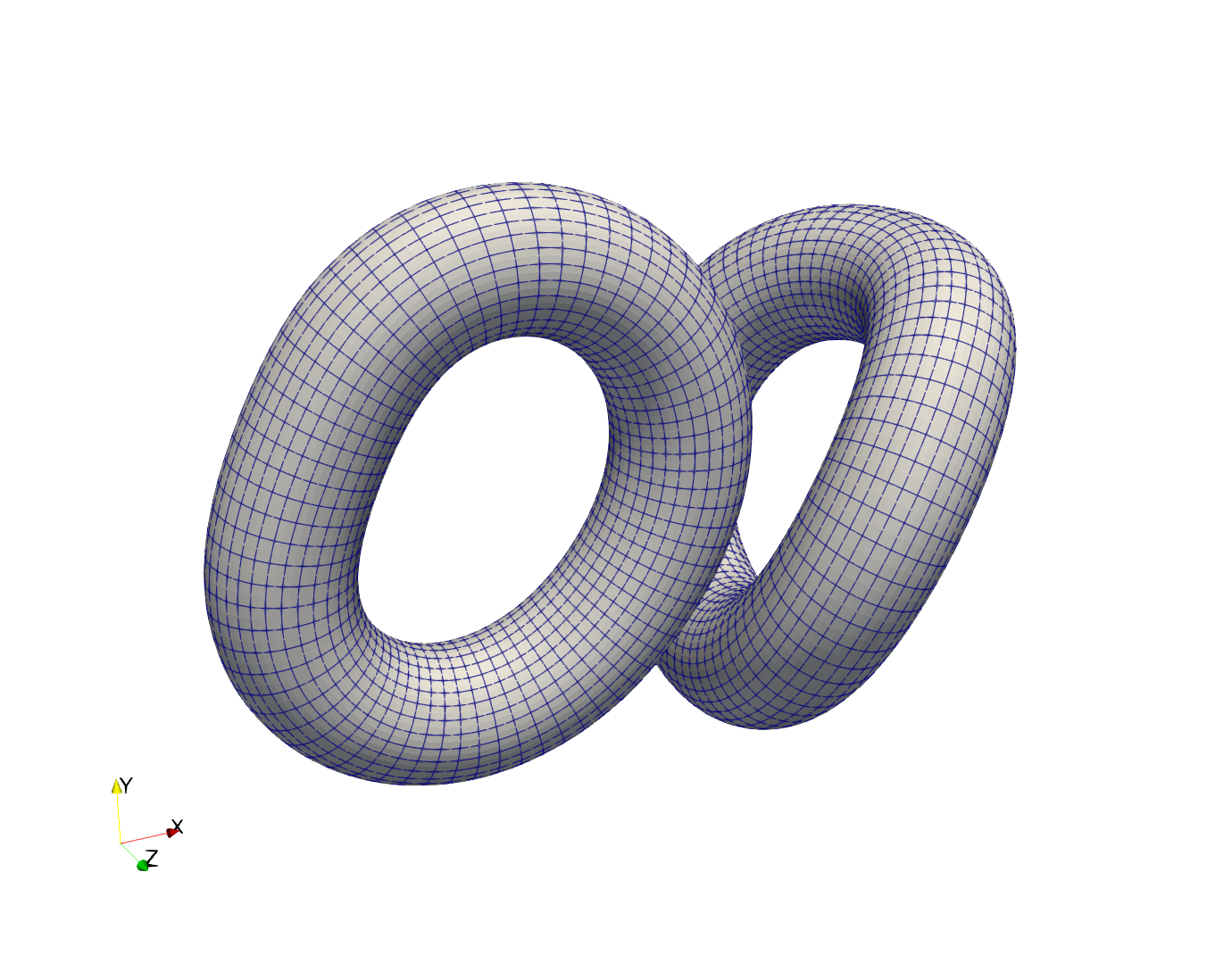}{$t=0.8$ }{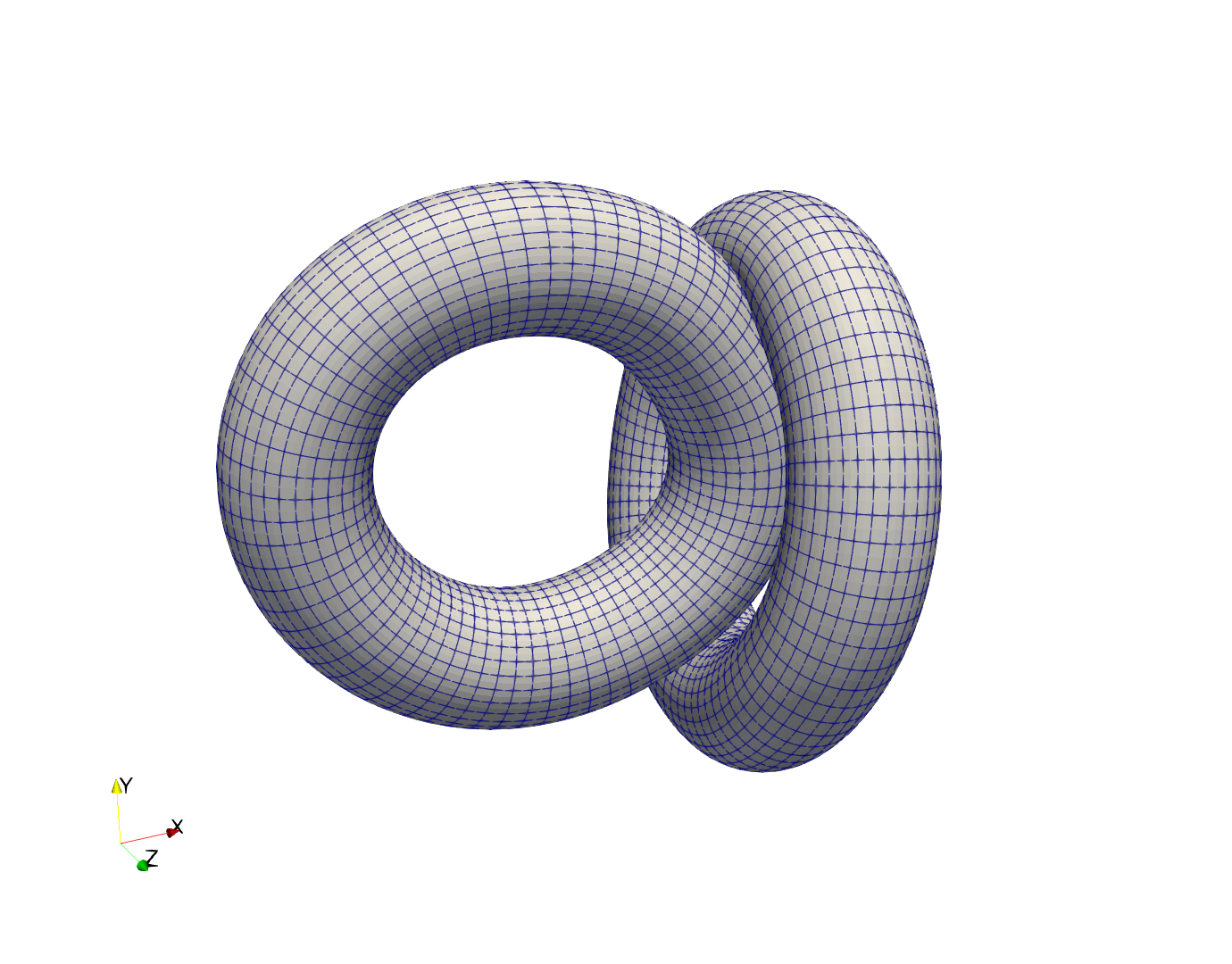}{$t=1.2$ }
  \imagerow{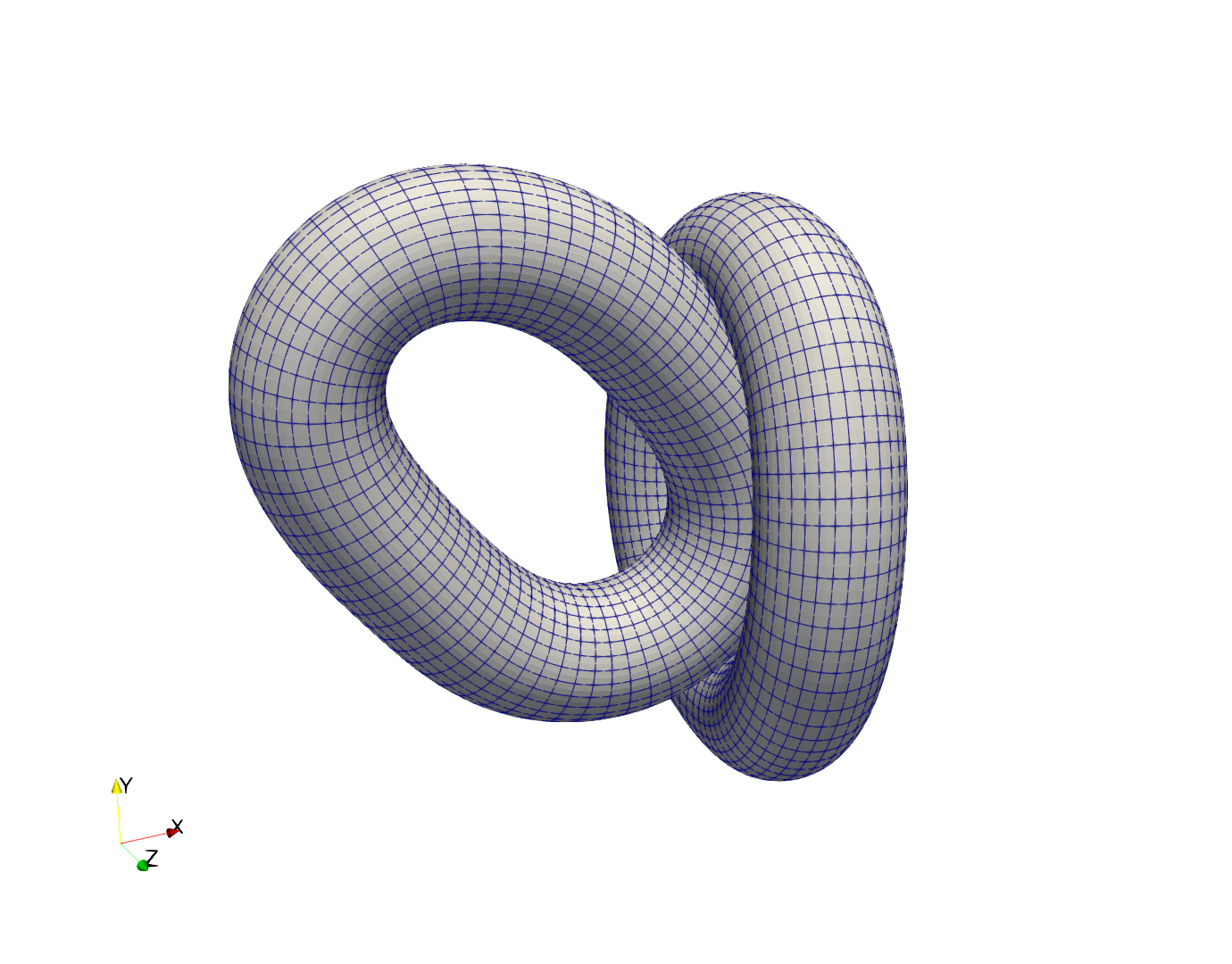}{$t=1.6$ }{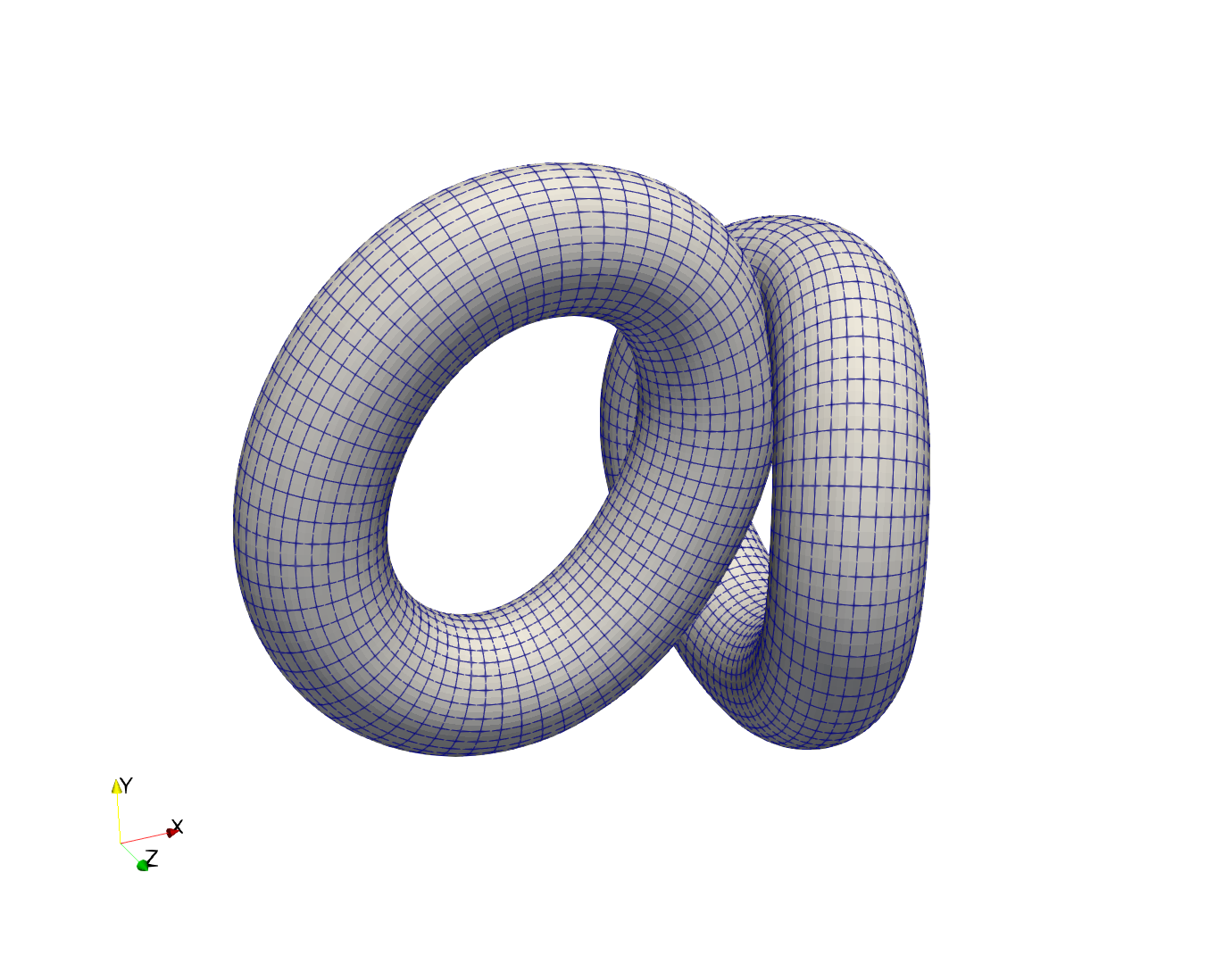}{$t=2.0$ }
  \caption{Two tori in contact - initial configuration and characteristic stages of deformation.}
  \label{fig:tori_contact}
\end{figure}

\begin{figure}
	\centering
	\imagerow{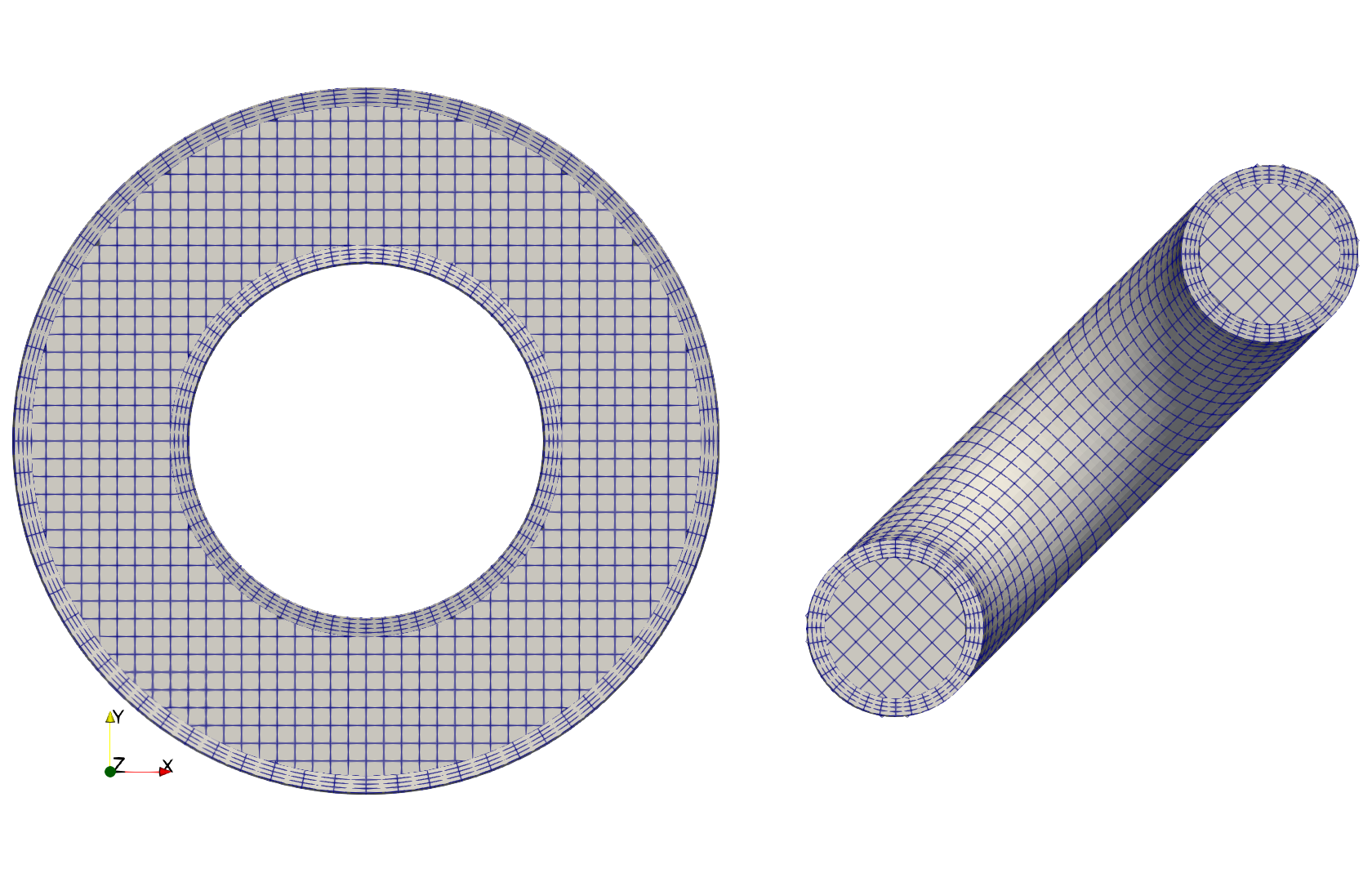}{$t=0.0$}{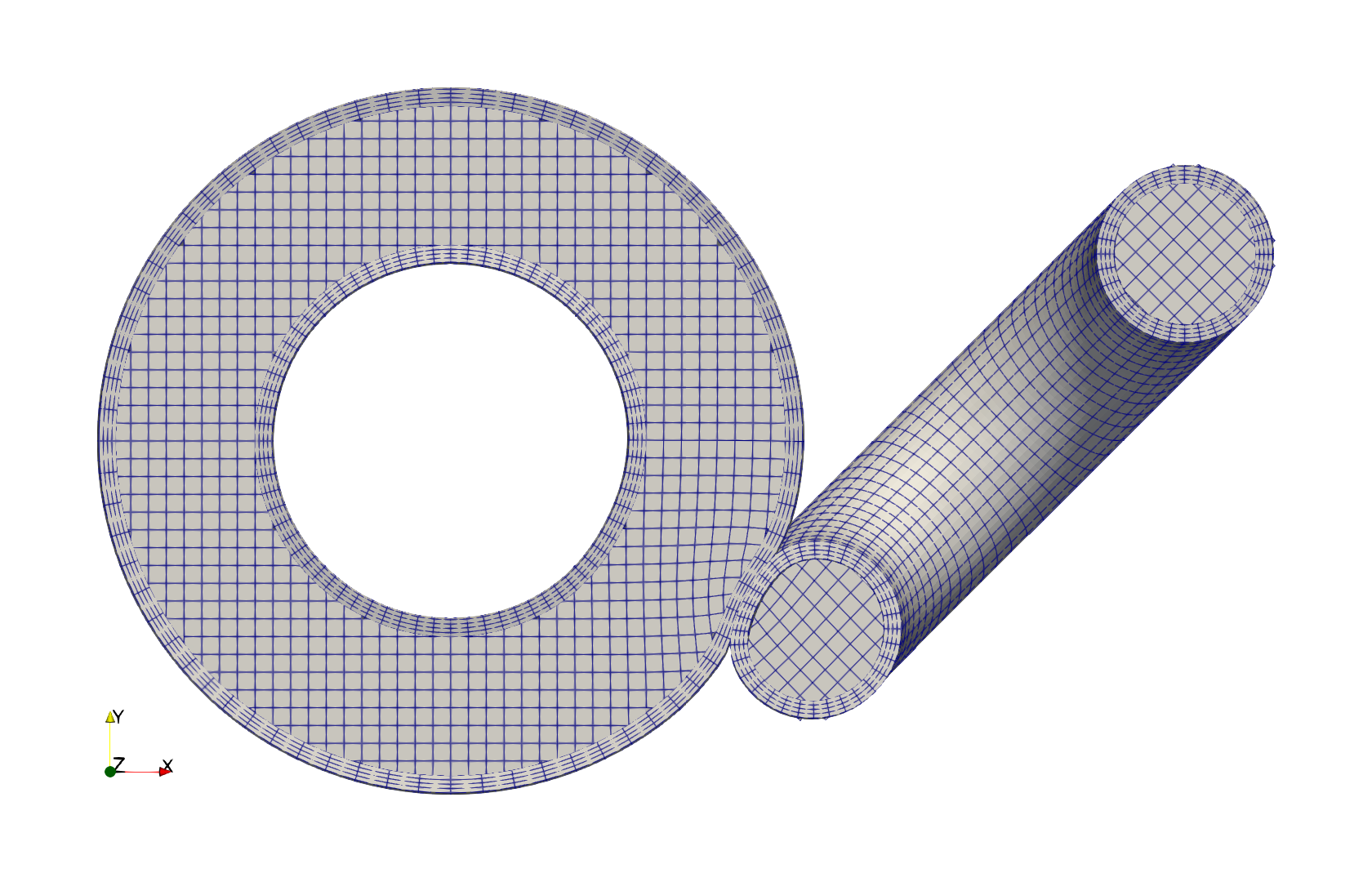}{$t=0.4$}
	\imagerow{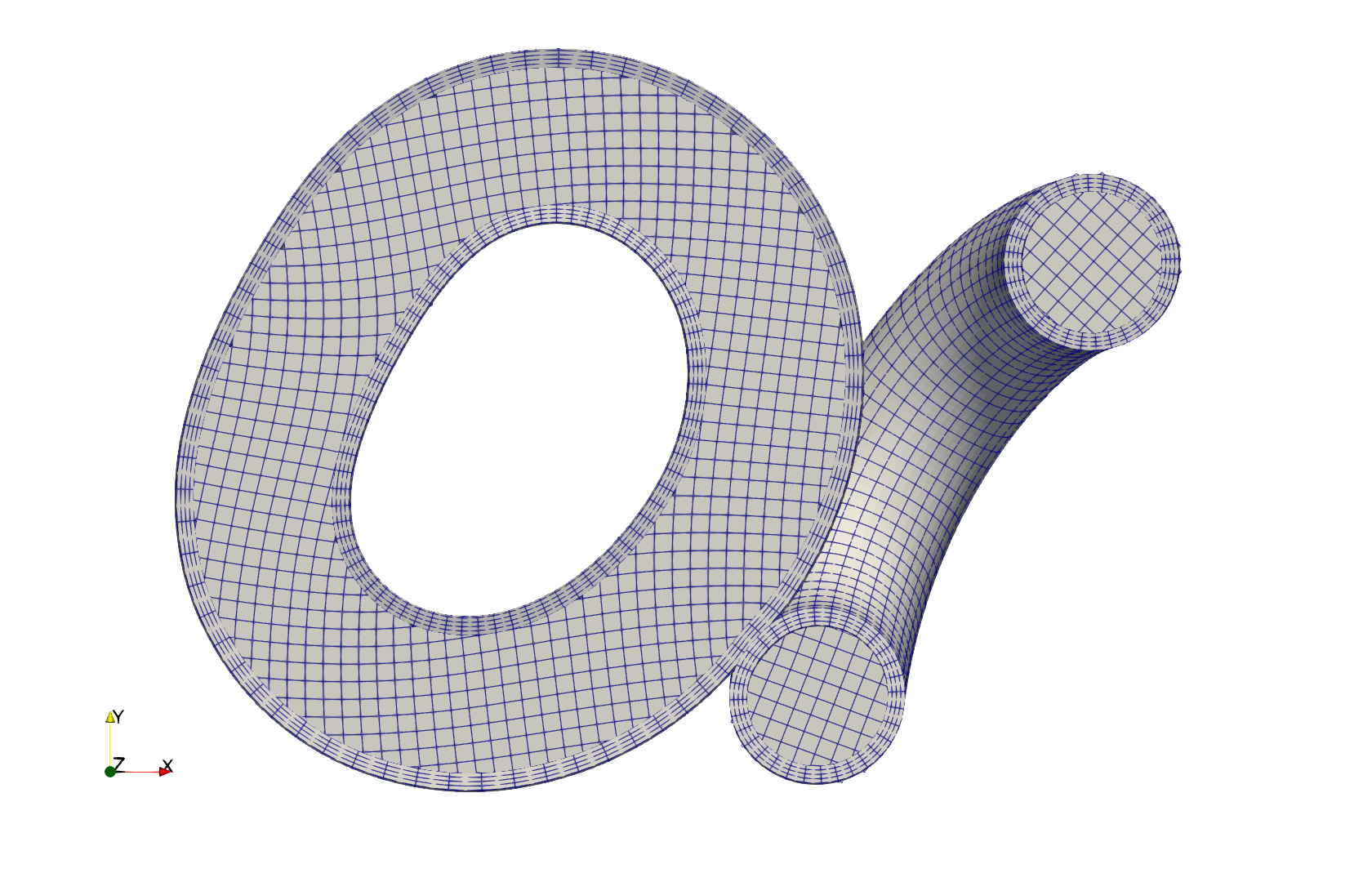}{$t=0.8$}{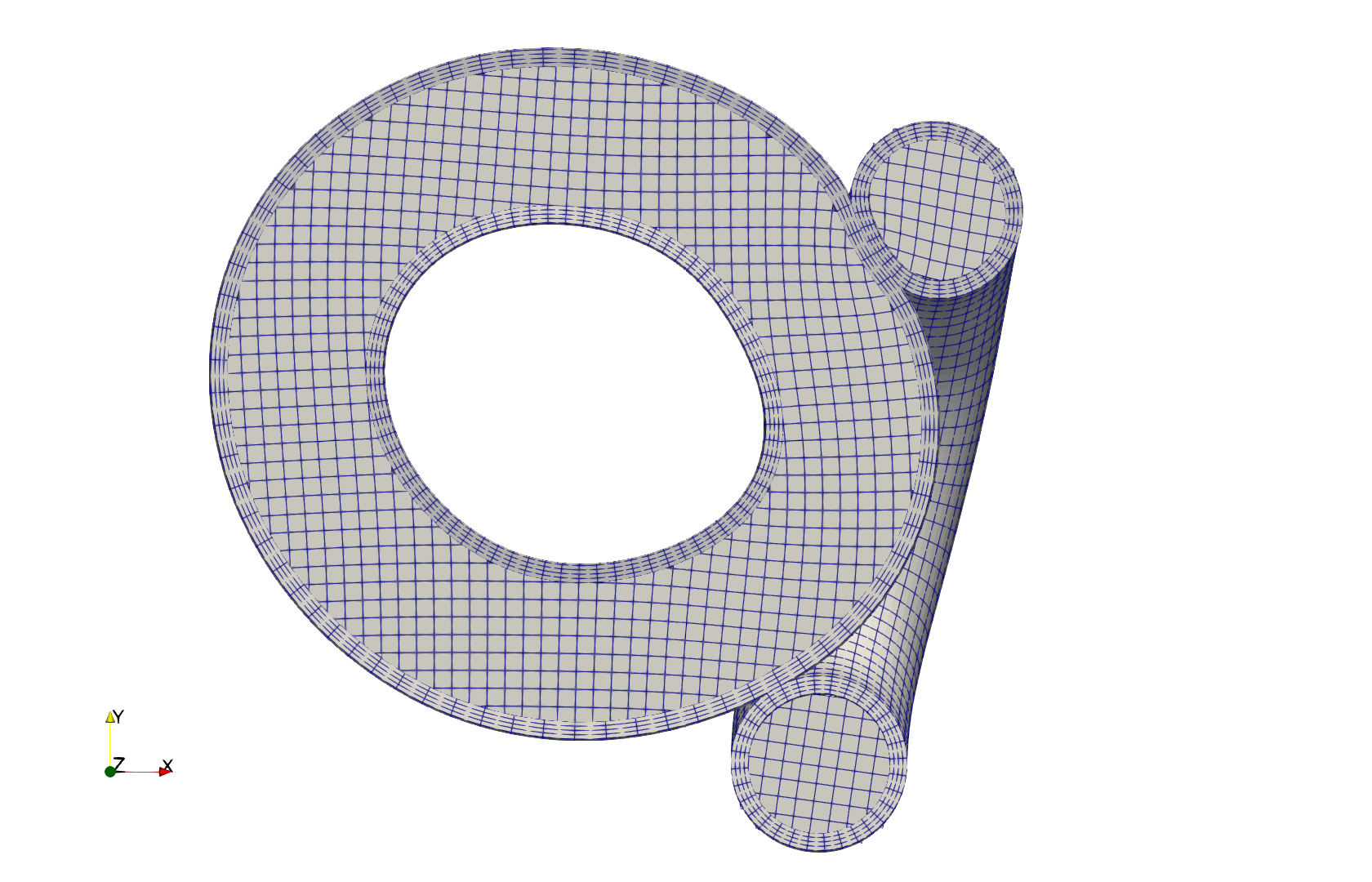}{$t=1.2$}
	\imagerow{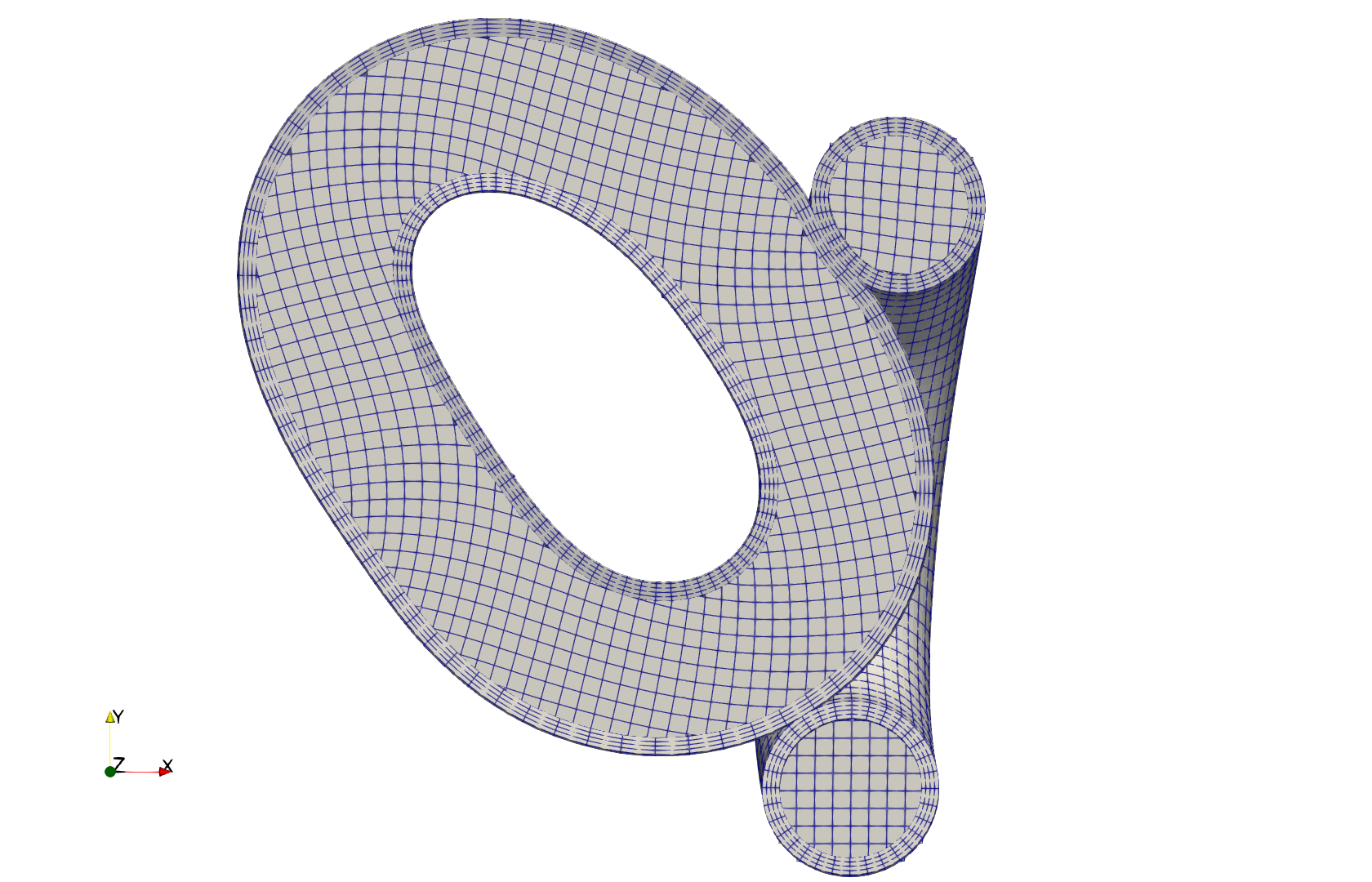}{$t=1.6$}{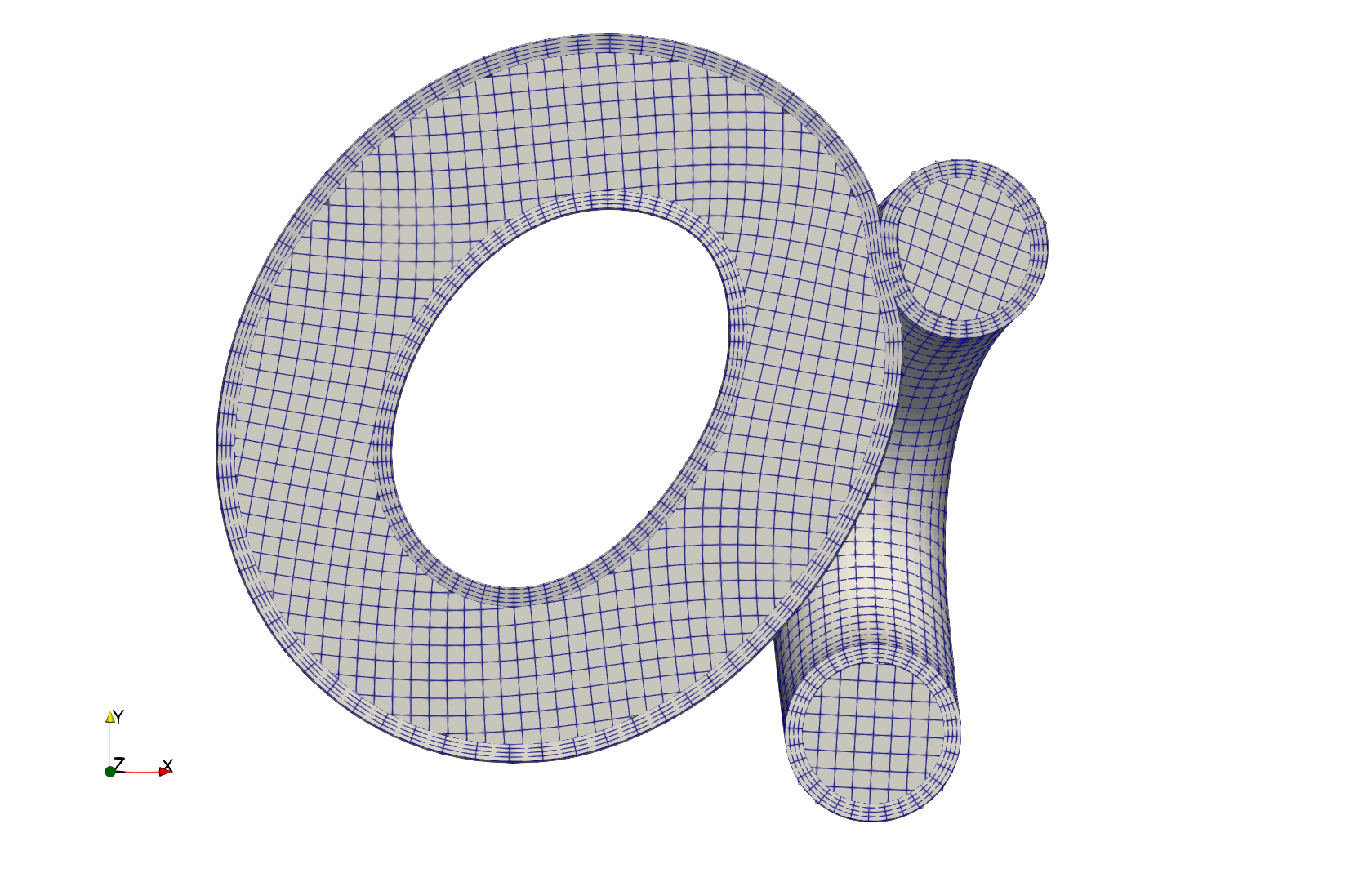}{$t=2.0$}
	\caption{Two tori in contact in side view - initial configuration and characteristic stages of deformation.}
	\label{fig:tori_contact_side_view}
  \end{figure}

%% file: fig/numerical_examples/patch_tests/boundary-conditions.eps_tex
\begingroup%
  \makeatletter%
  \providecommand\color[2][]{%
    \errmessage{(Inkscape) Color is used for the text in Inkscape, but the package 'color.sty' is not loaded}%
    \renewcommand\color[2][]{}%
  }%
  \providecommand\transparent[1]{%
    \errmessage{(Inkscape) Transparency is used (non-zero) for the text in Inkscape, but the package 'transparent.sty' is not loaded}%
    \renewcommand\transparent[1]{}%
  }%
  \providecommand\rotatebox[2]{#2}%
  \newcommand*\fsize{\dimexpr\f@size pt\relax}%
  \newcommand*\lineheight[1]{\fontsize{\fsize}{#1\fsize}\selectfont}%
  \ifx\svgwidth\undefined%
    \setlength{\unitlength}{239.37494111bp}%
    \ifx\svgscale\undefined%
      \relax%
    \else%
      \setlength{\unitlength}{\unitlength * \real{\svgscale}}%
    \fi%
  \else%
    \setlength{\unitlength}{\svgwidth}%
  \fi%
  \global\let\svgwidth\undefined%
  \global\let\svgscale\undefined%
  \makeatother%
  \begin{picture}(1,0.92572661)%
    \lineheight{1}%
    \setlength\tabcolsep{0pt}%
    \put(0,0){\includegraphics[width=\unitlength]{boundary-conditions.eps}}%
    \put(0.14077465,0.17614521){\makebox(0,0)[lt]{\lineheight{1.25}\smash{\begin{tabular}[t]{l}$\xaxis$\end{tabular}}}}%
    \put(0.00930982,0.32937287){\makebox(0,0)[lt]{\lineheight{1.25}\smash{\begin{tabular}[t]{l}$\yaxis$\end{tabular}}}}%
    \put(0.5379279,0.01){\makebox(0,0)[lt]{\lineheight{1.25}\smash{\begin{tabular}[t]{l}$a$\end{tabular}}}}%
    \put(0.91097257,0.4749145){\makebox(0,0)[lt]{\lineheight{1.25}\smash{\begin{tabular}[t]{l}$a$\end{tabular}}}}%
    \put(0.5379279,0.9){\makebox(0,0)[lt]{\lineheight{1.25}\smash{\begin{tabular}[t]{l}$p$\end{tabular}}}}%
  \end{picture}%
\endgroup%

%% file: fig/numerical_examples/patch_tests/straight_fine_2_nonlin_kinem_displacement_Y.tex
\begin{tikzpicture}
\node[anchor=south west,inner sep=-0.2pt] (image) at (0,0) {\includegraphics[scale=0.24]{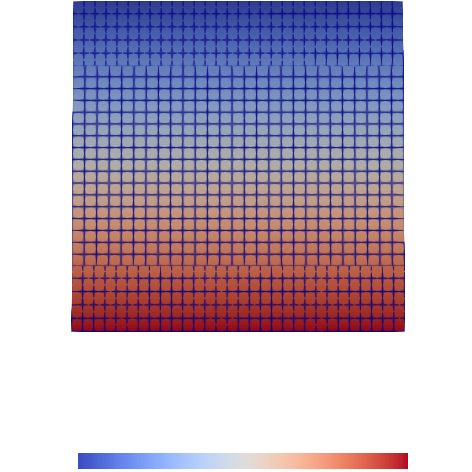}};
\begin{axis}[
scale only axis,
scaled x ticks=false,
scaled y ticks=false,
at={(0.6604cm,0.025400000000000002cm)},
tick label style={font=\footnotesize},
title=$d_{Y}$,
xticklabel=$\pgfmathprintnumber{\tick}$,
yticklabel=$\pgfmathprintnumber{\tick}$,
ymin=-0.030482408737122428,
ymax=0.0,
xmin=-0.030482408737122428,
xmax=0.0,
ytick=\empty,
height=0.13546666666666668cm,
width=2.7940000000000005cm,
xtick={-0.030482408737122428,0.0},
xtick pos=right,
xtick align=outside,
title style={yshift=10pt,},
]
\end{axis}
\end{tikzpicture}%

%% file: fig/numerical_examples/patch_tests/inclined_fine_2_nonlin_kinem_displacement_Y.tex
\begin{tikzpicture}
\node[anchor=south west,inner sep=-0.2pt] (image) at (0,0) {\includegraphics[scale=0.24]{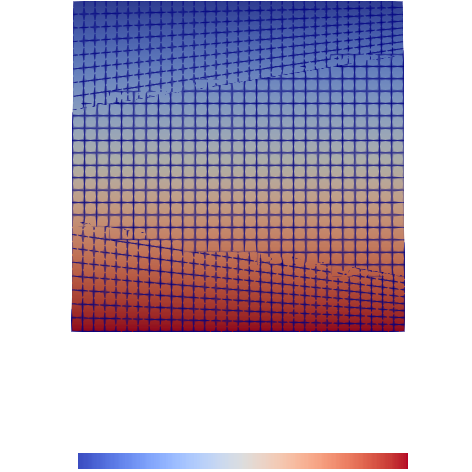}};
\begin{axis}[
scale only axis,
scaled x ticks=false,
scaled y ticks=false,
at={(0.6604cm,0.025400000000000002cm)},
tick label style={font=\footnotesize},
title=$d_{Y}$,
xticklabel=$\pgfmathprintnumber{\tick}$,
yticklabel=$\pgfmathprintnumber{\tick}$,
ymin=-0.03048213661559887,
ymax=0.0,
xmin=-0.03048213661559887,
xmax=0.0,
ytick=\empty,
height=0.13546666666666668cm,
width=2.7940000000000005cm,
xtick={-0.03048213661559887,0.0},
xtick pos=right,
xtick align=outside,
title style={yshift=10pt,},
]
\end{axis}
\end{tikzpicture}%

%% file: fig/numerical_examples/patch_tests/curved_fine_2_nonlin_kinem_displacement_Y.tex
\begin{tikzpicture}
\node[anchor=south west,inner sep=-0.2pt] (image) at (0,0) {\includegraphics[scale=0.24]{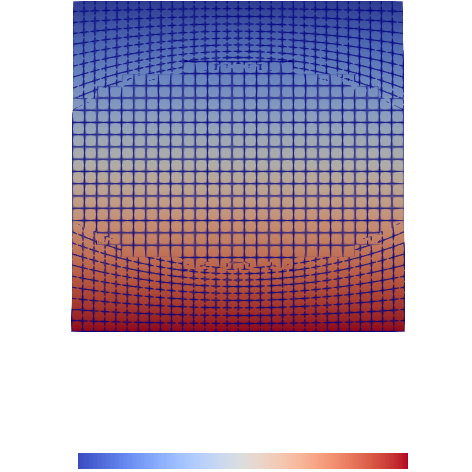}};
\begin{axis}[
scale only axis,
scaled x ticks=false,
scaled y ticks=false,
at={(0.6604cm,0.025400000000000002cm)},
tick label style={font=\footnotesize},
title=$d_{Y}$,
xticklabel=$\pgfmathprintnumber{\tick}$,
yticklabel=$\pgfmathprintnumber{\tick}$,
ymin=-0.03048348738564171,
ymax=0.0,
xmin=-0.03048348738564171,
xmax=0.0,
ytick=\empty,
height=0.13546666666666668cm,
width=2.7940000000000005cm,
xtick={-0.03048348738564171,0.0},
xtick pos=right,
xtick align=outside,
title style={yshift=10pt,},
]
\end{axis}
\end{tikzpicture}%

%% file: fig/numerical_examples/patch_tests/straight_fine_2_nonlin_kinem_stress_scale_element_cauchy_stresses_xyz_YY.tex
\begin{tikzpicture}
\node[anchor=south west,inner sep=-0.2pt] (image) at (0,0) {\includegraphics[scale=0.24]{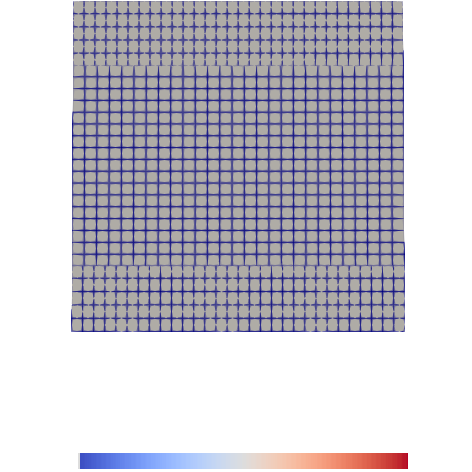}};
\begin{axis}[
/pgf/number format/.cd,fixed,precision=12,
scale only axis,
scaled x ticks=false,
scaled y ticks=false,
at={(0.6604cm,0.025400000000000002cm)},
tick label style={font=\scriptsize},
title=$\sigma_{YY}$,
ymin=0,
ymax=1,
xmin=0,
xmax=1,
ytick=\empty,
height=0.13546666666666668cm,
width=2.7940000000000005cm,
xtick={0,1},
xticklabels={-0.009939749712095,-0.009939749712095},
xtick pos=right,
xtick align=outside,
title style={yshift=10pt,},
]
\end{axis}
\end{tikzpicture}%

%% file: fig/numerical_examples/patch_tests/inclined_fine_2_nonlin_kinem_stress_scale_element_cauchy_stresses_xyz_YY.tex
\begin{tikzpicture}
\node[anchor=south west,inner sep=-0.2pt] (image) at (0,0) {\includegraphics[scale=0.24]{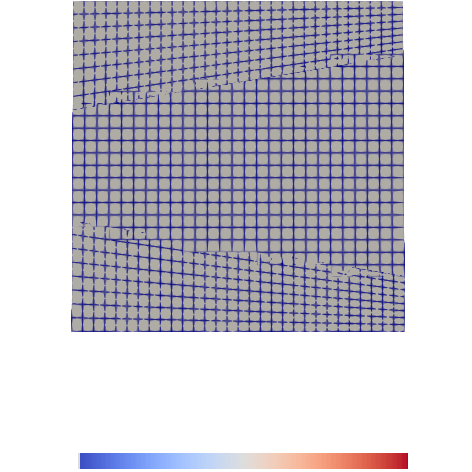}};
\begin{axis}[
/pgf/number format/.cd,fixed,precision=12,
scale only axis,
scaled x ticks=false,
scaled y ticks=false,
at={(0.6604cm,0.025400000000000002cm)},
tick label style={font=\scriptsize},
title=$\sigma_{YY}$,
ymin=0,
ymax=1,
xmin=0,
xmax=1,
ytick=\empty,
height=0.13546666666666668cm,
width=2.7940000000000005cm,
xtick={0,1},
xticklabels={-0.009939749712095,-0.009939749712095},
xtick pos=right,
xtick align=outside,
title style={yshift=10pt,},
]
\end{axis}
\end{tikzpicture}%

%% file: fig/numerical_examples/patch_tests/curved_fine_2_nonlin_kinem_element_cauchy_stresses_xyz_YY.tex
\begin{tikzpicture}
\node[anchor=south west,inner sep=-0.2pt] (image) at (0,0) {\includegraphics[scale=0.24]{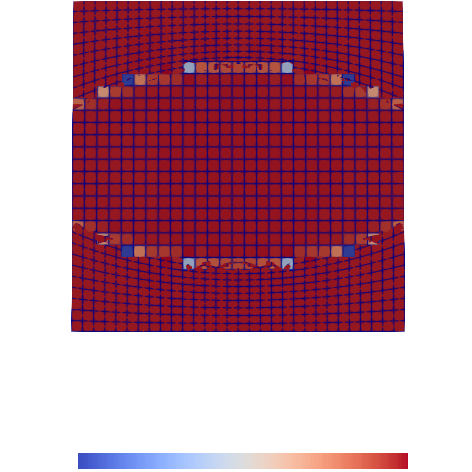}};
\begin{axis}[
scale only axis,
scaled x ticks=false,
scaled y ticks=false,
at={(0.6604cm,0.025400000000000002cm)},
tick label style={font=\footnotesize},
title=$\sigma_{YY}$,
xticklabel=$\pgfmathprintnumber{\tick}$,
yticklabel=$\pgfmathprintnumber{\tick}$,
ymin=-0.010121380935092683,
ymax=-0.00993609855157385,
xmin=-0.010121380935092683,
xmax=-0.00993609855157385,
ytick=\empty,
height=0.13546666666666668cm,
width=2.7940000000000005cm,
xtick={-0.010121380935092683,-0.00993609855157385},
xtick pos=right,
xtick align=outside,
title style={yshift=10pt,},
]
\end{axis}
\end{tikzpicture}%

%% file: fig/numerical_examples/bending_beam/bending_beam_2d_bcs.eps_tex
\begingroup%
  \makeatletter%
  \providecommand\color[2][]{%
    \errmessage{(Inkscape) Color is used for the text in Inkscape, but the package 'color.sty' is not loaded}%
    \renewcommand\color[2][]{}%
  }%
  \providecommand\transparent[1]{%
    \errmessage{(Inkscape) Transparency is used (non-zero) for the text in Inkscape, but the package 'transparent.sty' is not loaded}%
    \renewcommand\transparent[1]{}%
  }%
  \providecommand\rotatebox[2]{#2}%
  \newcommand*\fsize{\dimexpr\f@size pt\relax}%
  \newcommand*\lineheight[1]{\fontsize{\fsize}{#1\fsize}\selectfont}%
  \ifx\svgwidth\undefined%
    \setlength{\unitlength}{556.02161317bp}%
    \ifx\svgscale\undefined%
      \relax%
    \else%
      \setlength{\unitlength}{\unitlength * \real{\svgscale}}%
    \fi%
  \else%
    \setlength{\unitlength}{\svgwidth}%
  \fi%
  \global\let\svgwidth\undefined%
  \global\let\svgscale\undefined%
  \makeatother%
  \begin{picture}(1,0.65415283)%
    \lineheight{1}%
    \setlength\tabcolsep{0pt}%
    \put(0,0){\includegraphics[width=\unitlength]{bending_beam_2d_bcs.eps}}%
    \put(0.19853745,0.31176588){\makebox(0,0)[lt]{\lineheight{1.25}\smash{\begin{tabular}[t]{l}$\xaxis$\end{tabular}}}}%
    \put(0.50375099,0.68){\makebox(0,0)[lt]{\lineheight{1.25}\smash{\begin{tabular}[t]{l}$l$\end{tabular}}}}%
    \put(-0.01,0.32061394){\makebox(0,0)[lt]{\lineheight{1.25}\smash{\begin{tabular}[t]{l}$h$\end{tabular}}}}%
    \put(0.131,0.3787215){\makebox(0,0)[lt]{\lineheight{1.25}\smash{\begin{tabular}[t]{l}$\yaxis$\end{tabular}}}}%
    \put(0.05,0.38){\makebox(0,0)[lt]{\lineheight{1.25}\smash{\begin{tabular}[t]{l}$M$\end{tabular}}}}%
    \put(0.96,0.38){\makebox(0,0)[lt]{\lineheight{1.25}\smash{\begin{tabular}[t]{l}$M$\end{tabular}}}}%
  \end{picture}%
\endgroup%

%% file: fig/numerical_examples/bending_beam/reference_linear_kinem_element_cauchy_stresses_xyz_ZZ.tex
\begin{tikzpicture}
\node[anchor=south west,inner sep=-0.2pt] (image) at (0,0) {\includegraphics[scale=0.24]{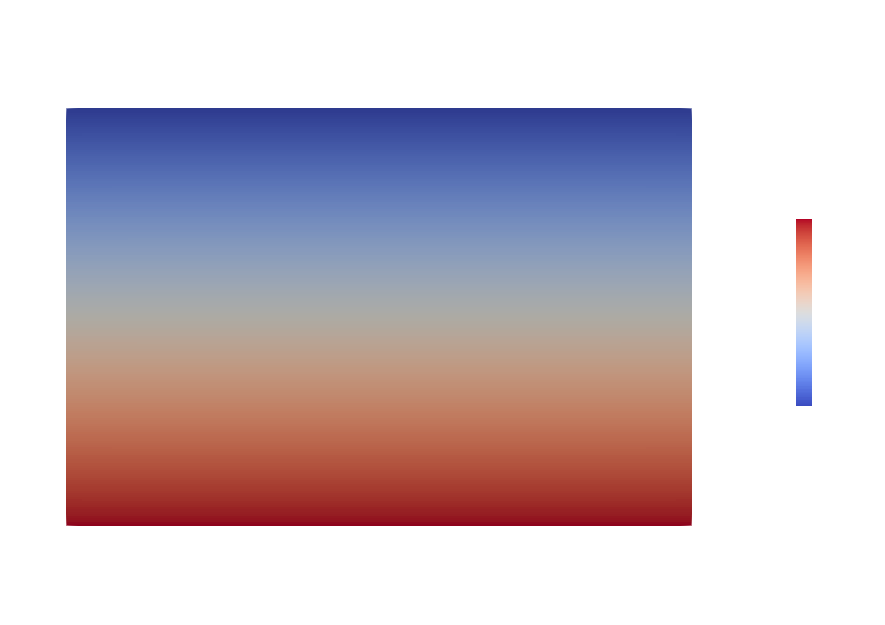}};
\begin{axis}[
scale only axis,
scaled x ticks=false,
scaled y ticks=false,
at={(6.739466666666667cm,1.8542cm)},
tick label style={font=\footnotesize},
title=$\sigma_{XX}$,
xticklabel=$\pgfmathprintnumber{\tick}$,
yticklabel=$\pgfmathprintnumber{\tick}$,
ymin=-0.1,
ymax=0.1,
xmin=-0.1,
xmax=0.1,
xtick=\empty,
height=1.5832666666666666cm,
width=0.13546666666666668cm,
ytick={-0.1,0.1},
ytick pos=right,
ytick align=outside,
]
\end{axis}
\end{tikzpicture}%

%% file: fig/numerical_examples/bending_beam/bending_beam_2d_discret.eps_tex
\begingroup%
  \makeatletter%
  \providecommand\color[2][]{%
    \errmessage{(Inkscape) Color is used for the text in Inkscape, but the package 'color.sty' is not loaded}%
    \renewcommand\color[2][]{}%
  }%
  \providecommand\transparent[1]{%
    \errmessage{(Inkscape) Transparency is used (non-zero) for the text in Inkscape, but the package 'transparent.sty' is not loaded}%
    \renewcommand\transparent[1]{}%
  }%
  \providecommand\rotatebox[2]{#2}%
  \newcommand*\fsize{\dimexpr\f@size pt\relax}%
  \newcommand*\lineheight[1]{\fontsize{\fsize}{#1\fsize}\selectfont}%
  \ifx\svgwidth\undefined%
    \setlength{\unitlength}{538.2003006bp}%
    \ifx\svgscale\undefined%
      \relax%
    \else%
      \setlength{\unitlength}{\unitlength * \real{\svgscale}}%
    \fi%
  \else%
    \setlength{\unitlength}{\svgwidth}%
  \fi%
  \global\let\svgwidth\undefined%
  \global\let\svgscale\undefined%
  \makeatother%
  \begin{picture}(1,0.660914)%
    \lineheight{1}%
    \setlength\tabcolsep{0pt}%
    \put(0,0){\includegraphics[width=\unitlength]{bending_beam_2d_discret.eps}}%
    \put(0.22201455,-0.02){\makebox(0,0)[lt]{\lineheight{1.25}\smash{\begin{tabular}[t]{l}$x = 0.625$\end{tabular}}}}%
    \put(0.23,0.3181414){\color[rgb]{0,0,0}\makebox(0,0)[lt]{\lineheight{1.25}\smash{\colorbox{white}{\begin{tabular}[t]{l}\boundarylayerdomainnumexample{}\end{tabular}}}}}%
    \put(0.6251164,0.3181414){\color[rgb]{0,0,0}\makebox(0,0)[lt]{\lineheight{1.25}\smash{\colorbox{white}{\begin{tabular}[t]{l}\bulkdomainnumexample{}\end{tabular}}}}}%
    \put(0.47271087,0.6344075){\makebox(0,0)[lt]{\lineheight{1.25}\smash{\begin{tabular}[t]{l}$\Gamma^{*}$\end{tabular}}}}%
  \end{picture}%
\endgroup%

%% file: fig/numerical_examples/bending_beam/case_2_linear_kinem_element_cauchy_stresses_xyz_ZZ.tex
\begin{tikzpicture}
\node[anchor=south west,inner sep=-0.2pt] (image) at (0,0) {\includegraphics[scale=0.24]{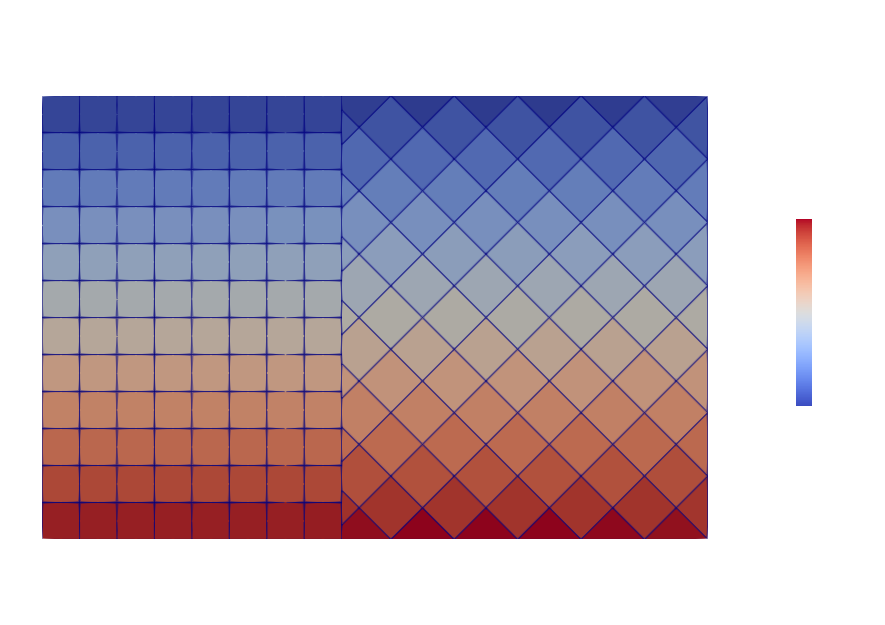}};
\begin{axis}[
scale only axis,
scaled x ticks=false,
scaled y ticks=false,
at={(6.739466666666667cm,1.8542cm)},
tick label style={font=\footnotesize},
title=$\sigma_{XX}$,
xticklabel=$\pgfmathprintnumber{\tick}$,
yticklabel=$\pgfmathprintnumber{\tick}$,
ymin=-0.09842078776298635,
ymax=0.09842078776298635,
xmin=-0.09842078776298635,
xmax=0.09842078776298635,
xtick=\empty,
height=1.5832666666666666cm,
width=0.13546666666666668cm,
ytick={-0.09842078776298635,0.09842078776298635},
ytick pos=right,
ytick align=outside,
]
\end{axis}
\end{tikzpicture}%

%% file: fig/numerical_examples/bending_beam/case_1_linear_kinem_element_cauchy_stresses_xyz_ZZ.tex
\begin{tikzpicture}
\node[anchor=south west,inner sep=-0.2pt] (image) at (0,0) {\includegraphics[scale=0.24]{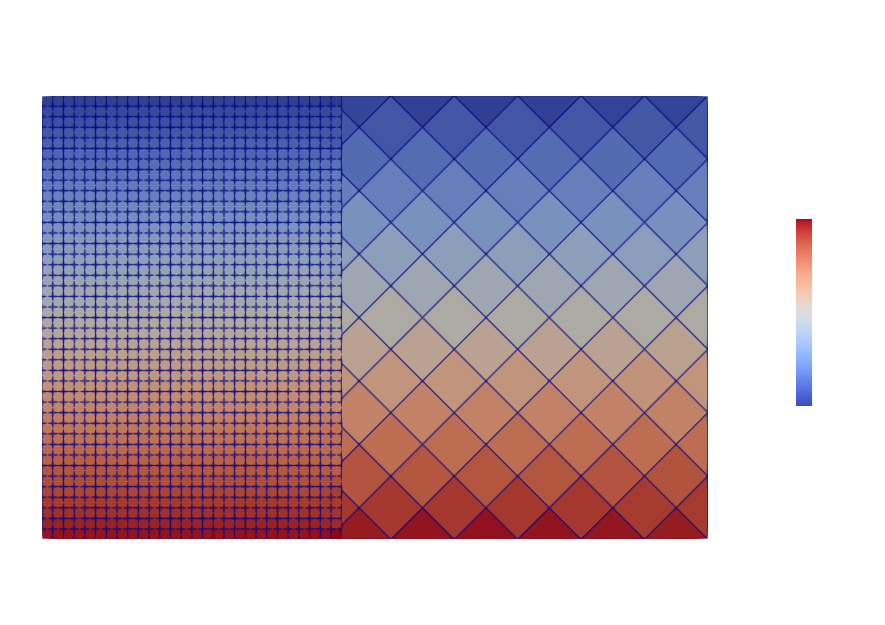}};
\begin{axis}[
scale only axis,
scaled x ticks=false,
scaled y ticks=false,
at={(6.739466666666667cm,1.8542cm)},
tick label style={font=\footnotesize},
title=$\sigma_{XX}$,
xticklabel=$\pgfmathprintnumber{\tick}$,
yticklabel=$\pgfmathprintnumber{\tick}$,
ymin=-0.10158576475242195,
ymax=0.10158566723445786,
xmin=-0.10158576475242195,
xmax=0.10158566723445786,
xtick=\empty,
height=1.5832666666666666cm,
width=0.13546666666666668cm,
ytick={-0.10158576475242195,0.10158566723445786},
ytick pos=right,
ytick align=outside,
]
\end{axis}
\end{tikzpicture}%

%% file: fig/numerical_examples/bending_beam/case_3_linear_kinem_element_cauchy_stresses_xyz_ZZ.tex
\begin{tikzpicture}
\node[anchor=south west,inner sep=-0.2pt] (image) at (0,0) {\includegraphics[scale=0.24]{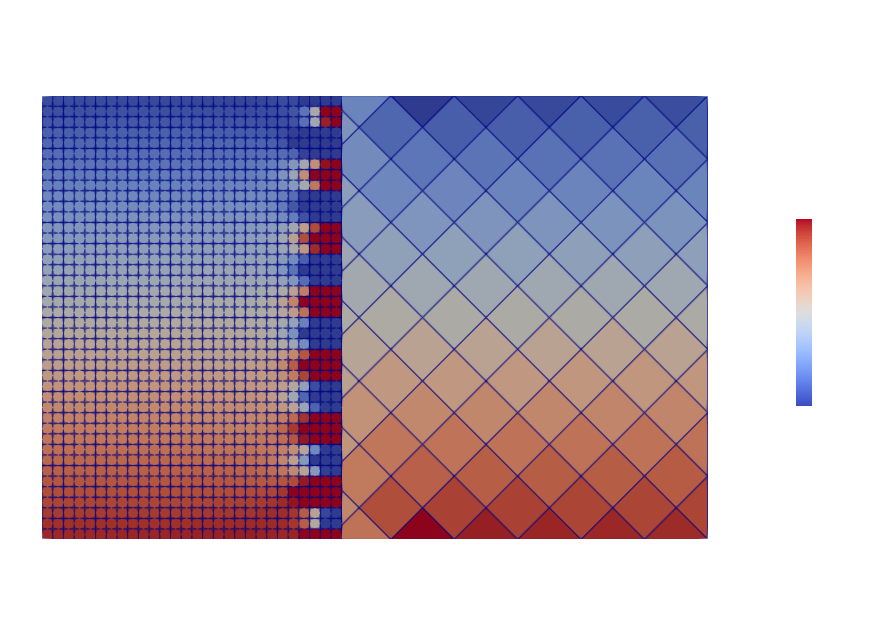}};
\begin{axis}[
scale only axis,
scaled x ticks=false,
scaled y ticks=false,
at={(6.739466666666667cm,1.8542cm)},
tick label style={font=\footnotesize},
title=$\sigma_{XX}$,
xticklabel=$\pgfmathprintnumber{\tick}$,
yticklabel=$\pgfmathprintnumber{\tick}$,
ymin=-0.10803579173192131,
ymax=0.10803579953835356,
xmin=-0.10803579173192131,
xmax=0.10803579953835356,
xtick=\empty,
height=1.5832666666666666cm,
width=0.13546666666666668cm,
ytick={-0.10803579173192131,0.10803579953835356},
ytick pos=right,
ytick align=outside,
]
\end{axis}
\end{tikzpicture}%

%% file: fig/numerical_examples/convergence_plot/single_block_contact_boundary_conditions.pdf_tex
\begingroup%
  \makeatletter%
  \providecommand\color[2][]{%
    \errmessage{(Inkscape) Color is used for the text in Inkscape, but the package 'color.sty' is not loaded}%
    \renewcommand\color[2][]{}%
  }%
  \providecommand\transparent[1]{%
    \errmessage{(Inkscape) Transparency is used (non-zero) for the text in Inkscape, but the package 'transparent.sty' is not loaded}%
    \renewcommand\transparent[1]{}%
  }%
  \providecommand\rotatebox[2]{#2}%
  \newcommand*\fsize{\dimexpr\f@size pt\relax}%
  \newcommand*\lineheight[1]{\fontsize{\fsize}{#1\fsize}\selectfont}%
  \ifx\svgwidth\undefined%
    \setlength{\unitlength}{289.13385827bp}%
    \ifx\svgscale\undefined%
      \relax%
    \else%
      \setlength{\unitlength}{\unitlength * \real{\svgscale}}%
    \fi%
  \else%
    \setlength{\unitlength}{\svgwidth}%
  \fi%
  \global\let\svgwidth\undefined%
  \global\let\svgscale\undefined%
  \makeatother%
  \begin{picture}(1,0.70588235)%
    \lineheight{1}%
    \setlength\tabcolsep{0pt}%
    \put(0,0){\includegraphics[width=\unitlength,page=1]{single_block_contact_boundary_conditions.pdf}}%
    \put(0.11616622,0.04703582){\color[rgb]{0,0,0}\makebox(0,0)[lt]{\lineheight{1.25}\smash{\begin{tabular}[t]{l}$\xaxis$\end{tabular}}}}%
    \put(0.01169136,0.15677381){\color[rgb]{0,0,0}\makebox(0,0)[lt]{\lineheight{1.25}\smash{\begin{tabular}[t]{l}$\yaxis$\end{tabular}}}}%
    \put(0,0){\includegraphics[width=\unitlength,page=2]{single_block_contact_boundary_conditions.pdf}}%
    \put(0.72855983,0.39863441){\color[rgb]{0,0,0}\makebox(0,0)[lt]{\lineheight{1.25}\smash{\begin{tabular}[t]{l}$u_x = 0$\end{tabular}}}}%
    \put(0,0){\includegraphics[width=\unitlength,page=3]{single_block_contact_boundary_conditions.pdf}}%
    \put(0.84522397,0.15184503){\color[rgb]{0,0,0}\makebox(0,0)[lt]{\lineheight{1.25}\smash{\begin{tabular}[t]{l}$master$\end{tabular}}}}%
    \put(0,0){\includegraphics[width=\unitlength,page=4]{single_block_contact_boundary_conditions.pdf}}%
    \put(0.78016132,0.27267565){\color[rgb]{0,0,0}\makebox(0,0)[lt]{\lineheight{1.25}\smash{\begin{tabular}[t]{l}$slave$\end{tabular}}}}%
    \put(0,0){\includegraphics[width=\unitlength,page=5]{single_block_contact_boundary_conditions.pdf}}%
    \put(0.70035531,0.61465535){\color[rgb]{0,0,0}\makebox(0,0)[lt]{\lineheight{1.25}\smash{\begin{tabular}[t]{l}$f_y = 0.1\xaxis^4$\end{tabular}}}}%
  \end{picture}%
\endgroup%

%% file: fig/numerical_examples/convergence_plot/single_block_contact_domain_subdivision.pdf_tex
\begingroup%
  \makeatletter%
  \providecommand\color[2][]{%
    \errmessage{(Inkscape) Color is used for the text in Inkscape, but the package 'color.sty' is not loaded}%
    \renewcommand\color[2][]{}%
  }%
  \providecommand\transparent[1]{%
    \errmessage{(Inkscape) Transparency is used (non-zero) for the text in Inkscape, but the package 'transparent.sty' is not loaded}%
    \renewcommand\transparent[1]{}%
  }%
  \providecommand\rotatebox[2]{#2}%
  \newcommand*\fsize{\dimexpr\f@size pt\relax}%
  \newcommand*\lineheight[1]{\fontsize{\fsize}{#1\fsize}\selectfont}%
  \ifx\svgwidth\undefined%
    \setlength{\unitlength}{289.13385827bp}%
    \ifx\svgscale\undefined%
      \relax%
    \else%
      \setlength{\unitlength}{\unitlength * \real{\svgscale}}%
    \fi%
  \else%
    \setlength{\unitlength}{\svgwidth}%
  \fi%
  \global\let\svgwidth\undefined%
  \global\let\svgscale\undefined%
  \makeatother%
  \begin{picture}(1,0.70588235)%
    \lineheight{1}%
    \setlength\tabcolsep{0pt}%
    \put(0,0){\includegraphics[width=\unitlength,page=1]{single_block_contact_domain_subdivision.pdf}}%
    \put(0.73282818,0.27297992){\color[rgb]{0,0,0}\makebox(0,0)[lt]{\lineheight{1.25}\smash{\begin{tabular}[t]{l}$h =3$\end{tabular}}}}%
    \put(0.39,0.53352792){\color[rgb]{0,0,0}\makebox(0,0)[lt]{\lineheight{1.25}\smash{\begin{tabular}[t]{l}$w =3$\end{tabular}}}}%
    \put(0,0){\includegraphics[width=\unitlength,page=2]{single_block_contact_domain_subdivision.pdf}}%
    \put(0.07,0.14025871){\color[rgb]{0,0,0}\makebox(0,0)[lt]{\lineheight{1.25}\smash{\begin{tabular}[t]{l}$\ell=1$\end{tabular}}}}%
    \put(0,0){\includegraphics[width=\unitlength,page=3]{single_block_contact_domain_subdivision.pdf}}%
    \put(0.44023609,0.1149337){\rotatebox{0.54797288}{\makebox(0,0)[lt]{\lineheight{1.25}\smash{\begin{tabular}[t]{l}\footnotesize{\boundarylayerdomain{}}\end{tabular}}}}}%
    \put(0,0){\includegraphics[width=\unitlength,page=4]{single_block_contact_domain_subdivision.pdf}}%
    \put(0.44001416,0.31911993){\rotatebox{0.03727542}{\makebox(0,0)[lt]{\lineheight{1.25}\smash{\begin{tabular}[t]{l}\footnotesize{\bulkdomain{}}\end{tabular}}}}}%
    \put(0,0){\includegraphics[width=\unitlength,page=5]{single_block_contact_domain_subdivision.pdf}}%
    \put(0.83567642,0.04332881){\color[rgb]{0,0,0}\makebox(0,0)[lt]{\lineheight{1.25}\smash{\begin{tabular}[t]{l}\footnotesize{\interfacestar{}{}}\end{tabular}}}}%
  \end{picture}%
\endgroup%

%% file: fig/numerical_examples/convergence_plot/single_block_contact_mesh.pdf_tex
\begingroup%
  \makeatletter%
  \providecommand\color[2][]{%
    \errmessage{(Inkscape) Color is used for the text in Inkscape, but the package 'color.sty' is not loaded}%
    \renewcommand\color[2][]{}%
  }%
  \providecommand\transparent[1]{%
    \errmessage{(Inkscape) Transparency is used (non-zero) for the text in Inkscape, but the package 'transparent.sty' is not loaded}%
    \renewcommand\transparent[1]{}%
  }%
  \providecommand\rotatebox[2]{#2}%
  \newcommand*\fsize{\dimexpr\f@size pt\relax}%
  \newcommand*\lineheight[1]{\fontsize{\fsize}{#1\fsize}\selectfont}%
  \ifx\svgwidth\undefined%
    \setlength{\unitlength}{289.13385827bp}%
    \ifx\svgscale\undefined%
      \relax%
    \else%
      \setlength{\unitlength}{\unitlength * \real{\svgscale}}%
    \fi%
  \else%
    \setlength{\unitlength}{\svgwidth}%
  \fi%
  \global\let\svgwidth\undefined%
  \global\let\svgscale\undefined%
  \makeatother%
  \begin{picture}(1,0.70588235)%
    \lineheight{1}%
    \setlength\tabcolsep{0pt}%
    \put(0,0){\includegraphics[width=\unitlength,page=1]{single_block_contact_mesh.pdf}}%
  \end{picture}%
\endgroup%

%% file: fig/numerical_examples/convergence_plot/convergence_plot_nurbs27-hex8.tex
\begin{tikzpicture}
  \begin{loglogaxis}[
    width=0.5\linewidth,
    xlabel={$h$},
    ylabel={$\|\mathbf{u} - \mathbf{u}_h\|_{E}$},
    grid=both,
    legend style={
      at={(0.5,-0.2)}, 
      anchor=north,    
      draw=black,       
      line width=0.5pt,
      font=\small,
      legend columns=2, 
      column sep=10pt,
    },
  ]
    \addplot[
      thick,
      mark=*,
      color=blue
    ] table[
      x index=0,
      y index=1
    ]{fig/numerical_examples/convergence_plot/plot_coordinates_nurbs27-hex8_embedded_mesh_curved_without_deletion_cells_nonmatching_meshes_epsilon_10000_cubature_1_epsilon_10000_ref_sol_nurbs27-hex20_size_3_128.txt};
    \addlegendentry{$\nurbssecondordertwodim{}-\lagrangefirstordertwodim{}$}

    \addplot[
      thick,
      mark=square*,
      color=red
    ] table[
      x index=0,
      y index=1
    ]{fig/numerical_examples/convergence_plot/plot_coordinates_nurbs27-hex20_embedded_mesh_curved_without_deletion_cells_nonmatching_meshes_epsilon_10000_cubature_1_epsilon_10000_ref_sol_nurbs27-hex20_size_3_128.txt};
    \addlegendentry{$\nurbssecondordertwodim{}-\lagrangeseconddordertwodim{}$}

    \addplot[
      domain=0.1:1.5,    
      thick,
      dashed,
      color=black
    ] {x^(1)/2};
    \addlegendentry{$\mathcal{O}(h^1)$}

    \addplot[
      domain=0.1:1.5,    
      thick,
      dotted,
      color=black
    ] {x^(2)/6};
    \addlegendentry{$\mathcal{O}(h^2)$}

  \end{loglogaxis}
\end{tikzpicture}

%% file: fig/numerical_examples/2d_hertzian_contact/2d_hertzian_contact_boundary.pdf_tex
\begingroup%
  \makeatletter%
  \providecommand\color[2][]{%
    \errmessage{(Inkscape) Color is used for the text in Inkscape, but the package 'color.sty' is not loaded}%
    \renewcommand\color[2][]{}%
  }%
  \providecommand\transparent[1]{%
    \errmessage{(Inkscape) Transparency is used (non-zero) for the text in Inkscape, but the package 'transparent.sty' is not loaded}%
    \renewcommand\transparent[1]{}%
  }%
  \providecommand\rotatebox[2]{#2}%
  \newcommand*\fsize{\dimexpr\f@size pt\relax}%
  \newcommand*\lineheight[1]{\fontsize{\fsize}{#1\fsize}\selectfont}%
  \ifx\svgwidth\undefined%
    \setlength{\unitlength}{147.4015748bp}%
    \ifx\svgscale\undefined%
      \relax%
    \else%
      \setlength{\unitlength}{\unitlength * \real{\svgscale}}%
    \fi%
  \else%
    \setlength{\unitlength}{\svgwidth}%
  \fi%
  \global\let\svgwidth\undefined%
  \global\let\svgscale\undefined%
  \makeatother%
  \begin{picture}(1,0.71153846)%
    \lineheight{1}%
    \setlength\tabcolsep{0pt}%
    \put(0,0){\includegraphics[width=\unitlength,page=1]{2d_hertzian_contact_boundary.pdf}}%
    \put(0.47628641,0.64601952){\makebox(0,0)[lt]{\lineheight{1.25}\smash{\begin{tabular}[t]{l}$p$\end{tabular}}}}%
    \put(0,0){\includegraphics[width=\unitlength,page=2]{2d_hertzian_contact_boundary.pdf}}%
    \put(0.63053148,0.0575557){\makebox(0,0)[lt]{\lineheight{1.25}\smash{\begin{tabular}[t]{l}$x$\end{tabular}}}}%
    \put(0.84633999,0.21714452){\makebox(0,0)[lt]{\lineheight{1.25}\smash{\begin{tabular}[t]{l}$slave$\end{tabular}}}}%
    \put(0,0){\includegraphics[width=\unitlength,page=3]{2d_hertzian_contact_boundary.pdf}}%
    \put(0.23258246,0.41885741){\makebox(0,0)[lt]{\lineheight{1.25}\smash{\begin{tabular}[t]{l}$R$\end{tabular}}}}%
    \put(0,0){\includegraphics[width=\unitlength,page=4]{2d_hertzian_contact_boundary.pdf}}%
    \put(0.10571143,0.46872326){\makebox(0,0)[lt]{\lineheight{1.25}\smash{\begin{tabular}[t]{l}$\offsetdistance$\end{tabular}}}}%
    \put(0,0){\includegraphics[width=\unitlength,page=5]{2d_hertzian_contact_boundary.pdf}}%
    \put(0.04953862,0.18161806){\makebox(0,0)[lt]{\lineheight{1.25}\smash{\begin{tabular}[t]{l}$master$\end{tabular}}}}%
    \put(0,0){\includegraphics[width=\unitlength,page=6]{2d_hertzian_contact_boundary.pdf}}%
  \end{picture}%
\endgroup%

%% file: fig/numerical_examples/2d_hertzian_contact/convergence_plot_hetzian_contact.tex
\pgfmathsetmacro{\p}{0.5}   
\pgfmathsetmacro{\plow}{0.3}   
\pgfmathsetmacro{\R}{10}   
\pgfmathsetmacro{\E}{250}   
\pgfmathsetmacro{\nu}{0.0}   
\pgfmathsetmacro{\x}{0.0}   
\pgfmathsetmacro{\bhigh}{2*((2*\R*\R*\p*(1-\nu*\nu)/(\E*pi)))^(1/2)} 
\pgfmathsetmacro{\blow}{2*((2*\R*\R*\plow*(1-\nu*\nu)/(\E*pi)))^(1/2)} 
\pgfmathsetmacro{\maxpressurehigh}{((4*\R*\p)/(pi*\bhigh*\bhigh))*(\bhigh*\bhigh-\x*\x)^(1/2)}   
\pgfmathsetmacro{\maxpressurelow}{((4*\R*\plow)/(pi*\blow*\blow))*(\blow*\blow-\x*\x)^(1/2)}   


\begin{tikzpicture}
    \begin{axis}[
      width=0.5\linewidth,
      xlabel={$1/h$},
      ylabel={$p_{max}$},
      grid=both,
      legend style={
        at={(0.5,-0.2)}, 
        anchor=north,    
        draw=black,       
        line width=0.5pt,
        font=\small,
        legend columns=2, 
        column sep=10pt,
      },
    ]
      \addplot[
        thick,
        mark=*,
        color=red
      ] table[
        x index=0,
        y index=1
      ]{fig/numerical_examples/2d_hertzian_contact/refinements_levels_maximum_contact_pressure_linear_kinem_p_0_3.txt};
      \addlegendentry{$\nurbssecondordertwodim{}-\lagrangefirstordertwodim{}: p = 0.3$}
      
      \addplot[
        domain=1:16,    
        thick,
        dashed,
        color=black
      ] {\maxpressurelow};
      \addlegendentry{$\text{linear theory}$} 

      \addplot[
        thick,
        mark=*,
        color=blue
      ] table[
        x index=0,
        y index=1
      ]{fig/numerical_examples/2d_hertzian_contact/refinements_levels_maximum_contact_pressure_linear_kinem_p_0_5.txt};
      \addlegendentry{$\nurbssecondordertwodim{}-\lagrangefirstordertwodim{}: p = 0.5$}
      
      \addplot[
        domain=1:16,    
        thick,
        dashed,
        color=black
      ] {\maxpressurehigh};
  
    \end{axis}
  \end{tikzpicture}
  

%% file: fig/numerical_examples/2d_hertzian_contact/pressure_plot_hertzian_contact_mesh_1_16.tex
\pgfmathsetmacro{\p}{0.3}   
\pgfmathsetmacro{\R}{10}   
\pgfmathsetmacro{\E}{250}   
\pgfmathsetmacro{\nu}{0.0}   
\pgfmathsetmacro{\b}{2*((2*\R*\R*\p*(1-\nu*\nu)/(\E*pi)))^(1/2)}


\begin{tikzpicture}
    \begin{axis}[
      width=0.5\linewidth,
      xlabel={$\text{radial position}$},
      ylabel={$\text{traction}$},
      grid=both,
      legend style={
        at={(0.5,-0.2)}, 
        anchor=north,    
        draw=black,       
        line width=0.5pt,
        font=\small,
        legend columns=2, 
        column sep=10pt,
      },
    ]

      \addplot[
        red,
        restrict x to domain=-1:1,
      ] table[
        col sep=comma,
        x=arc_length,
        y=norcontactstress_Magnitude
      ] {fig/numerical_examples/2d_hertzian_contact/resample_points_contact_linear_kinem_p_0_3_1_16.csv};
      \addlegendentry{$\text{numerical solution}$}
      
      \addplot[
        domain=-\b:\b,    
        thick,
        dashed,
        color=black
      ] {((4*\R*\p)/(pi*\b*\b))*(\b*\b-x*x)^(1/2)};
      \addlegendentry{$\text{linear theory}$}

      \addplot[
        domain=-1:1,    
        thick,
        color=black
      ] {0}; 
  
    \end{axis}
  \end{tikzpicture}
  

%% file: fig/numerical_examples/2d_hertzian_contact/refined_mesh.eps_tex
\begingroup%
  \makeatletter%
  \providecommand\color[2][]{%
    \errmessage{(Inkscape) Color is used for the text in Inkscape, but the package 'color.sty' is not loaded}%
    \renewcommand\color[2][]{}%
  }%
  \providecommand\transparent[1]{%
    \errmessage{(Inkscape) Transparency is used (non-zero) for the text in Inkscape, but the package 'transparent.sty' is not loaded}%
    \renewcommand\transparent[1]{}%
  }%
  \providecommand\rotatebox[2]{#2}%
  \newcommand*\fsize{\dimexpr\f@size pt\relax}%
  \newcommand*\lineheight[1]{\fontsize{\fsize}{#1\fsize}\selectfont}%
  \ifx\svgwidth\undefined%
    \setlength{\unitlength}{1038.75bp}%
    \ifx\svgscale\undefined%
      \relax%
    \else%
      \setlength{\unitlength}{\unitlength * \real{\svgscale}}%
    \fi%
  \else%
    \setlength{\unitlength}{\svgwidth}%
  \fi%
  \global\let\svgwidth\undefined%
  \global\let\svgscale\undefined%
  \makeatother%
  \begin{picture}(1,0.566787)%
    \lineheight{1}%
    \setlength\tabcolsep{0pt}%
    \put(0,0){\includegraphics[width=\unitlength]{refined_mesh.eps}}%
  \end{picture}%
\endgroup%

%% file: fig/numerical_examples/2d_hertzian_contact/pressure_plot_hertzian_contact_mesh_optimal_refinement.tex
\pgfmathsetmacro{\p}{0.3}   
\pgfmathsetmacro{\R}{10}   
\pgfmathsetmacro{\E}{250}   
\pgfmathsetmacro{\nu}{0.0}   
\pgfmathsetmacro{\b}{2*((2*\R*\R*\p*(1-\nu*\nu)/(\E*pi)))^(1/2)}


\begin{tikzpicture}[scale=0.6]
    \begin{axis}[
      xlabel={$\text{radial position}$},
      ylabel={$\text{traction}$},
      grid=both,
      legend style={
        at={(0.5,-0.2)}, 
        anchor=north,    
        draw=black,       
        line width=0.5pt,
        font=\small,
        legend columns=1, 
        column sep=10pt,
      },
    ]

      \addplot[
        red,
        restrict x to domain=-1:1,
      ] table[
        col sep=comma,
        x=arc_length,
        y=contact_normal_stresses_Magnitude
      ] {fig/numerical_examples/2d_hertzian_contact/resample_points_contact_linear_kinem_p_0_3_sides_1_center_1_25_cartesian_1_thick_1_4.csv};
      \addlegendentry{$\text{proposed method - selectively refined (940 \dofs{})}$}

      \addplot[
        black,
        restrict x to domain=-1:1,
      ] table[
        col sep=comma,
        x=arc_length,
        y=contact_normal_stresses_Magnitude
      ] {fig/numerical_examples/2d_hertzian_contact/resample_points_contact_linear_kinem_p_0_3_fem_reference_solution_1_32_lambda.csv};
      \addlegendentry{$\text{uniformly refined FEM (1.2M \dofs{})}$}

      \addplot[
        domain=-1:1,    
        thick,
        color=black
      ] {0}; 
  
    \end{axis}
  \end{tikzpicture}
  

%% file: fig/numerical_examples/2d_hertzian_contact/convergence_plot_penalty.tex
\begin{tikzpicture}
    \begin{axis}[
      xmode=log,
      width=\textwidth,
      xlabel={$\epsilon$},
      ylabel={$\Pi$},
      grid=both,
      legend style={
        at={(0.5,-0.2)}, 
        anchor=north,    
        draw=black,       
        line width=0.5pt,
        font=\small,
        legend columns=2, 
        column sep=10pt,
      },
    ]
      \addplot[
        thick,
        mark=*,
        color=red
      ] table[
        x index=0,
        y index=1
      ]{fig/numerical_examples/2d_hertzian_contact/epsilon_energy.txt};
  
    \end{axis}
\end{tikzpicture}
  

%% file: fig/numerical_examples/2d_hertzian_contact/convergence_plot_penalty_error.tex
\begin{tikzpicture}
    \begin{axis}[
      xmode=log,
      ymode=log,
      width=\textwidth,
      xlabel={$\epsilon$},
      ylabel={$e_E$},
      grid=both,
      legend style={
        at={(0.5,-0.2)}, 
        anchor=north,    
        draw=black,       
        line width=0.5pt,
        font=\small,
        legend columns=2, 
        column sep=10pt,
      },
    ]
      \addplot[
        thick,
        mark=*,
        color=red
      ] table[
        x index=0,
        y index=1
      ]{fig/numerical_examples/2d_hertzian_contact/epsilon_energy_error.txt};
  
    \end{axis}
\end{tikzpicture}
  

%% file: section/concluding_remarks.tex
\section{Conclusion} \label{section:concluding_remarks}

In this work, we proposed a novel discretization approach for contact 
problems with NURBS boundary layers and presented the algorithmic building blocks of the method. Starting 
from the mesh generation, a simplified strategy was introduced based on 
the construction of a boundary layer mesh that contains the NURBS \brep{} 
representation of the contact body. The remaining bulk domain 
of the contact body is described by a nonconforming Cartesian background mesh. 
This results in a configuration in which the boundary layer mesh overlaps 
the background discretization. 
Although several techniques can be employed to generate the boundary layer mesh, 
its construction via an offset operation proves to be a particularly natural 
and intuitive choice. Three approaches 
for generating offset geometries that preserve the original 
parametric representation of the boundary layer were compared. Special cases may arise when the local radius 
of curvature is smaller than the offset thickness, which can lead to self-intersections; 
the treatment of such situations will be addressed in future work.

To couple the overlapping meshes, a standard mortar-type method was implemented. 
Numerical experiments demonstrated its usability, showing that the method 
remains stable when both meshes share identical material parameters, independently 
of the mesh size ratio. Several numerical examples further validated the proposed 
embedded mesh approach. In addition, the spatial convergence of the 
discretization were verified for contact problems. Finally, a three-dimensional 
large-deformation contact example was presented.

Overall, the proposed discretization approach constitutes an attractive framework for contact 
problems, as it combines the high continuity and smoothness of NURBS 
B-reps at contact interfaces with a straightforward and flexible mesh generation strategy for the bulk domain.

Future work will focus on coupling techniques that are fully insensitive to 
mesh locking, such as Nitsche’s method. Furthermore, ghost element stabilization 
techniques for configurations involving small cut-element volumes will be 
investigated. Finally, the capabilities of the proposed approach will be 
explored in the context of more complex three-dimensional geometries.

%% file: section/acknowledgements.tex
\section*{Acknowledgements}

This work was supported by the Deutsche Forschungsgemeinschaft (DFG) - Project number 446494172.

%% file: section/declaration_generative_ai.tex
\section*{Declaration of generative AI and AI-assisted technologies in the manuscript preparation process.}

During the preparation of this work the author(s) used ChatGPT in order to improve 
the readability and language of the present script. After using this tool, the author(s) reviewed 
and edited the content as needed and take(s) full responsibility for the content of the published article.